%% file: main.tex
\documentclass[acmsmall,nonacm]{acmart}
\usepackage{url}
\usepackage{amsmath,amsfonts}
\usepackage{mathtools}
\usepackage{xcolor}
\usepackage{soul}
\usepackage{tabularx}
\usepackage{booktabs}
\usepackage{subcaption}
\usepackage{array}
\usepackage{multirow}
\usepackage{graphicx}
\usepackage{float}
\usepackage{pgfplots}
\usepackage{filecontents}

\pgfplotsset{compat=1.15}
\usepgfplotslibrary{statistics, colorbrewer, groupplots}
\usetikzlibrary{patterns}

\definecolor{plotColor1}{HTML}{2B3840}
\definecolor{plotColor2}{HTML}{D90718}
\definecolor{plotColor3}{HTML}{2A871E}
\definecolor{plotColor4}{HTML}{F2790F}
\definecolor{plotColor5}{HTML}{AF34DB}

\makeatletter
\tikzset{nomorepostaction/.code=\let\tikz@postactions\pgfutil@empty}
\makeatother

\newcommand{\ctext}[3][RGB]{%
  \begingroup
  \definecolor{hlcolor}{#1}{#2}\sethlcolor{hlcolor}%
  \hl{#3}%
  \endgroup
}
\newcommand{\green}[1]{\ctext[RGB]{168,221,181}{#1}}

\newcommand{\sr}{\textsc{Select-And-Rank}}

\newcommand{\trecdl}{\textsc{TREC-DL}}
\newcommand{\core}{\textsc{Core17}}
\newcommand{\clueweb}{\textsc{ClueWeb09}}

\newcommand{\berts}{\textsc{BERT-3S}}
\newcommand{\doclabeled}{\textsc{Doc-Labeled}}
\newcommand{\bertcls}{\textsc{BERT-CLS}}
\newcommand{\matchpyramid}{\textsc{MatchPyramid}}
\newcommand{\copacrr}{\textsc{CO-PACRR}}
\newcommand{\convknrm}{\textsc{CONV-KNRM}}
\newcommand{\tklsmall}{\textsc{TKL-2k}}

\newcommand{\sratt}{\textsc{S\&R-ATT}}
\newcommand{\srlin}{\textsc{S\&R-LIN}}
\newcommand{\plrnd}{\textsc{PL-RND}}
\newcommand{\plbert}{\textsc{PL-BERT}}
\newcommand{\pllstm}{\textsc{PL-LSTM}}
\newcommand{\plbm}{\textsc{PL-BM25}}
\newcommand{\plsem}{\textsc{PL-SEM}}

\setcopyright{acmcopyright}
\copyrightyear{2018}
\acmYear{2018}
\acmDOI{10.1145/1122445.1122456}

\acmConference[Woodstock '18]{Woodstock '18: ACM Symposium on Neural
  Gaze Detection}{June 03--05, 2018}{Woodstock, NY}
\acmBooktitle{Woodstock '18: ACM Symposium on Neural Gaze Detection,
  June 03--05, 2018, Woodstock, NY}
\acmPrice{15.00}
\acmISBN{978-1-4503-XXXX-X/18/06}

\begin{document}

\title{Extractive Explanations for Interpretable Text Ranking}

\author{Jurek Leonhardt}
\affiliation{
  \institution{L3S Research Center}
  \city{Hannover}
  \country{Germany}
}
\email{leonhardt@L3S.de}

\author{Koustav Rudra}
\affiliation{
  \institution{Indian Institute of Technology (ISM) Dhanbad}
  \city{Dhanbad}
  \country{India}
}
\email{koustav@iitism.ac.in}

\author{Avishek Anand}
\affiliation{
  \institution{Delft University of Technology}
  \city{Delft}
  \country{Netherlands}
}
\email{avishek.anand@tudelft.nl}


\begin{abstract}
Neural document ranking models perform impressively well due to superior language understanding gained from pre-training tasks. However, due to their complexity and large number of parameters, these (typically transformer-based) models are often non-interpretable in that ranking decisions can not be clearly attributed to specific parts of the input documents.

In this paper we propose ranking models that are inherently interpretable by generating explanations as a by-product of the prediction decision. We introduce the \sr{} paradigm for document ranking, where we first output an explanation as a selected subset of sentences in a document. Thereafter, we solely use the explanation or selection to make the prediction, making explanations first-class citizens in the ranking process. Technically, we treat sentence selection as a latent variable trained jointly with the ranker from the final output. To that end, we propose an end-to-end training technique for \sr{} models utilizing \emph{reparameterizable subset sampling} using the \emph{Gumbel-max trick}.

We conduct extensive experiments to demonstrate that our approach is competitive to state-of-the-art methods. Our approach is broadly applicable to numerous ranking tasks and furthers the goal of building models that are \emph{interpretable by design}. Finally, we present real-world applications that benefit from our sentence selection method.
\end{abstract}



\maketitle
\input{intro}
\input{related-work}
\input{approach}
\input{setup}
\input{results}
\input{applications}
\input{conclusion}

\begin{acks}
This work is supported by the European Union -- Horizon 2020 Program under the scheme ``INFRAIA-01-2018-2019 -- Integrating Activities for Advanced Communities'', Grant Agreement n.871042, ``SoBigData++: European Integrated Infrastructure for Social Mining and Big Data Analytics'' (\url{http://www.sobigdata.eu}).

Further, this work is supported in part by the Science and Engineering Research Board, Department of Science and Technology, Government of India, under Project SRG/2022/001548. Koustav Rudra is a recipient of the DST-INSPIRE Faculty Fellowship [DST/INSPIRE/04/2021/003055] in the year 2021 under Engineering Sciences.
\end{acks}

\bibliographystyle{ACM-Reference-Format}
\bibliography{references}

\newpage
\appendix
\input{appendix}

\end{document}

%% file: intro.tex
\section{Introduction}
\label{sec:intro}
Information prioritization is an essential and important problem to reduce information overload in a large multitude of Web based tasks like \textit{question-answering}~\cite{roy2021question}, \textit{fact verification}~\cite{thorne18Fever} and \textit{conversational search}~\cite{anand2020conversational}. Prioritizing information relies on retrieving a small set of highly relevant knowledge units from a large source of world knowledge contained in unstructured text collections with web and textual knowledge bases like Wikipedia that contain documents and articles. The most common way to address information prioritization is to cast it as a ranking problem, i.e.\ inducing a ranking over all documents in the collection and inspect only the top-ranked document(s) to satisfy the information need. This is called the \emph{document ranking problem} and is a central task in web search and information retrieval. The objective of the document ranking task is to rank documents relevant to a user-specified query. Consequently, tasks that require access to world knowledge rely on effective document ranking techniques for superior performance. This makes document ranking one of the primitive operations for a large number of knowledge-intensive tasks. Recent advances in document ranking have been dominated by over-parameterized contextual models based on tuning pre-trained contextual models like BERT~\cite{dai2019deeper,macavaney2019cedr,yilmaz2019cross, leonhardt2022efficient}, indicating that better language understanding~\cite{vanaken2019how} leads to better document understanding. However, such models are inherently non-interpretable, as they automatically extract latent and complex query-document features from large training sets, leading to opaque decision-making that limits understanding in case of failures or undesirable results. In this paper we focus on proposing interpretable models for document ranking that have a wide utility in numerous ranking tasks ranging from web search and question answering over fact checking to argument and entity retrieval.

\begin{figure}
    \small
    \begin{tabularx}{\linewidth}{X}
        \toprule
        \textbf{Result Document} \\
        \midrule
        \textit{What makes Bikram yoga unique is its focus on practicing yoga in a room heated to 105 degrees Fahrenheit with 40 percent humidity. In Bikram yoga, be prepared to sweat profusely and come armed with a towel and lots of water. To practice Bikram at home, you'll need a space heater and access to the pose sequence. \green{On a general basis, you need to hold the yoga poses for about 10-12 breaths.} \green{With practice, you can also go up to 30 breaths.} We chatted for a few moments, and found that we came to completely different conclusions. She finds Bikram more difficult because of the intense heat (about 5-10 degrees hotter than a hot vinyasa class) and lack of breaks in the standing series. That is why Bikram is easier for me. It will help you hold the pose for around 3 minutes. It is best to count the time in breaths (one breath cycle is one deep inhalation followed by complete exhalation).} \\
        \bottomrule
    \end{tabularx}
    \caption{Sentence selection for the query \texttt{how long to hold bow in yoga}, taken from a document in the MS MARCO corpus marked as relevant by human annotators.}
    \label{fig:sample_example}
\end{figure}

Although interpretability of machine learning models has been popular, there are few approaches for document ranking, mostly focusing on post-hoc interpretability of rankers~\cite{singh2018posthoc,singh2019exs}. However, post-hoc approaches are limited in the sense that their explanations might not accurately reflect the true rationale underlying the model decisions~\cite{rudin2019stop}. Further, the collection of ground-truth data for evaluating explanations is often hindered by human bias~\cite{lage2019evaluation}, making the evaluation of interpretability techniques difficult. Due to this, post-hoc methods are unreliable and one cannot be sure about the correctness of the explanations.

Unlike post-hoc approaches, we are specifically interested in ranking models that are interpretable by design. We argue that an interpretable ranking model should help us understand which sentences or passages in the document are used for the relevance estimation. In this paper, we present a document ranking model where each prediction can be \emph{unambiguously attributed} to a reason or rationale that is both accurate and human-understandable. We define explanations as extractive pieces of text from the input document. An example explanation is shown in Figure~\ref{fig:sample_example}, where the highlighted sentences serve as an explanation for the relevance of the document. For a set of additional examples, we refer the reader to Table~\ref{tab:trecdl_examples} and Table~\ref{tab:fever_examples}.

\begin{figure}
    \centering
    \includegraphics[width=0.3\linewidth]{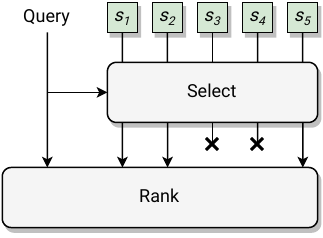}
    \caption{The \sr{} approach. The document is split into sentences $s_i$. The \textit{selector} assigns a score to each sentence. The scores determine which of the sentences are selected as input for the \textit{ranker}.}
    \label{fig:framework}
\end{figure}
This paper proposes a two-stage approach, which we refer to as \sr{}, for modeling long documents that addresses the above limitations. In the \textbf{selection phase}, we extract relevant sentences given a query. In the \textbf{ranking phase}, we perform the relevance estimation only on the extracted evidence. Our idea is based on the observation that not all sentences in a document are relevant; instead, the document's relevance signals are typically sparse (refer to Figure~\ref{fig:framework}). The selection phase essentially acts as a noise removal mechanism, resulting in a succinct, query-based document representation. As an added advantage, our sentence selection allows for choosing a concise query-based document representation as input into size limited models like BERT, in contrast to other heuristic truncation approaches~\cite{dai2019deeper}.

Within our modular framework, we consider \textit{joint models} that are trained end-to-end with gradient descent. Specifically, we allow the user to regulate the sparsity by setting the number of sentences $k$ to be selected. The selection is akin to sampling from a latent distribution over sentences in a document. We use a parameterized model to output such a distribution and apply the \textit{Gumbel-max} trick. Finally, we use \textit{relaxed subset sampling} to enforce the user-specified sparsity $k$, i.e.\ the number of sentences to be selected for the summary or explanation. This allows us to approximate \emph{hard masking}, i.e.\ the multiplication of the input with a boolean mask in order to remove certain parts, by using \emph{soft masking} (or \emph{continuous masks}), where a similar result is achieved, but the process remains fully differential and thus trainable end-to-end.

We conduct extensive empirical evaluation over three document ranking datasets -- \trecdl{}, \core{} and \clueweb{}. Our intention is not to achieve the best performance in document ranking. Instead, we aim to present a ranking model that is interpretable without compromising ranking performance. Firstly, we find that query-specific sparse document representation by sentence selection can improve the task performance over heuristic sentence selection approaches~\cite{dai2019deeper}. Secondly and more strikingly, our \sr{} models (with 20 selected sentences) perform on par with and sometimes outperform other document modeling approaches that model the entire document. We further show how \sr{} can be used to explain the decisions of BERT rankers that operate only on small parts of the input document.

Additionally, we conduct experiments on twelve diverse datasets provided by the BEIR benchmark~\cite{thakur2021beir}, including tasks such as passage ranking, fact checking and argument retrieval. Our experiments show that \sr{} models also perform well on corpora consisting of shorter documents and provide more interpretable results compared to standard approaches like BERT.

Finally, we highlight the utility of \sr{} to human users through a user study and present real-world applications by showing how the extractive explanations can be used to uncover model bias or bugs and illustrating their utility in search engines.

%% file: related-work.tex
\section{Related Work}
\label{sec:related-work}
We divide the related work into two major categories -- \textit{text ranking models} and \textit{interpretability approaches in IR}.

\subsection{Ranking Models for Text}
Classical approaches in information retrieval for ad-hoc document retrieval are probabilistic query likelihood (QL)~\cite{lavrenko2017relevance} and term frequency-based models such as BM25~\cite{robertson2009probabilistic} and BM25P~\cite{muntean2020weighting} models. More recently, neural models have entered the field of IR. Common approaches include semantic representation learning~\cite{shen2014latent,huang2013learning,shen2014learning}, query-document cross-interactions~\cite{xiong2017end,guo2016deep,pang2016study,nie2018empirical,nie2018multi} or the exploitation of positional information~\cite{hui2017pacrr,hui2018copacrr,mcdonald2018deep}. \cite{mitra2017learning} employs a combination of the aforementioned approaches. Nowadays, contextual self-attention-based models such as BERT achieve state-of-the-art performance in ranking tasks. \citet{macavaney2019cedr} were the first to replace static word embeddings in existing document retrieval models by contextualized token embeddings output by BERT.

Since self-attention models have quadratic time complexity with respect to the input length, work has been done to address this limitation by splitting the input documents into either passages~\cite{dai2019deeper,wu2020leveraging,rudra2020distant} or sentences~\cite{yilmaz2019cross} and subsequently labeling those. \doclabeled{}~\cite{dai2019deeper} uses passage-level relevance scores from a fine-tuned BERT model to obtain relevance scores. However, this approach assumes that all passages inherit their relevance from the corresponding document, which might be problematic. \berts{}~\cite{yilmaz2019cross} works similarly, but on a per-sentence level using a cross-domain transfer model. This leads to a substantially slow inference.

Recently, researchers also focused on the efficiency aspect of document and passage retrieval along with the performance aspect. The major bottleneck of existing language model-based ranking models is the processing time required during inference phase. Some of the works use dual-encoder based models to alleviate the need of document processing during inference~\cite{khattab2020colbert,lin2021batch,hofstaetter2021intra,althammer2022parm}. \citet{zhuang2021tilde} proposed a term-independent likelihood model for passage ranking that relies on both query and document likelihood to rank the documents. This approach pre-computes and stores the likelihood of terms and thus removes the requirement of running deep language models during query processing. In recent times, some studies have focused on the mitigation of positional bias in passage ranking~\cite{hofstatter2021mitigating} and robustness against misspellings in document retrieval~\cite{sidiropoulos2022analysing}. However, these studies do not focus on the interpretability aspects of the ranking models. We believe that the selection model in our \sr{} approach is modular and can be used in any text ranking transformers.

\subsection{Interpretability of Ranking Models}
Interpretability of ranking models focuses on building models that either can be \textit{analysed for interpretability in a post-hoc fashion} or are \textit{interpretable by design}. However, post-hoc approaches suffer from the limitation that their explanations might not accurately reflect the true rationale underlying the model decisions~\cite{rudin2019stop}.

Different from classical feature selection, our aim is to select features from a document given a query, that is, we want to dynamically select sentences from a document based on the input query. Such instance-wise feature selection has been explored in the machine learning literature~\cite{yoon2018invase}, however, their applicability to modeling documents is limited.

In NLP, similar models have been studied for ensuring interpretability by design~\cite{lei2016rationalizing,lehman2019inferring}. In~\cite{li2019teach}, the authors use sentence selection to mimic human reading behavior to estimate the relevance of a document to a query. These works mainly differ in how they perform end-to-end training. Training has been done using REINFORCE~\cite{lei2016rationalizing,li2019teach}, actor-critic methods~\cite{yoon2018invase}, pipeline approaches~\cite{zhang2021explain} or re-parameterization tricks~\cite{bastings2019interpretable}. \citet{lehman2019inferring} use a decoupled rationale generator and predictor. In~\cite{zhong2019fine}, additional human annotations are used for task supervision. However, we do not have any explicit training data to train the selector network and rather use the task supervision signal to update its parameters. Finally, \cite{li2021keybld} pursues an idea similar to ours, however, the authors only use non-trainable selectors and do not consider an end-to-end trainable model.

There exists lots of work on post-hoc analysis of trained neural models on different tasks. Such kinds of analysis use different methods like probing tasks~\cite{singh2020bertnesia}, attention weights~\cite{bahdanau2014neural, cheng2016long, martins2016softmax, cui2016attention, xu2015multi, yang2016hierarchical} or state activation~\cite{hermans2013training, karpathy2015visualizing,li2016understanding}. In previous works, researchers tried to learn attention weights of different tokens to judge their contribution towards a task prediction and mark the tokens that got higher attention scores as rationales or explanations. However, recent studies showed that attention weights are not explanations~\cite{jain2019attention,wiegreffe2019attention} and models are able to maintain the same prediction accuracy even in the absence of those tokens. Sometimes, tokens that get high attention scores do not correlate well with the human annotated rationales. As recent language models are contextual, it is very difficult to disentangle the importance of token inputs. Finally, there has been recent work on devising decoy datasets to measure the utility of explanation methods for NLP models~\cite{idahl2021towards}. Recent approaches also tried to de-bias masked language models with automated bias prompts~\cite{guo2022auto}. A major bottleneck of interpretability studies is the availability of annotated benchmark datasets. In recent times, many interpretability evaluation benchmark datasets have been introduced for neural NLP tasks~\cite{deyoung2020eraser,wang2022fine}. In this paper, our objective is to extend this interpretability aspect towards the document ranking task.

For the ranking task, most of the work has focused on post-hoc interpretability of text rankers~\cite{singh2019exs,singh2020model,fernando2019study,volske2021towards} and learning-to-rank models~\cite{singh2018posthoc,singh2021extracting}. In contrast, our \sr{} models are interpretable by design. The closest to our work is \cite{hofstaetter2021intra}, where the authors use cascading rankers after retrieval. However, cascading rankers differ from our approach in the style of optimization and the type of interpretability they provide.

%% file: approach.tex
\section{\sr{}}
\label{sec:approach}
In this section we formally define the problem of document ranking (Section~\ref{sec:problem}). We then give a high-level overview of our \sr{} framework that aims to generate an \textit{extractive sentence-level summary} from the document prior to ranking (cf.\ Figure~\ref{fig:framework}). Finally, we present our algorithmic contribution that aims to train the \textit{selector} and \textit{ranker} models using gradient-based optimization in a joint manner (Section~\ref{sec:end_to_end}).

\subsection{Problem Statement}
\label{sec:problem}
The usual ranking pipeline consists of two stages: First, given a query, an inexpensive term-frequency based retriever retrieves a set of documents from the complete, usually very large, collection. Afterwards, a more involved, expensive model \textbf{re-ranks} the result of the first-stage retrieval.

Our objective is to learn a parameterized model for document re-ranking. Specifically, given a training set of triples $\left\{q^{(i)}, d^{(i)}, y^{(i)} \right\}_{i=1}^{N}$, where $q^{(i)}$ is a query, $d^{(i)}$ is a document and $y^{(i)}$ is a relevance label, our goal is to learn a model that predicts relevance scores $\hat{y} \in \mathbb{R}$ for query-document pairs $(q, d)$. We denote the set of documents retrieved in the first stage for the query $q$ as $D_q$. The resulting predictions are then used to obtain a ranking of all documents $d \in D_q$. Finally, the rankings corresponding to all queries are evaluated using appropriate ranking metrics.

We model each document as a sequence of sentences, i.e. $d = \left(s_1, s_2, ..., s_{|d|} \right)$.
Our \sr{} approach assumes that only a subset of the constituent sentences actually contribute towards the relevance estimation. Based on this assumption, the model consists of two components: The \textbf{selector} $\Psi$ defines a distribution $p \left( s \,|\, q, d \right)$ over sentences in $d$ given the input query $q$, encoding the relevance of the sentence given the query. This distribution is used to select an extractive, query-dependent summary $\hat{d} \subseteq d$. The \textbf{ranker} $\Phi$ is a relatively involved relevance estimation model that generates a relevance label $\hat{y}$ given the query and an extractive document summary $\hat{d}$, thus taking only parts of the document into account.

The selector $\Psi$ is a parameterized model that takes the query and sentences as input and outputs a score or weight $w_i$ for each sentence, representing its relevance to the query, i.e.
\begin{equation*}
    \left(w_1, ..., w_{|d|} \right) = \Psi \left(q, d \right).
\end{equation*}
The logit weights $w_i$ are normalized using the softmax function, defining a distribution over the sentences:
\begin{equation*}
    p \left(s_i \mid q, d \right) = \frac{\exp \left(w_i \right)}{\sum_{j=1}^{|d|} \exp \left(w_j \right)}
\end{equation*}
Using this distribution, a document summary $\hat{d} \subseteq d$ is created as a subset of the document's sentences based on the selector's scores, i.e.\ by dropping some of the lower scoring sentences. The ranker takes as input the query and the document summary to compute the query-document relevance $\hat{y} = \Phi \left(q, \hat{d} \right)$.

Since the selector and ranker are in principle independent models, it is possible to train them in one of two ways:
\begin{enumerate}
    \item Both models are trained separately; the selector is trained to extract a summary from a document with respect to a query, while the ranker is trained on a ranking dataset. The models are then applied consecutively to a query-document pair. We refer to this family of approaches as \textit{pipeline approaches}.
    \item The models are trained jointly in an end-to-end fashion, where the gradients are propagated directly from the final outputs back to the selector network. Since this approach includes a non-differentiable selection operation ($\text{arg max}$), it requires approximated differentiable subset sampling.
\end{enumerate}
In this paper we analyze and compare the approaches above; furthermore, we implement different selector models and compare them. Section~\ref{sec:pipeline} describes the pipeline approach, Section~\ref{sec:end_to_end} describes the end-to-end approach.

\subsection{Pipeline Approach}
\label{sec:pipeline}
In this section we apply the \sr{} framework in the aforementioned pipeline setting. Concretely, this means that the selector and ranker are trained independently of each other. For sentence selection, we consider multiple approaches from simple term matching to rather complex auto-regressive language models:
\begin{enumerate}
    \item \textbf{Term-matching-based selectors}: We use tf-idf scores between the query $q$ and sentences $s_i$ to determine the best sentences.
    \item \textbf{Embedding-based selectors}: We use semantic similarity scores between the query $q$ and sentences $s_i$ to determine the best sentences. Both the query and sentence are represented as average over the constituent word embeddings.
    \item \textbf{Neural non-contextual selectors}: We build a neural network to define a distribution over the sentences $s_i$.
    \item \textbf{Contextual selectors}: We use BERT to define a distribution over the sentences $s_i$.
\end{enumerate}
Term- and embedding-based selectors are non-parameterized. As the other selectors (neural and contextualized models) are parameterized and need to be trained, we follow a transfer learning approach and use the MS MARCO passage re-ranking dataset~\cite{nguyen2016ms} to train each selector on a passage ranking task. Specifically, the models learn to predict a relevance score given a query and a passage (or sentence). This task in itself is very similar to document summarization, supported by the fact that the passages in this particular dataset were created by splitting documents. We do not consider summarized documents in the training phase of the ranker. During inference, the pipeline approach may be described as follows: The selector is applied to the query and document, outputting a score for each sentence in the document. Along with the query, the $k$ highest scoring sentences then form the input to the ranker, maintaining their original order as in the source document. The ranker outputs the final score $\hat{y}$ that is used to rank the document.

\subsection{End-to-End Approach}
\label{sec:end_to_end}
Existing approaches rely on sampling from a stochastic distribution using the REINFORCE algorithm, resulting in a boolean mask over the sentences. An alternative way to achieve end-to-end training instead is by allowing a continuous mask over the sentences. This is akin to using a soft-attention mechanism that is arguably easier to train. However, this approach does not allow for a reduction of the input sequence length, which can be problematic, especially with Transformer-based rankers. Additionally, during inference, one would still need a selection of $k$ sentences given a soft-selection model. This in particular is ineffective, given that soft-selection models still rely on all sentences for more effective predictions. We therefore propose an approach based on the Gumbel-max trick~\cite{maddison2014sampling}, that enables gradient flow in models where discrete variables must be sampled.

\subsubsection{Feature Attribution and Masking}
In interpretability, explaining the model output in terms of the input features is called \emph{feature attribution}. Feature attribution (or \emph{saliency}) methods create explanations in terms of input feature importance for individual predictions. In our case of text ranking, a \emph{feature} refers to a subset of the input, such as a sentence or a passage in the document. Feature attributions can be \emph{soft} or \emph{hard}. Soft attributions are scalar values representing importance that are assigned to each input feature. The output of an attribution method is typically a vector of the same dimension as the input with either scalar or boolean values, called a \emph{mask}. A soft mask is an output of soft attributions that can be viewed as a distribution of word-level or sentence-level relevance over the document text. However, it has been shown that, for large input length or large input spaces, humans find it hard to make sense of soft masks and prefer boolean or hard masks instead. Hard masks are sparse and have no ambiguity or uncertainty in terms of the presence or absence of a word or sentence in an explanation.

\subsubsection{The Gumbel-Max Trick}
The Gumbel-max trick provides a simple and efficient way to parameterize a discrete distribution and draw samples from it. Let $X$ be a random variable. We wish to parameterize a categorical distribution such that $P \left(X=i \right) \propto w_i$, where $w_i$ is a weight associated to the $i$-th category. Using the Gumbel-max trick, we can simply draw a sample as
\begin{equation*}
    X = \text{arg max}_i \left(\log w_i + g_i \right),
\end{equation*}
where $g_i = -\log \left(-\log u_i \right)$ is called a \textit{Gumbel random variable} and $u_i \sim \text{Uniform} \left(0, 1 \right)$. The resulting sample is parameterized by the weights $w$. In order to completely relax the sampling process and allow for the propagation of gradients (i.e.\ end-to-end training), the trick is commonly extended, replacing $\text{arg max}$ with softmax (\textit{Gumbel-softmax trick}). In detail, the Gumbel-softmax estimator gives an approximate one-hot sample $y$ with
\begin{equation*}
    y_i = \frac{\exp \left( \left(\log w_i + g_i \right) / t \right)}{\sum_{j=1}^{k} \exp\left(\left(\log w_j + g_j \right) / t \right)} \quad \text {for } i = 1, ..., k,
\end{equation*}
where $t$ is a temperature. By using the Gumbel-softmax estimator, one can generate samples $y = (y_1, ..., y_k)$ to approximate the categorical distribution. Furthermore, as the randomness $g$ is independent of $w$, which is usually defined by a set of parameters, the reparameterization trick can be used to optimize the model's parameters using standard backpropagation algorithms.

\subsubsection{Relaxed Subset Sampling}
\label{sec:subset_sampling}
Since we are interested in sampling a subset, i.e. drawing a number of samples (in our case sentences) without replacement, we employ a relaxed subset sampling algorithm proposed in~\cite{xie2019reparameterizable} that makes use of the aforementioned Gumbel-max trick. Let a set of items $x_1, ..., x_n$ have associated weights $w_i$ and Gumbel variables $g_i$ as above. In order to sample a subset, a \textit{Gumbel-max key} 
\begin{equation*}
    \hat{r}_i = \log w_i + g_i
\end{equation*}
is computed for each item. Since $\hat{r}_i$ is a monotonic transformation of $w_i$ (fixing $u_i$), a relaxed subset sample of the items can be drawn by applying a relaxed top-$k$ procedure directly on $\hat{r}$. The procedure proposed in~\cite{plotz2018neural} defines
\begin{equation*}
    \alpha_i^1 := \hat{r}_i \qquad \alpha_i^{j+1} := \alpha_i^j + \log \left(1 - p \left(a_i^j = 1 \right) \right),
\end{equation*}
where $p \left(a_i^j = 1 \right)$ is the expectation of the distribution
\begin{equation*}
    p \left(a_i^j = 1 \right) = \frac{\exp \left(\alpha_i^j / t \right)}{\sum_{m=1}^n \exp \left(\alpha_m^j / t \right)}
\end{equation*}
and $t$ is a temperature. Finally, a relaxed $k$-hot vector is computed as
\begin{equation*}
    v = \left(v_1, ..., v_n \right) \qquad v_i = \sum_{j=1}^k p \left(a_i^j = 1 \right).
\end{equation*}

\subsubsection{Training and Inference}
In order to train both selector and ranker jointly, we make use of the relaxed subset sampling as described in Section~\ref{sec:subset_sampling}. We start by obtaining query and document representations $q^{\text{emb}}$ and $d^{\text{emb}}$ from a shared embedding $E$:
\begin{equation*}
    q^{\text{emb}} = E \left(q \right) \qquad d^{\text{emb}} = E \left(d \right)
\end{equation*}
The selector then operates on these representations and computes a weight $w_i$ for each sentence $s_i$, i.e. 
\begin{equation*}
    \left(w_1, ..., w_{|d|}\right) = \Psi \left(q^{\text{emb}}, d^{\text{emb}} \right).
\end{equation*}
We now draw a relaxed $k$-hot sample $\hat{w}$ (cf.\ Section~\ref{sec:subset_sampling}) from the set of sentences using the weights $w$ and a temperature $t$ as
\begin{equation*}
    \left( \hat{w}_1, ..., \hat{w}_{|d|} \right) = \mathtt{SubsetSample} \left(w, k, t \right).
\end{equation*}
Finally, the document summary $\hat{d}$ is selected as the $k$ highest scoring sentences according to $\hat{w}$. The ranker only operates on the document summary $\hat{d}$ and discards all other sentences. This means that the ranker needs to assemble its new inputs during the training process. The ordering of the sentences is maintained irrespectively of their scores. Since our goal is to train both models jointly, we have to preserve the gradients of the selector (i.e.\ $\hat{w}$) by combining them with the ranker inputs in a differentiable way. Let $t_1, ..., t_n$ denote the embedded tokens corresponding to some sentence $s_i \in \hat{d}$. We compute the actual input tokens for the ranker as
\begin{equation*}
    \hat{t}_j = t_j \odot w_i.
\end{equation*}
Note that $t_j$ is a vector and $w_i$ is a scalar. We use $\odot$ to denote the multiplication of each element in the vector with the scalar. This multiplication changes the input representations, which is undesirable. We mitigate this by making use of the \textit{straight-through} estimator~\cite{bengio2013estimating}. The idea is to use $\hat{t}_j$  \textbf{only} during the backward pass, i.e.\ when computing gradients. The forward pass simply ignores $w_i$ and considers just $t_j$.

During inference, we do not use relaxed subset sampling. Instead, we simply select the $k$ highest scoring sentences.

\subsubsection{Selectors}
\label{sec:selectors}
\begin{figure}
    \begin{subfigure}{.49\linewidth}
      \centering
      \includegraphics[width=0.5\linewidth]{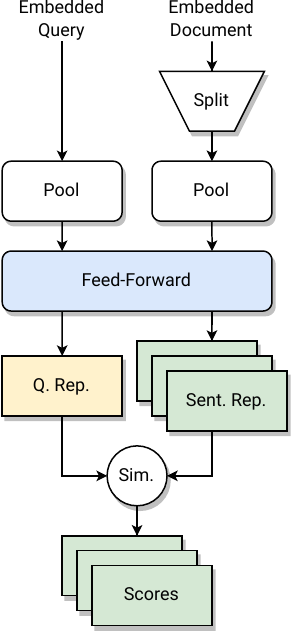}
      \caption{Linear selector}
      \label{fig:linear_selector}
    \end{subfigure}
    \begin{subfigure}{.49\linewidth}
      \centering
      \includegraphics[width=0.5\linewidth]{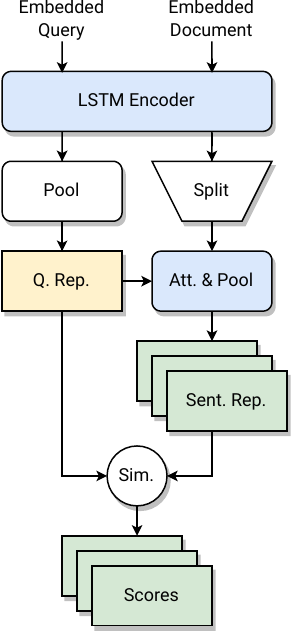}
      \caption{Attentive LSTM selector}
      \label{fig:lstm_selector}
    \end{subfigure}
    \caption{The selectors used in the end-to-end approach. The linear selector represents sentences as the average of their embedded tokens and applies a linear layer. The LSTM selector uses a simple attention mechanism.}
    \label{fig:selectors}
\end{figure}
In this section we present the selector networks we use in the end-to-end approach. Figure~\ref{fig:selectors} illustrates the two selectors.

\paragraph{Linear Selector}
The \textbf{linear selector} (Figure~\ref{fig:linear_selector}) simply represents a sequence as the average of its token embeddings. Query and sentence representation are fed through a single feed-forward layer. The score is computed as the dot product.

\paragraph{Attentive LSTM Selector}
The \textbf{attentive LSTM selector} (Figure~\ref{fig:lstm_selector}) is inspired by the QA-LSTM model proposed in~\cite{tan2016improved}. Query and document are passed through a shared, bidirectional many-to-many LSTM. On the query side, we obtain the representation $\hat{q}$ by max-pooling over all LSTM outputs. On the document side, we split the LSTM outputs into sequences that correspond to the sentences. Let $h_j^i$ denote the LSTM output corresponding to the $j$-th token of the $i$-th sentence. Prior to max-pooling, we apply a simple token-level attention mechanism as
\begin{align*}
    m_j^i &= W_1 h_j^i + W_2 \hat{q} \\
    \hat{h}_j^i &= h_j^i \exp \left(W_3 \tanh \left(m_j^i \right) \right)
\end{align*}
where $W_1$, $W_2$ and $W_3$ are trainable parameters. We finally compute the sentence representation $\hat{s}_i$ by max-pooling over all $\hat{h}_j^i$. The score of each sentence is the cosine similarity of its representation to the query representation.

\subsection{Ranker}
\label{sec:ranker}
Throughout all of our experiments, we use a $\text{BERT}_{\text{base}}$ model as the ranker. The model is fine-tuned according to~\cite{nogueira2019passage}: For a query $q = \left(q_1, ..., q_n \right)$ and a document summary $\hat{d} = \left(\hat{d}_1, ..., \hat{d}_m \right)$ (produced by the selector), where $q_i$ and $\hat{d}_i$ denote input tokens, the ranker input is
\begin{equation*}
    \label{eq:bert_in}
    \texttt{[CLS]}, q_1, ..., q_n, \texttt{[SEP]}, \hat{d}_1, ..., \hat{d}_m, \texttt{[SEP]}.
\end{equation*}
We impose a limit of $512$ input tokens, i.e. $n + m + 3 \leq 512$. Consequently, long documents are truncated to fit within this limit. We take the output $o$ of BERT, which corresponds to the \texttt{[CLS]} input token, and discard the rest. It is fed through dropout and a single feed-forward layer that outputs the final score
\begin{equation*}
    \hat{y} = \sigma \left(Wo + b \right).
\end{equation*}
$W$ and $b$ denote the trainable parameters of the feed-forward layer and $\sigma$ is the sigmoid function.

%% file: setup.tex
\section{Experimental Setup}
\label{sec:setup}
In this section we describe our datasets, baselines and evaluation procedure.

\subsection{Datasets}
\label{sec:datasets}
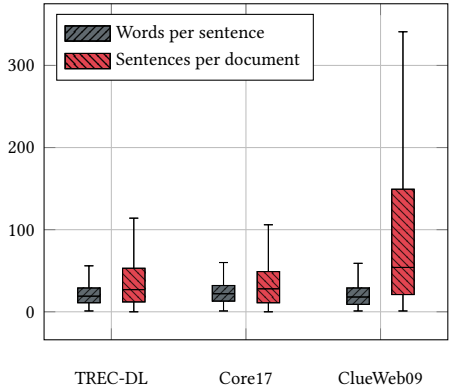
\begin{figure}
    \centering
    \input{plots/trec_stats}
    \caption{The distribution of document and sentence lengths of the TREC datasets. Outliers omitted.}
    \label{fig:dataset_stats}
\end{figure}
First, we consider the following diverse TREC datasets with varying properties:
\begin{enumerate}
    \item \textbf{\trecdl{}}: The \trecdl{} document ranking task uses the MS MARCO document corpus. We use the test set from 2019 for our experiments. Our models use training and validation data from the MS MARCO document ranking task. For each of the 43 queries in the \trecdl{} test set, we re-rank the top-$100$ retrieved documents.
    \item \textbf{\clueweb{}}: We consider the \clueweb{} dataset shared in \cite{dai2019deeper}. The dataset contains 200 queries distributed uniformly in five folds and the top-$100$ documents for each query are retrieved using QL~\cite{strohman2005indri}.
    \item \textbf{\core{}}: The \core{} dataset contains 50 queries with sub-topics and descriptions. Queries are accompanied by a collection of 1.8M documents. We retrieve the top-$1000$ documents for each query using QL.
\end{enumerate}
Characteristics in terms of document and sentence lengths of these datasets are illustrated in Figure~\ref{fig:dataset_stats}. We observed that the distribution of the number of tokens per sentence is almost identical among all three datasets. In particular, approximately $50\%$ of all sentences have less than 25 tokens, and $90\%$ of all sentences have less than 50 tokens. We use these findings to choose $k=20$ for our experiments, based on the rough estimation that in this way, all $512$ available input tokens of the BERT ranker will be used in most cases, while in the remaining cases, the number of inputs does not exceed the limit by a lot.

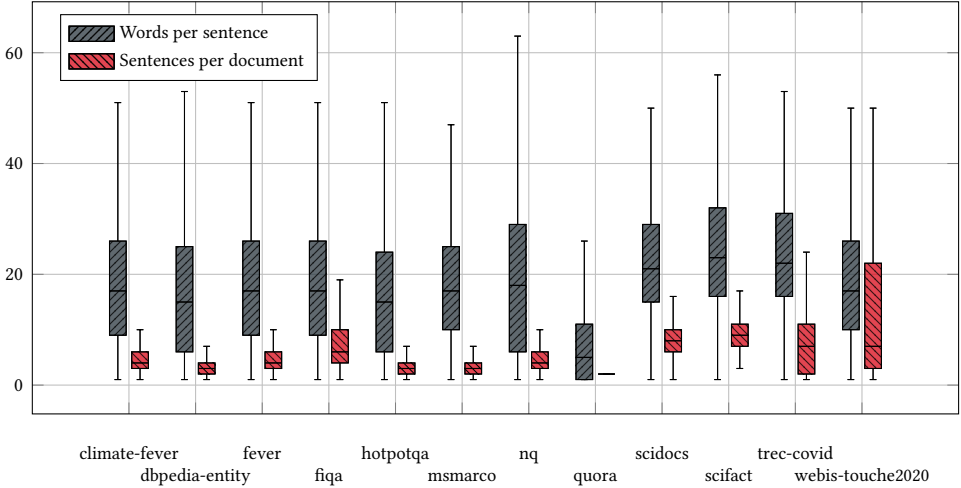
\begin{figure}
    \centering
    \input{plots/beir/beir_stats}
    \caption{The distribution of document and sentence lengths of the datasets from the BEIR benchmark. Outliers omitted.}
    \label{fig:beir_stats}
\end{figure}
Second, we consider a wide variety of additional IR datasets provided by the \textbf{BEIR benchmark}~\cite{thakur2021beir}. These include classical ranking datasets, such as MS MARCO (passage ranking), fact checking tasks, such as FEVER or SciFact, and others. In contrast to the experiments on the TREC ranking datasets, we perform zero-shot evaluation, i.e.\ we train a single model on the MS MARCO training set provided by BEIR and use it to evaluate on each test set. As before, we set $k = 20$ for training. An important difference compared to the ranking datasets above is the average length of the documents (or passages). Figure~\ref{fig:beir_stats} shows plots of the distribution of the number of words per sentence and the number of sentences per document for each of the datasets. Overall, the documents are shorter compared to the web retrieval corpora. This means that the limitation of the input length of BERT-based models does not always apply here. We conduct experiments to analyze how sentence selection within those short passages influences both performance and interpretability.

\subsection{Baselines and Competitors}
\label{sec:baselines}
Since prior studies~\cite{yilmaz2019cross,macavaney2019cedr} already established the effectiveness of contextual neural rankers over non-contextual ones, we consider the following contextual language model-based rankers as our baselines:
\begin{enumerate}
    \item \textbf{\doclabeled}~\cite{dai2019deeper} splits the documents into passages of 150 words with an overlap of 75 words between consecutive passages and considers 30 passages (first, last and 28 random passages). The relevance label of a query-document pair is then transferred to each of its query-passage pairs. This setup is used to train the models with passage-level annotation, and finally, passage-level scores are aggregated to come up with document-level scores during inference.
    \item \textbf{\berts}~\cite{yilmaz2019cross} is a BERT-based transfer model trained on MS MARCO and Microblog\footnote{We only consider MS MARCO to ensure a fair comparison.} to compute the scores of query-sentence pairs. The query-document level score and the top-three query-sentences score are taken into account to compute the final relevance score of that query-document pair.
    \item \textbf{\bertcls}~\cite{nogueira2019passage} uses a vanilla BERT model to rank the documents, which are truncated to 512 tokens.
\end{enumerate}
Additionally, the first-stage retrieval model, the query likelihood model~\cite{lavrenko2017relevance}, is also considered as a ranking baseline.

\subsection{Training Details}
\label{sec:training}
We train and validate using consistent and common experimental design. The neural models are trained using a pairwise max-margin loss; we consider triples $\left(q, d^+, d^- \right)$ of a query and two documents, where $d^+$ is more relevant to $q$ than $d^-$. The loss is computed as
\begin{equation*}
    \mathcal{L} = \max \left\{0, m - R \left(q, d^+ \right) + R \left(q, d^- \right) \right\},
\end{equation*}
where $m$ is the margin and $R$ is the model. Training triples are sampled in a balanced way such that each query is represented evenly in the training set. We train the models using the AdamW optimizer~\cite{loshchilov2018decoupled} with linear warmup during the first 1000 batches (10000 on \trecdl{}). Validation is performed using MAP over the validation set to choose the best model. We use a fixed random seed for all experiments.

\subsubsection{Hyperparameters}
In our experiments we use hyperparameters commonly found in earlier works; the ranker is an uncased $768$-dimensional $\text{BERT}_{\text{base}}$ model with a maximum sequence length of $512$. We use a learning rate of $3 \times 10^{-5}$, dropout of $0.1$ and a batch size of $32$. The selectors (cf.\ Section~\ref{sec:selectors}) use $256$-dimensional hidden representations throughout.

For performance reasons, we restrict the maximum number of query tokens to 50 and the maximum number of document tokens to 5000. Similarly, no more than the first 500 sentences in a single document are considered by the selector. We set the loss margin to $m = 0.2$ and the temperature to $t = 1.0$. As described in Section~\ref{sec:datasets}, we set $k = 20$ for training and inference.

%% file: plots/trec_stats.tex
\begin{tikzpicture}[font=\scriptsize]
    \begin{axis}
        [
            width=0.5\textwidth,
            boxplot/draw direction=y,
            boxplot={box extend=0.5, whisker extend=0.2},
            every axis plot/.append style={fill opacity=0.75},
            xtick={0.5, 3.5, 6.5},
            xticklabels={TREC-DL, Core17, ClueWeb09},
            xticklabel style={align=center, text height=3ex},
            grid=major,
            area legend,
            legend entries={Words per sentence, Sentences per document},
            legend cell align={left},
            legend pos=north west,
            cycle list={
                    {draw=black, fill=plotColor1, solid, postaction={
                                    pattern=north east lines
                                }},
                    {draw=black, fill=plotColor2, solid, postaction={
                                    pattern=north west lines
                                }}},
        ]
        \addplot+[
            boxplot prepared={
                    draw position=0,
                    median=19.0,
                    upper quartile=29.0,
                    lower quartile=11.0,
                    upper whisker=56,
                    lower whisker=1,
                },
            every path/.style={
                    postaction={
                            nomorepostaction,
                            pattern=north east lines
                        },
                },
        ] coordinates {};
        \addplot+[
            boxplot prepared={
                    draw position=1,
                    median=27.0,
                    upper quartile=53.0,
                    lower quartile=12.0,
                    upper whisker=114,
                    lower whisker=0,
                },
            every path/.style={
                    postaction={
                            nomorepostaction,
                            pattern=north west lines
                        },
                },
        ] coordinates {};
        \addplot+[
            boxplot prepared={
                    draw position=3,
                    median=22.0,
                    upper quartile=32.0,
                    lower quartile=13.0,
                    upper whisker=60,
                    lower whisker=1,
                },
            every path/.style={
                    postaction={
                            nomorepostaction,
                            pattern=north east lines
                        },
                },
        ] coordinates {};
        \addplot+[
            boxplot prepared={
                    draw position=4,
                    median=28.0,
                    upper quartile=49.0,
                    lower quartile=11.0,
                    upper whisker=106,
                    lower whisker=0,
                },
            every path/.style={
                    postaction={
                            nomorepostaction,
                            pattern=north west lines
                        },
                },
        ] coordinates {};
        \addplot+[
            boxplot prepared={
                    draw position=6,
                    median=18.0,
                    upper quartile=29.0,
                    lower quartile=9.0,
                    upper whisker=59,
                    lower whisker=1,
                },
            every path/.style={
                    postaction={
                            nomorepostaction,
                            pattern=north east lines
                        },
                },
        ] coordinates {};
        \addplot+[
            boxplot prepared={
                    draw position=7,
                    median=54.0,
                    upper quartile=149.25,
                    lower quartile=21.0,
                    upper whisker=341,
                    lower whisker=1,
                },
            every path/.style={
                    postaction={
                            nomorepostaction,
                            pattern=north west lines
                        },
                },
        ] coordinates {};
    \end{axis}
\end{tikzpicture}

%% file: plots/beir/beir_stats.tex
\begin{tikzpicture}[font=\scriptsize]
    \begin{axis}
        [
            width=\textwidth,
            height=7cm,
            boxplot/draw direction=y,
            boxplot={box extend=0.75, whisker extend=0.3},
            every axis plot/.append style={fill opacity=0.75},
            xtick={0.5, 3.5, 6.5, 9.5, 12.5, 15.5, 18.5, 21.5, 24.5, 27.5, 30.5, 33.5},
            xticklabels={climate-fever\\, dbpedia-entity, fever\\, fiqa, hotpotqa\\, msmarco, nq\\, quora, scidocs\\, scifact, trec-covid\\, webis-touche2020},
            xticklabel style={align=center, text height=5ex},
            grid=major,
            area legend,
            legend entries={Words per sentence, Sentences per document},
            legend cell align={left},
            legend pos=north west,
            cycle list={
                    {draw=black, fill=plotColor1, solid, postaction={
                                    pattern=north east lines
                                }},
                    {draw=black, fill=plotColor2, solid, postaction={
                                    pattern=north west lines
                                }}},
        ]
        \addplot+[
            boxplot prepared={
                    draw position=0,
                    median=17.0,
                    upper quartile=26.0,
                    lower quartile=9.0,
                    upper whisker=51,
                    lower whisker=1,
                },
            every path/.style={
                    postaction={
                            nomorepostaction,
                            pattern=north east lines
                        },
                },
        ] coordinates {};
        \addplot+[
            boxplot prepared={
                    draw position=1,
                    median=4.0,
                    upper quartile=6.0,
                    lower quartile=3.0,
                    upper whisker=10,
                    lower whisker=1,
                },
            every path/.style={
                    postaction={
                            nomorepostaction,
                            pattern=north west lines
                        },
                },
        ] coordinates {};
        \addplot+[
            boxplot prepared={
                    draw position=3,
                    median=15.0,
                    upper quartile=25.0,
                    lower quartile=6.0,
                    upper whisker=53,
                    lower whisker=1,
                },
            every path/.style={
                    postaction={
                            nomorepostaction,
                            pattern=north east lines
                        },
                },
        ] coordinates {};
        \addplot+[
            boxplot prepared={
                    draw position=4,
                    median=3.0,
                    upper quartile=4.0,
                    lower quartile=2.0,
                    upper whisker=7,
                    lower whisker=1,
                },
            every path/.style={
                    postaction={
                            nomorepostaction,
                            pattern=north west lines
                        },
                },
        ] coordinates {};
        \addplot+[
            boxplot prepared={
                    draw position=6,
                    median=17.0,
                    upper quartile=26.0,
                    lower quartile=9.0,
                    upper whisker=51,
                    lower whisker=1,
                },
            every path/.style={
                    postaction={
                            nomorepostaction,
                            pattern=north east lines
                        },
                },
        ] coordinates {};
        \addplot+[
            boxplot prepared={
                    draw position=7,
                    median=4.0,
                    upper quartile=6.0,
                    lower quartile=3.0,
                    upper whisker=10,
                    lower whisker=1,
                },
            every path/.style={
                    postaction={
                            nomorepostaction,
                            pattern=north west lines
                        },
                },
        ] coordinates {};
        \addplot+[
            boxplot prepared={
                    draw position=9,
                    median=17.0,
                    upper quartile=26.0,
                    lower quartile=9.0,
                    upper whisker=51,
                    lower whisker=1,
                },
            every path/.style={
                    postaction={
                            nomorepostaction,
                            pattern=north east lines
                        },
                },
        ] coordinates {};
        \addplot+[
            boxplot prepared={
                    draw position=10,
                    median=6.0,
                    upper quartile=10.0,
                    lower quartile=4.0,
                    upper whisker=19,
                    lower whisker=1,
                },
            every path/.style={
                    postaction={
                            nomorepostaction,
                            pattern=north west lines
                        },
                },
        ] coordinates {};
        \addplot+[
            boxplot prepared={
                    draw position=12,
                    median=15.0,
                    upper quartile=24.0,
                    lower quartile=6.0,
                    upper whisker=51,
                    lower whisker=1,
                },
            every path/.style={
                    postaction={
                            nomorepostaction,
                            pattern=north east lines
                        },
                },
        ] coordinates {};
        \addplot+[
            boxplot prepared={
                    draw position=13,
                    median=3.0,
                    upper quartile=4.0,
                    lower quartile=2.0,
                    upper whisker=7,
                    lower whisker=1,
                },
            every path/.style={
                    postaction={
                            nomorepostaction,
                            pattern=north west lines
                        },
                },
        ] coordinates {};
        \addplot+[
            boxplot prepared={
                    draw position=15,
                    median=17.0,
                    upper quartile=25.0,
                    lower quartile=10.0,
                    upper whisker=47,
                    lower whisker=1,
                },
            every path/.style={
                    postaction={
                            nomorepostaction,
                            pattern=north east lines
                        },
                },
        ] coordinates {};
        \addplot+[
            boxplot prepared={
                    draw position=16,
                    median=3.0,
                    upper quartile=4.0,
                    lower quartile=2.0,
                    upper whisker=7,
                    lower whisker=1,
                },
            every path/.style={
                    postaction={
                            nomorepostaction,
                            pattern=north west lines
                        },
                },
        ] coordinates {};
        \addplot+[
            boxplot prepared={
                    draw position=18,
                    median=18.0,
                    upper quartile=29.0,
                    lower quartile=6.0,
                    upper whisker=63,
                    lower whisker=1,
                },
            every path/.style={
                    postaction={
                            nomorepostaction,
                            pattern=north east lines
                        },
                },
        ] coordinates {};
        \addplot+[
            boxplot prepared={
                    draw position=19,
                    median=4.0,
                    upper quartile=6.0,
                    lower quartile=3.0,
                    upper whisker=10,
                    lower whisker=1,
                },
            every path/.style={
                    postaction={
                            nomorepostaction,
                            pattern=north west lines
                        },
                },
        ] coordinates {};
        \addplot+[
            boxplot prepared={
                    draw position=21,
                    median=5.0,
                    upper quartile=11.0,
                    lower quartile=1.0,
                    upper whisker=26,
                    lower whisker=1,
                },
            every path/.style={
                    postaction={
                            nomorepostaction,
                            pattern=north east lines
                        },
                },
        ] coordinates {};
        \addplot+[
            boxplot prepared={
                    draw position=22,
                    median=2.0,
                    upper quartile=2.0,
                    lower quartile=2.0,
                    upper whisker=2,
                    lower whisker=2,
                },
            every path/.style={
                    postaction={
                            nomorepostaction,
                            pattern=north west lines
                        },
                },
        ] coordinates {};
        \addplot+[
            boxplot prepared={
                    draw position=24,
                    median=21.0,
                    upper quartile=29.0,
                    lower quartile=15.0,
                    upper whisker=50,
                    lower whisker=1,
                },
            every path/.style={
                    postaction={
                            nomorepostaction,
                            pattern=north east lines
                        },
                },
        ] coordinates {};
        \addplot+[
            boxplot prepared={
                    draw position=25,
                    median=8.0,
                    upper quartile=10.0,
                    lower quartile=6.0,
                    upper whisker=16,
                    lower whisker=1,
                },
            every path/.style={
                    postaction={
                            nomorepostaction,
                            pattern=north west lines
                        },
                },
        ] coordinates {};
        \addplot+[
            boxplot prepared={
                    draw position=27,
                    median=23.0,
                    upper quartile=32.0,
                    lower quartile=16.0,
                    upper whisker=56,
                    lower whisker=1,
                },
            every path/.style={
                    postaction={
                            nomorepostaction,
                            pattern=north east lines
                        },
                },
        ] coordinates {};
        \addplot+[
            boxplot prepared={
                    draw position=28,
                    median=9.0,
                    upper quartile=11.0,
                    lower quartile=7.0,
                    upper whisker=17,
                    lower whisker=3,
                },
            every path/.style={
                    postaction={
                            nomorepostaction,
                            pattern=north west lines
                        },
                },
        ] coordinates {};
        \addplot+[
            boxplot prepared={
                    draw position=30,
                    median=22.0,
                    upper quartile=31.0,
                    lower quartile=16.0,
                    upper whisker=53,
                    lower whisker=1,
                },
            every path/.style={
                    postaction={
                            nomorepostaction,
                            pattern=north east lines
                        },
                },
        ] coordinates {};
        \addplot+[
            boxplot prepared={
                    draw position=31,
                    median=7.0,
                    upper quartile=11.0,
                    lower quartile=2.0,
                    upper whisker=24,
                    lower whisker=1,
                },
            every path/.style={
                    postaction={
                            nomorepostaction,
                            pattern=north west lines
                        },
                },
        ] coordinates {};
        \addplot+[
            boxplot prepared={
                    draw position=33,
                    median=17.0,
                    upper quartile=26.0,
                    lower quartile=10.0,
                    upper whisker=50,
                    lower whisker=1,
                },
            every path/.style={
                    postaction={
                            nomorepostaction,
                            pattern=north east lines
                        },
                },
        ] coordinates {};
        \addplot+[
            boxplot prepared={
                    draw position=34,
                    median=7.0,
                    upper quartile=22.0,
                    lower quartile=3.0,
                    upper whisker=50,
                    lower whisker=1,
                },
            every path/.style={
                    postaction={
                            nomorepostaction,
                            pattern=north west lines
                        },
                },
        ] coordinates {};
    \end{axis}
\end{tikzpicture}

%% file: results.tex
\section{Results}
\label{sec:experiments}
In this section we analyze the effectiveness and interpretability of our approaches. We first conduct extensive evaluation of the different selectors, including both pipeline and end-to-end models. Next, we highlight the benefits of our proposed end-to-end modeling scheme (\srlin{} and \sratt{}). Our experiments aim to answer the following questions:
\begin{enumerate}
    \item How well do \sr{} models perform in document and passage ranking tasks (Section~\ref{sec:sen_selector} and Section~\ref{sec:performance})?
    \item How comprehensive are explanations from \sr{} models, i.e.\ how important are the selected sentences for the model decision (Section~\ref{sec:sr_comprehensiveness})?
    \item How faithful are \sr{} explanations and what is their utility to human users (Section ~\ref{sec:sr_faithfulness})?
    \item Does sentence selection lead to sparsification of the input documents, resulting in more interpretable ranking decisions (Section~\ref{sec:bert_token_limitation} and Section~\ref{sec:explaining_bert})?
    \item Can \sr{} models be used to explain rankers that focus only on the head of the documents due to limitations, such as BERT (Section~\ref{sec:explaining_bert})?
\end{enumerate}

\subsection{Variation of Selectors}
\label{sec:sen_selector}
\begin{table*}
    \centering
    \begin{tabular}{lccccccccc}
        \toprule
                & \multicolumn{3}{c}{\trecdl}
                & \multicolumn{3}{c}{\core}
                & \multicolumn{3}{c}{\clueweb}                                                                                         \\
        \cmidrule(lr){2-4}
        \cmidrule(lr){5-7}
        \cmidrule(lr){8-10}
                & MAP                          & nDCG@20       & MRR   & MAP   & nDCG@20       & MRR   & MAP   & nDCG@20       & MRR   \\
        \midrule
        \plrnd  & 0.231                        & $0.492^{*\#}$ & 0.754 & 0.173 & $0.345^{*\#}$ & 0.649 & 0.138 & $0.236^{*\#}$ & 0.495 \\
        \plbert & 0.237                        & $0.501^{*\#}$ & 0.822 & 0.200 & 0.399         & 0.759 & 0.169 & 0.294         & 0.529 \\
        \pllstm & 0.257                        & $0.558^{*}$   & 0.827 & 0.194 & 0.399         & 0.788 & 0.166 & 0.289         & 0.552 \\
        \plbm   & 0.264                        & 0.568         & 0.893 & 0.196 & 0.412         & 0.727 & 0.171 & 0.297         & 0.555 \\
        \plsem  & 0.265                        & 0.571         & 0.920 & 0.207 & 0.414         & 0.768 & 0.167 & 0.286         & 0.534 \\
        \midrule
        \srlin  & 0.269                        & 0.597         & 0.946 & 0.203 & 0.411         & 0.710 & 0.174 & 0.303         & 0.535 \\
        \sratt  & 0.271                        & 0.590         & 0.924 & 0.205 & 0.403         & 0.714 & 0.168 & 0.292         & 0.518 \\
        \bottomrule
    \end{tabular}
    \caption{Retrieval performance with $k=20$. \plrnd{} refers to the selection of $k$ random sentences. Significant improvements (nDCG@20) at a level of $95\%$ are indicated by $*$ (\srlin{}) and $\#$ (\sratt{}).}
    \label{tab:model_sel_eval}
\end{table*}
In this section we first briefly describe four different hard selection strategies used by the pipeline models. Next, we compare the pipeline strategies and the two proposed end-to-end variants (cf.\ Section~\ref{sec:selectors}) of our approach.

The hard selection approaches are described as follows:
\begin{enumerate}
    \item \textbf{\plbert}: The similarity or relevance between a query and a sentence is computed using the approach proposed by \citet{yilmaz2019cross}. The model is trained on the MS MARCO passage re-ranking dataset according to \citet{nogueira2019passage}. Finally, it is used to infer query-sentence level relevance score for each query-document pair.
    \item \textbf{\pllstm}: It is similar to \plbert{}, but uses an LSTM instead of BERT. We limit the input to 1000 words for this configuration, similar to BERT's limit of 512 tokens. The model is trained on the MS MARCO passage re-ranking dataset and the trained model is used to infer the relevance score of query-sentence pairs.
    \item \textbf{\plbm}: We use a simple BM25-based term matching function\footnote{\url{https://pypi.org/project/rank-bm25/}} to obtain the score between the query and the sentence.
    \item \textbf{\plsem}: Semantic similarity score between query and sentence is computed using 300-dimensional GloVe embeddings.
\end{enumerate}
These models apply the selection strategy only during the inference phase, i.e.\ trained models are used to predict the relevance of pairs of queries and summarized documents. The ranker itself is simply trained without any selection, i.e.\ documents are truncated to fit. We refer to these strategies as \textit{pipeline} (PL). They can be seen as an implementation of the method proposed in \cite{li2021keybld}.

To analyze the effectiveness of the above-mentioned selection approaches, we also measure the performance of a simple strategy, \plrnd{}, where we randomly select $k$ sentences from the document. Table~\ref{tab:model_sel_eval} shows the results at $k = 20$ and highlights the effectiveness of the proposed approaches over random selection on the TREC datasets. We also tried other values, but $k = 20$ gives consistent performance for all three datasets. This may be attributed to the token limitation of the ranker.

It is interesting to note that our lightweight selection strategies such as \plbm{} and \plsem{} perform better than heavy parameterized and time-consuming neural selection models such as \plbert{} and \pllstm{}. \plsem{} shows the best or comparable performance for all three datasets. \plbm{}, while slightly worse, also shows promising performance. This compact representation of documents also helps in developing computationally efficient ranking models and reducing noise.

Our end-to-end models, \srlin{} and \sratt{}, show improvements over the pipeline models in most cases. Surprisingly, the linear, more lightweight selector often matches or exceeds the performance of the attention-based one.

We also perform statistical pairwise t-tests~\cite{gallagher2019pairwise} for nDCG@20 between pipeline approaches and \srlin{} and \sratt{}. We do not observe significant improvements for \core{} and \clueweb{}. However, end-to-end models perform significantly better than \plbert{} and \pllstm{}.

\subsection{Performance of \sr{}}
\label{sec:performance}
\begin{table*}
    \small
    \centering
    \begin{tabular}{lccccccccc}
        \toprule
                    & \multicolumn{3}{c}{\trecdl}
                    & \multicolumn{3}{c}{\core}
                    & \multicolumn{3}{c}{\clueweb}                                                                             \\
        \cmidrule(lr){2-4}
        \cmidrule(lr){5-7}
        \cmidrule(lr){8-10}
                    & MAP                          & nDCG@20       & MRR   & MAP   & nDCG@20 & MRR   & MAP   & nDCG@20 & MRR   \\
        \midrule
        QL          & 0.237                        & $0.487^{*\#}$ & 0.785 & 0.203 & 0.395   & 0.686 & 0.165 & 0.277   & 0.487 \\
        \midrule
        \doclabeled & 0.203                        & $0.434^{*\#}$ & 0.731 & 0.237 & 0.437   & 0.742 & 0.165 & 0.284   & 0.503 \\
        \berts      & 0.245                        & $0.519^{*\#}$ & 0.799 & 0.204 & 0.406   & 0.694 & 0.178 & 0.306   & 0.544 \\
        \midrule
        \bertcls    & 0.260                        & 0.581         & 0.874 & 0.196 & 0.419   & 0.749 & 0.178 & 0.313   & 0.572 \\
        \plsem      & 0.265                        & 0.571         & 0.920 & 0.207 & 0.414   & 0.768 & 0.167 & 0.286   & 0.534 \\
        \midrule
        \srlin      & 0.269                        & 0.597         & 0.946 & 0.203 & 0.411   & 0.710 & 0.174 & 0.303   & 0.535 \\
        \sratt      & 0.271                        & 0.590         & 0.924 & 0.205 & 0.403   & 0.714 & 0.168 & 0.292   & 0.518 \\
        \bottomrule
    \end{tabular}
    \caption{Retrieval performance. \sr{} models use $k = 20$. For \doclabeled{}, we report the best strategy (FirstP, MaxP, AvgP for \trecdl{}, \core{} and \clueweb{} respectively). Significant improvements (nDCG@20) at a level of $95\%$ are indicated by $*$ (\srlin{}) and $\#$ (\sratt{}).}
    \label{tab:model_eval}
\end{table*}
\begin{table}
    \centering
    \begin{tabular}{lcc}
        \toprule
                      & MAP   & nDCG@10 \\
        \midrule
        \matchpyramid & 0.232 & 0.567   \\
        \copacrr      & 0.231 & 0.550   \\
        \convknrm     & 0.241 & 0.565   \\
        \tklsmall     & 0.264 & 0.634   \\
        \midrule
        \srlin        & 0.269 & 0.646   \\
        \sratt        & 0.271 & 0.639   \\
        \bottomrule
    \end{tabular}
    \caption{Neural baselines on \trecdl{}. \sr{} models use $k = 20$. Results are taken from \cite{hofstaetter2020local}. \tklsmall{} refers to TKL operating on 2000 tokens.}
    \label{tab:baselines}
\end{table}
In this section we compare the performance of our proposed models to state-of-the-art models. We further compare our end-to-end approaches, \srlin{} and \sratt{}, to a simple BERT baseline, denoted by \bertcls{}, which uses truncation of the document instead of sentence selection. First, each model is trained (fine-tuned) and evaluated on \trecdl{}, as it offers an abundance of training data. For \core{} and \clueweb{}, we use the model from the \trecdl{} experiment as initialization. This helps us to properly train the selector, as, unlike the BERT ranker, it does not start from a pre-trained model.

The results are illustrated in Table~\ref{tab:model_eval}. Table~\ref{tab:baselines} shows additional neural baselines~\cite{pang2016text,hui2018copacrr,zhuyun2018convolutional,hofstaetter2020local} evaluated on \trecdl{}. The pipeline model works quite well on the \core{} dataset, but falls short on \trecdl{} and \clueweb{} compared to the end-to-end models. In the pipeline model, the selection phase is independent of the ranking phase; hence, the selection strategy does not receive any feedback from the ranking phase. It is evident from Table~\ref{tab:model_eval} that the end-to-end approach improves the ranking process.

By selecting sentences from the complete document, our approaches perform similarly to stand-alone rankers that operate only on the head of the documents, specifically \bertcls{}. This indicates that most documents contain redundant information (likely in the form of summaries) near the beginning that \bertcls{} is able to exploit. We confirm this in Section~\ref{sec:bert_token_limitation} by showing that there is little overlap between the document head and the sentences selected by \srlin{}. Thus, in Section~\ref{sec:explaining_bert} we use a \sr{} model to explain the predictions of \bertcls{} by specifically selecting sentences from just the head of the documents.

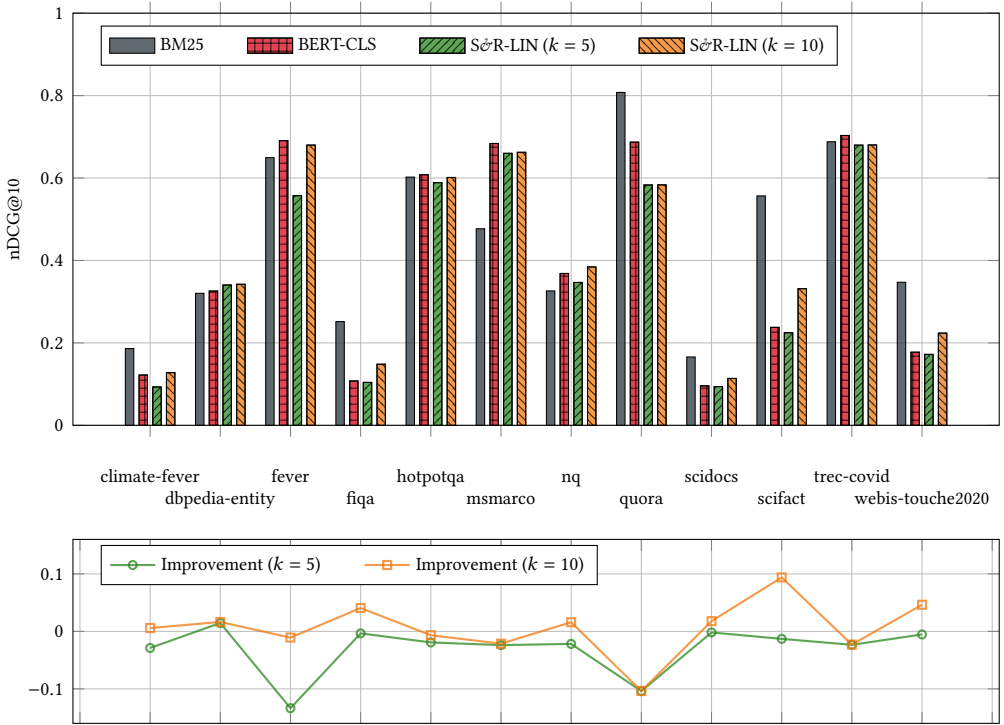
\begin{figure}
    \centering
    \input{plots/beir/beir}
    \caption{Ranking results (nDCG@10) on datasets from the BEIR framework using zero-shot evaluation with models trained on the MS MARCO dataset. The lines show the difference between \srlin{} and \bertcls{} performance. Positive improvement indicates that \srlin{} performs better, negative improvement indicates the opposite.}
    \label{fig:beir_results}
\end{figure}
Next, we evaluate our \srlin{} model on various datasets provided by the BEIR framework and compare it to sparse retrieval methods (BM25) and a standard \bertcls{} model. The models are trained on MS MARCO, i.e.\ only the results on that dataset are in-domain, whereas the other datasets are evaluated in a zero-shot fashion. The contextual models are used to re-rank the top-$100$ BM25 results, which we retrieved using Elasticsearch. The results are illustrated in Figure~\ref{fig:beir_results}. It is evident that \srlin{} and \bertcls{} show similar ranking performance on most datasets with only few exceptions. It is interesting to note that, in some cases, contextual re-ranking models fail to improve BM25 ranking. We assume the reason for this to be lack of domain knowledge due to the zero-shot setup.

\begin{figure*}
    \begin{subfigure}{.32\linewidth}
        \centering
        \input{plots/beir/beir_k_fever}
        \caption{FEVER}
    \end{subfigure}
    \begin{subfigure}{.32\linewidth}
        \centering
        \input{plots/beir/beir_k_hotpotqa}
        \caption{HotpotQA}
    \end{subfigure}
    \begin{subfigure}{.32\linewidth}
        \centering
        \input{plots/beir/beir_k_scifact}
        \caption{SciFact}
    \end{subfigure}
    \caption{Ranking results (nDCG@10) using \srlin{} with decreasing number of selected sentences ($k$).}
    \label{fig:beir_results_varying_k}
\end{figure*}
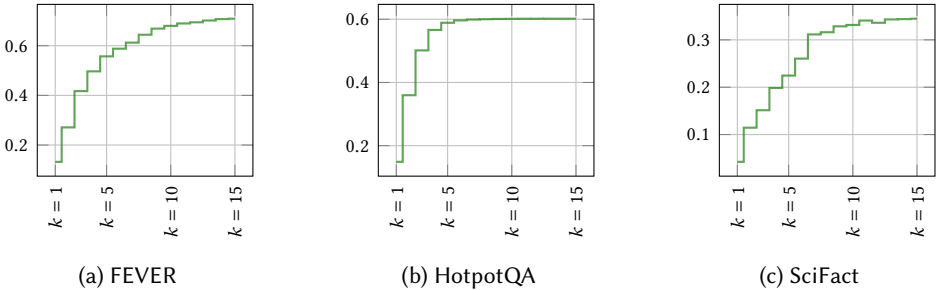
As shown in Section~\ref{sec:datasets}, the datasets contain mostly short documents (or passages). As a consequence, selecting as many as $k = 10$ sentences might already select the complete document in some cases. We thus perform additional experiments on some of the datasets, decreasing $k$ all the way to a single selected sentence. Figure~\ref{fig:beir_results_varying_k} shows the results. This experiment nicely illustrates the controllable trade-off between performance and interpretability: On each of the datasets, the performance plateaus once a certain number of selected sentences is reached, which depends on the document lengths of the dataset. On the other hand, the performance drops when the number of selected sentences is decreased, which in turn makes the ranking decision more interpretable.

\subsection{On the Comprehensiveness of \sr{}}
\label{sec:sr_comprehensiveness}
\begin{figure}
    \centering
    \input{plots/comprehensiveness}
    \caption{Ranking results (nDCG@20) on \trecdl{} using \srlin{} with $k = 20$, where $N$ sentences are removed (leaving $k - N$ sentences). We compare removing random sentences and the highest scoring sentences.}
    \label{fig:sr_comprehensiveness}
\end{figure}
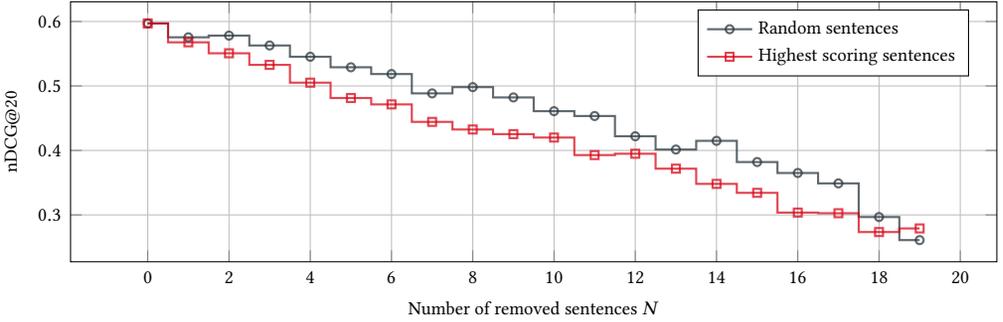
\emph{Comprehensiveness}~\cite{deyoung2020eraser} is a metric that evaluates the quality of \emph{rationales}, i.e.\ parts of the model input that aim to explain the corresponding output. Specifically, a \emph{contrast example} $\Tilde{x}_i = x_i \setminus r_i$ is constructed for each input $x_i$, where the rationales $r_i$ are removed. Comprehensiveness is then computed as
\begin{equation*}
    \operatorname{Comp}(x_i) = m(x_i)_j - m(\Tilde{x}_i)_j,
\end{equation*}
where $m(\cdot)_j$ is the model output or prediction corresponding to the class $j$.

Intuitively, comprehensiveness measures the degree of influence the rationales have on the final prediction by computing how much worse the model performs without them. In or case, $x_i$ is a query-document pair. However, due to the length of the documents and limitation of ranking models, the ranker does not see the complete document in the vast majority of cases. Thus, computing the exact comprehensiveness is difficult. Instead, we use a proxy to get an idea about the comprehensiveness of \sr{} models: During evaluation, we remove the $N$ highest scoring (as assigned by the selector) sentences (out of $k$ selected sentences) from the input and observe the drop in performance. We then compare the results to
\begin{enumerate}
    \item the performance with all $k$ sentences and
    \item the performance when $N$ random sentences are removed instead.
\end{enumerate}
The results on \trecdl{} with $k = 20$ are illustrated in Figure~\ref{fig:sr_comprehensiveness}. They show that removing high-scoring sentences has a higher impact on overall performance than removing random sentences. This suggests that higher scoring sentences have a higher impact on the model predictions.

\subsection{On the Faithfulness and Utility of \sr{} for Human Users}
\label{sec:sr_faithfulness}
\begin{figure*}
    \begin{subfigure}{.49\linewidth}
        \centering
        \input{plots/study/study_acc_ctrlf}
        \caption{The accuracy of relevance judgments and usage of the web browser's search function.}
    \end{subfigure}
    \begin{subfigure}{.49\linewidth}
        \centering
        \input{plots/study/study_time}
        \caption{The time taken to complete a single query-document relevance judgment. Outliers omitted.}
    \end{subfigure}
    \caption{The results of our user study to determine the faithfulness of \sr{} explanations. Participants were presented with a query-document pair, where the document was either unaltered (\emph{full}) or $k \in [5, 10, 15]$ sentences were selected using a the selector of a trained \srlin{} model. The query-document pairs originate from \trecdl{}. We plot the average accuracy, the fraction of instances where participants used their browser search and the time taken to complete a single query-document relevance judgment.}
    \label{fig:sr_faithfulness}
\end{figure*}
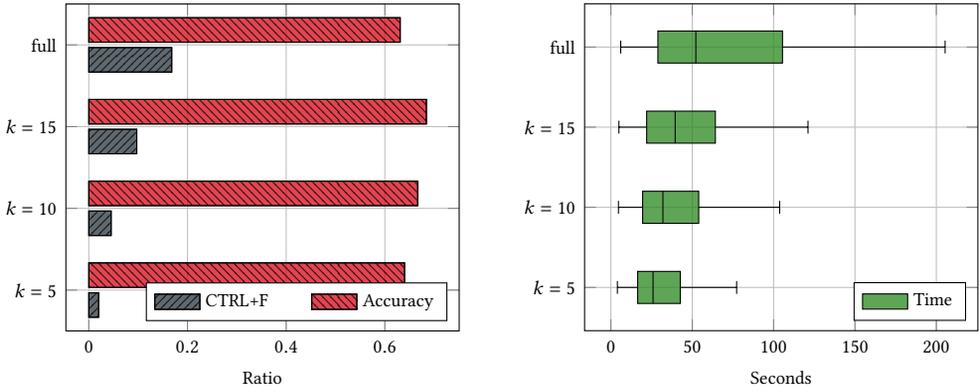
Generally speaking, the \emph{faithfulness} of interpretations refers to the degree to which they accurately represent the reasoning of the model~\cite{jacovi2020towards}. In the case of \sr{}, this corresponds to the question \emph{how well the selected sentences represent the document they originate from}. This problem is closely tied to the actual utility and usefulness of \sr{} models for \emph{human users}; the idea is that the users should be able to comprehend a ranking decision solely based on the selected sentences, i.e.\ the explanation. In order to assess how faithful the explanations are for human users, we have conducted a study which is described in this section.

\subsubsection{Study Setup}
We randomly selected 30 queries from the \trecdl{} test set for our study. For each of these queries, we randomly sampled one relevant and one irrelevant document from the official query relevance judgments. We used the selector of a trained \srlin{} model (from Section~\ref{sec:performance}) to select $k \in \{5, 10, 15\}$ sentences for each document (with respect to the corresponding query), resulting in four variations of each query-document pair. In total, we ended up with 240 $(q, d, k)$ instances (where $k$ can be \texttt{null}, representing no sentence selection). We employed 80 participants for the study, each of which judged 12 individual $(q, d, k)$ instances (i.e.\ 960 relevance judgments in total). Thus, each instance was judged approximately four times.\footnote{Due to some participants never finishing the study, this number can vary in rare cases; however, each instance has been judged at least three times.} Instances were allocated to participants randomly, making sure that no participant ever saw two instances with the same query and document.

The user interface presents the query at the top and the document just below. Within the document, a line break is inserted after every sentence. At the bottom, the participant is asked to indicate
\begin{enumerate}
    \item whether or not the document is relevant to the query and
    \item whether they used their browser's integrated search function for this instance.
\end{enumerate}
We further measure and record the time taken for each relevance judgment. After each instance, an intermediate page prompts the participant to take a break before the next instance if necessary, such that the recorded times are less noisy.

Our study is implemented using the \texttt{oTree} framework~\cite{chen2016otree} and was conducted on the Prolific\footnote{\url{https://www.prolific.co/}} platform. Additional details can be found in Appendix~\ref{app:study_details}.

\subsubsection{Study Results}
The results are illustrated in Figure~\ref{fig:sr_faithfulness}. It is apparent that the longer the documents are (in terms of number of sentences), the more the average time taken to judge the relevance of a single query-document pair increases. Additionally, participants resort to the usage of their browser's search function more often, but this is not enough to compensate for the increased length and keep the time down. Moreover, the participants' accuracy remains roughly stable across all settings, peaking at $k = 15$. We assume the drop in accuracy for the full documents could be caused by participants relying too much on term matching provided by their browsers rather than reading the complete documents.

Overall, our study highlights the utility of \sr{} models to humans: The sentences extracted by our approach serve as \emph{faithful} explanations to users, as is apparent from the accuracy. At the same time, it enables them to judge documents more quickly using only a small subset of sentences.

\subsection{The Effect of Token Limitation}
\label{sec:bert_token_limitation}
\begin{figure}
    \centering
    \input{plots/token_limitation}
    \caption{CDF of tokens on \trecdl{} that would have been missed without sentence selection.}
    \label{fig:missing_token_dist}
\end{figure}
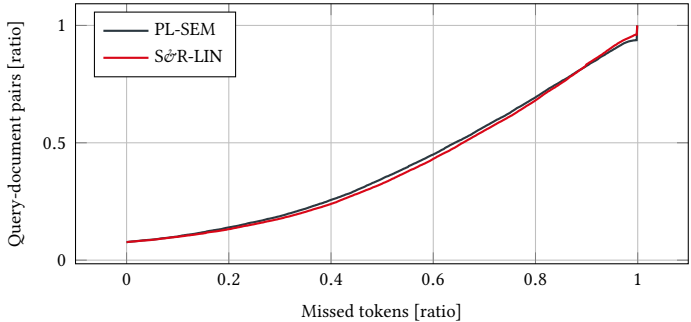
In this section we analyze the the token limitation that is inherent to the BERT ranker and further the role the selection strategy has in mitigating that limitation. In other words, we answer the following question: \textit{How many input tokens of the selected sentences would not have been seen by BERT without selection due to length restrictions?} In general, existing research assumes that most of the information relevant to the query is present in the first part of the document~\cite{nogueira2019passage}. The \bertcls{} baseline also works based on that assumption. However, recent strategies~\cite{hofstaetter2020interpretable} show that some information also exists beyond this token limit. In \cite{dai2019deeper}, the authors try to handle this by selecting the first, last and 28 random passages in their \doclabeled{} approach, but this heuristic does not always work. To that end, we choose the top-$20$ sentences based on \plsem{} and \srlin{} and measure what fraction of these tokens exceeds the usable BERT input, i.e.\ is lost when we only consider the head of a document. Figure~\ref{fig:missing_token_dist} shows the cumulative distribution of the ratio of missed tokens for \trecdl. The distribution pattern is similar for both methods: Less than $10\%$ of the query-document pairs do not miss any of the selected tokens. Given the performance of the models shown in Section~\ref{sec:performance}, this suggests that relevant information is repeated within the documents, such that multiple selections exists which result in similar performance.

\subsection{Explaining \bertcls{}}
\label{sec:explaining_bert}
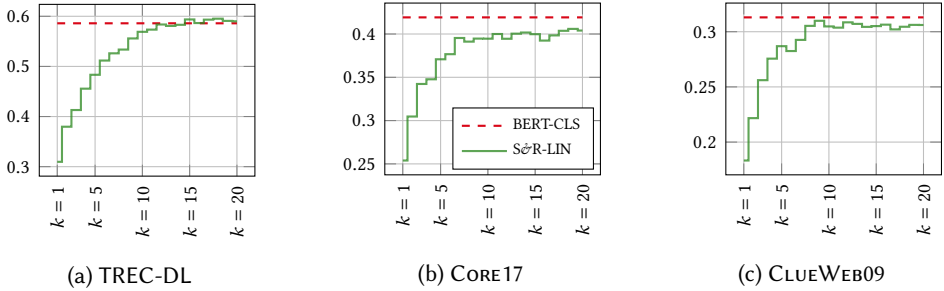
\begin{figure*}
    \begin{subfigure}{.32\linewidth}
        \centering
        \input{plots/explaining_bert/explaining_bert_trecdl}
        \caption{\trecdl{}}
    \end{subfigure}
    \begin{subfigure}{.32\linewidth}
        \centering
        \input{plots/explaining_bert/explaining_bert_core}
        \caption{\core{}}
    \end{subfigure}
    \begin{subfigure}{.32\linewidth}
        \centering
        \input{plots/explaining_bert/explaining_bert_clueweb}
        \caption{\clueweb{}}
    \end{subfigure}
    \caption{The performance of \srlin{} (nDCG@20) applied only to the first 20 sentences of each document, which approximates selecting sentences from the input of \bertcls{}.}
    \label{fig:explaining_bert}
\end{figure*}
\begin{figure*}
    \begin{subfigure}{.49\linewidth}
        \centering
        \input{plots/explaining_bert/explaining_bert_sr}
        \caption{\sr{}}
    \end{subfigure}
    \begin{subfigure}{.49\linewidth}
        \centering
        \input{plots/explaining_bert/explaining_bert_trunc}
        \caption{Truncation to $T$ input tokens}
    \end{subfigure}
    \caption{Ranking performance (nDCG@20) on \trecdl{}, where the length of the input to the ranker is limited in two ways. On the left, \sr{} is used to select $k$ sentences from the head of the document (i.e.\ from the first 20 sentences). On the right, inputs are simply truncated, i.e.\ the ranker only sees $T$ tokens in total, which includes the query and the first part of the document. For \bertcls{}, $T = 512$ is the default setting and is consistent with the results in Table~\ref{tab:model_eval} and Figure~\ref{fig:explaining_bert}.}
    \label{fig:explaining_bert_token_limit}
\end{figure*}
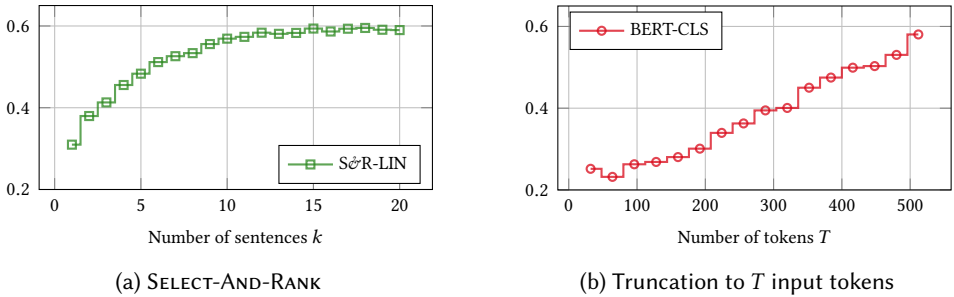
In Section~\ref{sec:bert_token_limitation}, we showed that \srlin{} and the standard \bertcls{} model operate on different parts of the input documents, yet they achieve comparable performance (cf.\ Table~\ref{tab:model_eval}). In this section we explore whether \sr{} models can be used to further sparsify the head of a document and thus explain the predictions of \bertcls{}.

To that end, we conduct a set of experiments where we limit the available sentences for the selector to choose from to the first $20$ of each documents based on our length estimation (cf.\ Section~\ref{sec:datasets}). We then vary $k$ to compare the performance with respect to sparsity. The results are illustrated in Figure~\ref{fig:explaining_bert} in terms of nDCG@20. We observe that the performance plateaus for roughly $k \geq 10$ (slightly later for \trecdl{}) and approximately matches \bertcls{}. For lower values of $k$, the performance drops.

In addition, we compare the above result to a simpler strategy, where, instead of using \sr{} to select sentences, we limit the length of the ranker input by simply truncating it to a constant number of tokens. This is identical to the \bertcls{} approach, but instead of 512 tokens, we use smaller numbers. Figure~\ref{fig:explaining_bert_token_limit} shows the comparison the the two methods (\srlin{} with $k$ sentences and \bertcls{} truncated to $T$ tokens). On the far right side of each of the plots, i.e.\ $k = 20$ and $T = 512$, there is no selection or truncation, thus both models have roughly the same performance. However, decreasing $k$ or $T$, respectively, it becomes evident that by selecting relevant sentences using \sr{}, substantially higher performance can be reached with similar numbers of input tokens. For example, comparing $k = 10$ and $T = 256$, both of which drop (roughly) half of the tokens, \sr{} achieves an nDCG value of 0.569, while \bertcls{} only reaches 0.363. This suggests that \sr{} is able to select representative summaries of the documents that are sufficient for the ranker to output similar performance. Truncation, on the other hand, does not have the same effect, which ultimately reflects in the performance.

Overall, these experiments show that sentence selection may be used even in combination with models that only operate on the head of documents to achieve interpretability while maintaining performance.

\subsection{The Effect of First-Stage Retrieval}
\label{sec:effect_rm3}
\begin{table}
    \centering
    \begin{tabular}{lcccc}
        \toprule
                    & \multicolumn{2}{c}{QL}
                    & \multicolumn{2}{c}{QL+RM3}                                         \\
        \cmidrule(lr){2-3}
        \cmidrule(lr){4-5}
                    & MAP                        & nDCG@20       & MAP   & nDCG@20       \\
        \midrule
        QL(+RM3)    & 0.237                      & $0.487^{*\#}$ & 0.272 & $0.513^{*\#}$ \\
        \midrule
        \doclabeled & 0.203                      & $0.434^{*\#}$ & 0.219 & $0.426^{*\#}$ \\
        \berts      & 0.245                      & $0.519^{*\#}$ & 0.281 & 0.539         \\
        \midrule
        \bertcls    & 0.260                      & 0.581         & 0.279 & 0.559         \\
        \plsem      & 0.265                      & 0.571         & 0.268 & 0.537         \\
        \midrule
        \srlin      & 0.269                      & 0.597         & 0.286 & 0.568         \\
        \sratt      & 0.271                      & 0.590         & 0.284 & 0.563         \\
        \bottomrule
    \end{tabular}
    \caption{Performance over two first stage retrieval models, QL and QL+RM3 at depth 100 on the \trecdl{} test set using $k=20$. Significant improvements (nDCG@20) at a level of $95\%$ are indicated by $*$ (\srlin{}) and $\#$ (\sratt{}).}
    \label{tab:rm3_effect}
\end{table}
From Section~\ref{sec:performance} and Section~\ref{sec:explaining_bert} it is evident that the performance of \srlin{} is on par with the baselines, while maintaining the interpretability aspect of the approach. However, the re-ranking performance of the models is computed over the top-$100$ documents per query, retrieved using a QL model. One obvious question is, whether this performance is lost with a better first stage retrieval system. To answer this question, we re-retrieve the top-$100$ documents with QL and RM3 and apply the models to that set. Table~\ref{tab:rm3_effect} shows the results on \trecdl{}. Note that the models are not re-trained, i.e.\ the models from previous experiments are used. There is no significant influence of RM3 on the performance of baselines; rather, performance drops to some extent in terms of nDCG. We assume that the reason for this is the fact that the models were not re-trained using the documents retrieved by QL and RM3.

\subsection{Anecdotal Examples}
\label{sec:examples}
\input{table-anecdotal}
In Table~\ref{tab:trecdl_examples} we present an anecdotal example of the top sentence for each document, selected by our \sr{} approaches in both pipeline and end-to-end variants. The documents marked as relevant are the ground-truth documents as assessed by TREC annotators. We see that the selected sentences already provide an insight into the what evidence is considered important by the overall ranking model. Specifically, the rank 5 prediction by \sratt{} happens because it mistakes \textit{bow pose} in yoga with bows and arrows. It is clear from the selected sentence of \plsem{} that it does not consider the duration aspect of the query. A key aspect of \sr{} is that the decision of the final ranker can be unambiguously attributed to these extracted sentences, providing interpretability to the model decision. Note that we cannot completely explain the decision making of the final ranker, since it could select a further subset of the selected sentences.

Moreover, we present examples from the FEVER dataset in Table~\ref{tab:fever_examples}. It shows the two relevant documents for a query (here: a fact), each split into sentences. These sentences are then ranked by their scores w.r.t.\ the query as assigned by the selector model (\srlin{}). Finally, the $5$ highest scoring sentences are selected as input for the ranker. This setting is consistent with the experiments in Section~\ref{sec:performance} and Figure~\ref{fig:beir_results}. The part corresponding to the first document depicts the case where the sentence selection works well: The sentence that contains the answer to the query is scored high and thus selected. The ranker receives the selected sentences and is able to rank the document high (rank 3). On the contrary, in the second relevant document, the selector misses the only relevant sentence and does not include it in the selection. Thus, the ranker does not see the relevant part of the document and consequently ranks it lower (rank 23). This example further illustrates how each ranking decision can be attributed to a small fraction of the input document.

%% file: plots/beir/beir.tex
\begin{tikzpicture}[font=\scriptsize]
    \pgfplotstableread[col sep=comma]{plots/beir/beir.csv}\data
    \begin{groupplot}[
            group style={
                    group size=1 by 2,
                    vertical sep=1.5cm
                },
            legend style={/tikz/every even column/.append style={column sep=0.5cm}},
        ]

        \nextgroupplot[
            ybar,
            width=\textwidth,
            bar width=0.12,
            height=7cm,
            every axis plot/.append style={fill, fill opacity=0.75},
            xtick={1, 2, 3, 4, 5, 6, 7, 8, 9, 10, 11, 12},
            xticklabels={climate-fever\\, dbpedia-entity, fever\\, fiqa, hotpotqa\\, msmarco, nq\\, quora, scidocs\\, scifact, trec-covid\\, webis-touche2020},
            xticklabel style={align=center, text height=5ex},
            ylabel={nDCG@10},
            ymax=1,
            ymin=0,
            grid=major,
            area legend,
            legend entries={BM25, \bertcls{}, \srlin{} ($k=5$), \srlin{} ($k=10$)},
            legend cell align={left},
            legend pos=north west,
            legend columns=-1,
            cycle list={{draw=black, fill=plotColor1, solid}, {draw=black, fill=plotColor2, solid}, {draw=black, fill=plotColor3, solid}, {draw=black, fill=plotColor4, solid}},
        ]
        \addplot+[
        ] table[y index=1] {\data};
        \addplot+[
            postaction={
                    pattern=grid
                },
        ] table[y index=2] {\data};
        \addplot+[
            postaction={
                    pattern=north east lines
                },
        ] table[y index=3] {\data};
        \addplot+[
            postaction={
                    pattern=north west lines
                },
        ] table[y index=4] {\data};

        \nextgroupplot[
            width=\textwidth,
            height=4cm,
            every axis plot/.append style={opacity=0.75, thick, mark size=1.5pt},
            xticklabels=\empty,
            ymax=0.16,
            ymin=-0.16,
            grid=major,
            legend entries={Improvement ($k=5$), Improvement ($k=10$)},
            legend cell align={left},
            legend pos=north west,
            legend columns=-1,
            cycle list={{plotColor3}, {plotColor4}},
        ]
        \pgfplotsset{cycle list shift=2}
        \addplot+[mark=o] table[y index=5] {\data};
        \addplot+[mark=square] table[y index=6] {\data};
    \end{groupplot}
\end{tikzpicture}

%% file: plots/beir/beir_k_fever.tex
\begin{tikzpicture}[font=\scriptsize]
    \pgfplotstableread[col sep=comma]{plots/beir/beir_k.csv}\data
    \begin{axis}[
            width=\textwidth,
            xtick={1, 5, 10, 15},
            xticklabels={$k=1$, $k=5$, $k=10$, $k=15$},
            x tick label style={rotate=90, anchor=east},
            grid=major,
        ]
        \addplot+[
            const plot mark mid,
            thick,
            mark=none,
            mark size=1.5pt,
            plotColor3,
            opacity=0.75,
        ] table[y index=2] {\data};
    \end{axis}
\end{tikzpicture}

%% file: plots/beir/beir_k_hotpotqa.tex
\begin{tikzpicture}[font=\scriptsize]
    \pgfplotstableread[col sep=comma]{plots/beir/beir_k.csv}\data
    \begin{axis}[
            width=\textwidth,
            xtick={1, 5, 10, 15},
            xticklabels={$k=1$, $k=5$, $k=10$, $k=15$},
            x tick label style={rotate=90, anchor=east},
            grid=major,
        ]
        \addplot+[
            const plot mark mid,
            thick,
            mark=none,
            mark size=1.5pt,
            plotColor3,
            opacity=0.75,
        ] table[y index=1] {\data};
    \end{axis}
\end{tikzpicture}

%% file: plots/beir/beir_k_scifact.tex
\begin{tikzpicture}[font=\scriptsize]
    \pgfplotstableread[col sep=comma]{plots/beir/beir_k.csv}\data
    \begin{axis}[
            width=\textwidth,
            xtick={1, 5, 10, 15},
            xticklabels={$k=1$, $k=5$, $k=10$, $k=15$},
            x tick label style={rotate=90, anchor=east},
            grid=major,
        ]
        \addplot+[
            const plot mark mid,
            thick,
            mark=none,
            mark size=1.5pt,
            plotColor3,
            opacity=0.75,
        ] table[y index=3] {\data};
    \end{axis}
\end{tikzpicture}

%% file: plots/comprehensiveness.tex
\begin{tikzpicture}[font=\scriptsize]
    \pgfplotstableread[col sep=comma]{plots/comprehensiveness.csv}\data
    \begin{axis}[
            width=\textwidth,
            height=5cm,
            xlabel={Number of removed sentences $N$},
            ylabel={nDCG@20},
            grid=major,
            legend entries={Random sentences, Highest scoring sentences},
            legend cell align={left},
            legend pos=north east,
        ]
        \addplot+[
            const plot mark mid,
            thick,
            mark=o,
            mark size=1.5pt,
            plotColor1,
            opacity=0.75,
        ] table[y index=1] {\data};
        \addplot+[
            const plot mark mid,
            thick,
            mark=square,
            mark size=1.5pt,
            plotColor2,
            opacity=0.75,
        ] table[y index=2] {\data};
    \end{axis}
\end{tikzpicture}

%% file: plots/study/study_acc_ctrlf.tex
\begin{tikzpicture}[font=\scriptsize]
    \pgfplotstableread[col sep=comma]{plots/study/study_acc_ctrlf.csv}\data
    \begin{axis}[
            xbar,
            width=\textwidth,
            bar width=0.3,
            every axis plot/.append style={fill, fill opacity=0.75},
            ytick={1, 2, 3, 4},
            yticklabels={$k = 5$, $k = 10$, $k = 15$, full},
            ytick align=inside,
            xlabel={Ratio},
            ymin=0.5,
            ymax=4.5,
            grid=major,
            area legend,
            legend entries={CTRL+F, Accuracy},
            legend cell align={left},
            legend pos=south east,
            legend columns=-1,
            legend style={/tikz/every even column/.append style={column sep=0.5cm}},
        ]
        \addplot+[
            plotColor1,
            draw=black,
            postaction={
                    pattern=north east lines
                },
        ] table[x=ctrlf, y index=0] {\data};
        \addplot+[
            plotColor2,
            draw=black,
            postaction={
                    pattern=north west lines
                },
        ] table[x=acc, y index=0] {\data};
    \end{axis}
\end{tikzpicture}

%% file: plots/study/study_time.tex
\begin{tikzpicture}[font=\scriptsize]
    \begin{axis}
        [
            width=\textwidth,
            boxplot/draw direction=x,
            boxplot={box extend=0.4, whisker extend=0.16},
            every axis plot/.append style={fill opacity=0.75},
            xlabel={Seconds},
            ytick={1, 2, 3, 4},
            yticklabels={$k = 5$, $k = 10$, $k = 15$, full},
            grid=major,
            area legend,
            legend entries={Time},
            legend cell align={left},
            legend pos=south east,
            cycle list={
                    {draw=black, fill=plotColor3, solid},
                    {draw=black, fill=plotColor3, solid}},
        ]
        \addplot+[
            boxplot prepared={
                    median=25.9124325,
                    upper quartile=42.6554125,
                    lower quartile=16.46201825,
                    upper whisker=77.286027,
                    lower whisker=3.983776,
                },
        ] coordinates {};
        \addplot+[
            boxplot prepared={
                    median=31.968845,
                    upper quartile=53.981310500000006,
                    lower quartile=19.513562999999998,
                    upper whisker=103.735101,
                    lower whisker=4.646334,
                },
        ] coordinates {};
        \addplot+[
            boxplot prepared={
                    median=39.511359,
                    upper quartile=64.22340650000001,
                    lower quartile=21.994775500000003,
                    upper whisker=121.149694,
                    lower whisker=4.938067,
                },
        ] coordinates {};
        \addplot+[
            boxplot prepared={
                    median=52.213253,
                    upper quartile=105.475054,
                    lower quartile=28.886381749999998,
                    upper whisker=205.403892,
                    lower whisker=6.044302,
                },
        ] coordinates {};
    \end{axis}
\end{tikzpicture}

%% file: plots/token_limitation.tex
\begin{tikzpicture}[font=\scriptsize]
    \pgfplotstableread[col sep=comma]{plots/token_limitation_cdf.csv}\data
    \begin{axis}[
            width=0.7\textwidth,
            height=5cm,
            xlabel={Missed tokens [ratio]},
            ylabel={Query-document pairs [ratio]},
            grid=major,
            legend entries={\plsem{}, \srlin{}},
            legend cell align={left},
            legend pos=north west,
        ]
        \addplot+[
            plotColor1,
            mark=none,
            thick,
        ] table[y=plsem] {\data};
        \addplot+[
            plotColor2,
            mark=none,
            thick,
        ] table[y=srlin] {\data};
    \end{axis}
\end{tikzpicture}

%% file: plots/explaining_bert/explaining_bert_trecdl.tex
\begin{tikzpicture}[font=\scriptsize]
    \pgfplotstableread[col sep=comma]{plots/explaining_bert/explaining_bert.csv}\data
    \begin{axis}[
            width=\textwidth,
            xtick={1, 5, 10, 15, 20},
            xticklabels={$k=1$, $k=5$, $k=10$, $k=15$, $k=20$},
            x tick label style={rotate=90, anchor=east},
            grid=major,
        ]
        \addplot[
            domain=1:20,
            samples=2,
            dashed,
            mark=none,
            thick,
            plotColor2,
        ] {0.586};
        \addplot+[
            const plot mark mid,
            thick,
            mark=none,
            mark size=1.5pt,
            plotColor3,
            opacity=0.75,
        ] table[y index=1] {\data};
    \end{axis}
\end{tikzpicture}

%% file: plots/explaining_bert/explaining_bert_core.tex
\begin{tikzpicture}[font=\scriptsize]
    \pgfplotstableread[col sep=comma]{plots/explaining_bert/explaining_bert.csv}\data
    \begin{axis}[
            width=\textwidth,
            xtick={1, 5, 10, 15, 20},
            xticklabels={$k=1$, $k=5$, $k=10$, $k=15$, $k=20$},
            x tick label style={rotate=90, anchor=east},
            grid=major,
            legend entries={\bertcls{}, \srlin{}},
            legend cell align={left},
            legend pos=south east,
            legend style={font=\tiny},
        ]
        \addplot[
            domain=1:20,
            samples=2,
            dashed,
            mark=none,
            thick,
            plotColor2,
        ] {0.4192};
        \addplot+[
            const plot mark mid,
            thick,
            mark=none,
            mark size=1.5pt,
            plotColor3,
            opacity=0.75,
        ] table[y index=2] {\data};
    \end{axis}
\end{tikzpicture}

%% file: plots/explaining_bert/explaining_bert_clueweb.tex
\begin{tikzpicture}[font=\scriptsize]
    \pgfplotstableread[col sep=comma]{plots/explaining_bert/explaining_bert.csv}\data
    \begin{axis}[
            width=\textwidth,
            xtick={1, 5, 10, 15, 20},
            xticklabels={$k=1$, $k=5$, $k=10$, $k=15$, $k=20$},
            x tick label style={rotate=90, anchor=east},
            grid=major,
        ]
        \addplot[
            domain=1:20,
            samples=2,
            dashed,
            mark=none,
            thick,
            plotColor2,
        ] {0.313};
        \addplot+[
            const plot mark mid,
            thick,
            mark=none,
            mark size=1.5pt,
            plotColor3,
            opacity=0.75,
        ] table[y index=3] {\data};
    \end{axis}
\end{tikzpicture}

%% file: plots/explaining_bert/explaining_bert_sr.tex
\begin{tikzpicture}[font=\scriptsize]
    \pgfplotstableread[col sep=comma]{plots/explaining_bert/explaining_bert_sr.csv}\data
    \begin{axis}[
            width=\textwidth,
            height=4cm,
            xlabel={Number of sentences $k$},
            ymin=0.2,
            ymax=0.65,
            grid=major,
            legend entries={\srlin{}},
            legend cell align={left},
            legend pos=south east,
        ]
        \addplot+[
            const plot mark mid,
            thick,
            mark=square,
            mark size=1.5pt,
            plotColor3,
            opacity=0.75,
        ] table[y index=1] {\data};
    \end{axis}
\end{tikzpicture}

%% file: plots/explaining_bert/explaining_bert_trunc.tex
\begin{tikzpicture}[font=\scriptsize]
    \pgfplotstableread[col sep=comma]{plots/explaining_bert/explaining_bert_trunc.csv}\data
    \begin{axis}[
            width=\textwidth,
            height=4cm,
            xlabel={Number of tokens $T$},
            ymin=0.2,
            ymax=0.65,
            grid=major,
            legend entries={\bertcls{}},
            legend cell align={left},
            legend pos=north west,
        ]
        \addplot+[
            const plot mark mid,
            thick,
            mark=o,
            mark size=1.5pt,
            plotColor2,
            opacity=0.75,
        ] table[y index=1] {\data};
    \end{axis}
\end{tikzpicture}

%% file: table-anecdotal.tex
\begin{table*}
    \centering
    \scriptsize
    \begin{tabularx}{\textwidth}{rll}
        \toprule
        \textbf{Rank} & \textbf{Document} & \textbf{Most Relevant Sentence} \\
        \midrule
        \multicolumn{3}{l}{\textbf{\sratt{}}} \\
        1 & \textcolor{blue}{D970461$^{+}$} & How long do I hold yoga poses? \\
        2 & \textcolor{blue}{D3378721$^{+}$} & How Long to Hold Bikram Yoga Poses. \\
        3 & \textcolor{blue}{D970460$^{+}$} & How Long You Should Hold A Yoga Posture? \\
        4 & \textcolor{blue}{D1211050$^{+}$} & How Long To Hold Yoga Pose To Gain All The Benefits? \\
        5 & \textcolor{red}{D337672$^{-}$} & One way to build strength and endurance is to pull your hunting bow [...] before releasing the arrow [...] \\
        6 & \textcolor{red}{D2587656$^{-}$} & Traditional Closing of a Yoga Practice [...] the teacher will say ``namaste'' \& bow to students. \\
        7 & \textcolor{red}{D1125612$^{-}$} & Consult your doctor before beginning these new flexibility exercises [...] \\
        8 & \textcolor{red}{D520508$^{-}$} & Yoga should be done with an open, gentle, and non-critical mind [...] working on one's limits \\
        \midrule
        \multicolumn{3}{l}{\textbf{\plsem{}}} \\
        1 & \textcolor{red}{D3378723$^{-}$} & [...] Bow Pose is an intermediate yoga backbend that deeply opens the chest and the front of the body. \\
        2 & \textcolor{blue}{D970458$^{+}$} & In the style of hatha yoga I teach there are longer holds in the poses. \\
        3 & \textcolor{red}{D3378725$^{-}$} & [...] After a brief break, you move into the last eight standing exercises [...] of the Bikram yoga sequence \\
        4 & \textcolor{blue}{D970461$^{+}$} & How long do I hold yoga poses? \\
        5 & \textcolor{blue}{D337672$^{+}$} & [...] isolate the muscles needed to pull the bow back and hold the bow up [...] \\
        6 & \textcolor{blue}{D2285733$^{+}$} & Straighten your legs, so that your body makes a ‘V’ shape and hold this position for 2 to 5 breaths. \\
        7 & \textcolor{red}{D1930297$^{-}$} & IF YOU ARE A BEGINNER, YOU OUGHT TO BEND YOUR KNEES SLIGHTLY TO ACCOMPLISH THIS. \\
        8 & \textcolor{red}{D520508$^{-}$} & Iyengar yoga can be good for physical therapy [...] easier for some people to get into the yoga postures. \\
        \midrule
        \multicolumn{3}{l}{\textbf{Query}: \texttt{how long to hold bow in yoga} (query ID 1132213)} \\
        \bottomrule
    \end{tabularx}
    \caption{Example rankings from the \trecdl{} dataset with the most relevant selected sentences. Document IDs have a suffix (+/-) indicating the relevance of the corresponding TREC judgments.}
    \label{tab:trecdl_examples}
\end{table*}

\begin{table*}
    \centering
    \scriptsize
    \begin{tabularx}{\textwidth}{crl}
        \toprule
        \multicolumn{2}{r}{\textbf{Rank}} & \textbf{Sentence} \\
        \midrule
        \multicolumn{3}{l}{\textbf{Document}: \texttt{Commodore\_(rank)}, \textbf{Rank}: 3} \\
        \cmidrule{1-2}
        \parbox[b]{0pt}{\multirow{5}{*}{\rotatebox[origin=c]{90}{selected}}}
        & \bf 1 & A commodore's ship is typically designated by the flying of a Broad pennant, as opposed to an admiral's flag. \\
        & \bf 2 & \green{Commodore is a naval rank used in many navies that is superior to a navy captain, but below a rear admiral.} \\
        & \bf 3 & It is sometimes abbreviated: as "Cdre" in British Royal Navy, "CDRE" in the US Navy [...] \\
        & \bf 4 & Commodore (rank). \\
        & \bf 5 & Non-English-speaking nations often use the rank of flotilla admiral or counter admiral [...] \\
        \cmidrule{1-2}
        & 6 & As an official rank, a commodore typically commands a flotilla or squadron of ships [...] \\
        & 7 & Traditionally, "commodore" is the title for any officer assigned to command more than one ship [...] \\
        & 8 & It is often regarded as a one-star rank with a NATO code of OF-6 [...] \\
        \midrule
        \multicolumn{3}{l}{\textbf{Document}: \texttt{Rear\_admiral}, \textbf{Rank}: 23} \\
        \cmidrule{1-2}
        \parbox[b]{0pt}{\multirow{5}{*}{\rotatebox[origin=c]{90}{selected}}}
        & \bf 1 & In the German Navy the rank is known as Konteradmiral, superior to the flotilla admiral (Commodore in other navies). \\
        & \bf 2 & In the Royal Netherlands Navy, this rank is known as schout-bij-nacht (lit. \\
        & \bf 3 & [...] and in the Canadian Forces' French rank translations, the rank of rear admiral is known as contre-amiral. \\
        & \bf 4 & In some European navies (e.g. \\
        & \bf 5 & In many navies it is referred to as a two-star rank (OF-7). \\
        \cmidrule{1-2}
        & ... & ... \\
        & 13 & \green{Rear admiral is a naval commissioned officer rank above that of a commodore and captain, and below [...]} \\
        & 14 & Each naval squadron would be assigned an admiral as its head, who would command from the centre vessel [...] \\
        \midrule
        \multicolumn{3}{l}{\textbf{Query}: \texttt{Commodore is ranked above a rear admiral.} (query ID 204575)} \\
        \bottomrule
    \end{tabularx}
    \caption{Example selections of sentences from relevant documents w.r.t.\ a query from the FEVER dataset by \srlin{}. The selector selects the highest scoring $k = 5$ sentences from each document. The final rank of a document is computed using only these sentences. The sentences that contain the answer are \green{highlighted}.}
    \label{tab:fever_examples}
\end{table*}

%% file: applications.tex
\section{Applications of \sr{}}
\label{sec:applications}
In this section we highlight several real-world applications of \sr{} models.

\subsection{Discovering Biased or Buggy Ranking Decisions}
\label{sec:application_bugs_bias}
\begin{figure}
    \small
    \begin{subfigure}{.49\linewidth}
        \centering
        \begin{tabularx}{.9\linewidth}{X}
            \toprule
            \textit{What makes Bikram yoga unique is its focus on practicing yoga in a room heated to 105 degrees Fahrenheit with 40 percent humidity. In Bikram yoga, be prepared to sweat profusely and come armed with a towel and lots of water. To practice Bikram at home, you'll need a space heater and access to the pose sequence. \green{On a general basis, you need to hold the yoga poses for about 10-12 breaths.} \green{With practice, you can also go up to 30 breaths.} We chatted for a few moments, and found that we came to completely different conclusions. [...]} \\
            \bottomrule
        \end{tabularx}
        \caption{Unaltered relevant document}
    \end{subfigure}
    \begin{subfigure}{.49\linewidth}
        \centering
        \begin{tabularx}{0.9\linewidth}{X}
            \toprule
            \textit{\textbf{THIS IS A BUG.} What makes Bikram yoga unique is its focus on practicing yoga in a room heated to 105 degrees Fahrenheit with 40 percent humidity. In Bikram yoga, be prepared to sweat profusely and come armed with a towel and lots of water. To practice Bikram at home, you'll need a space heater and access to the pose sequence. \green{On a general basis, you need to hold the yoga poses for about 10-12 breaths.} \green{With practice, you can also go up to 30 breaths.} We chatted for a few moments, and found that we came to completely different conclusions. [...]} \\
            \bottomrule
        \end{tabularx}
        \caption{Relevant document with label leakage}
    \end{subfigure}
    \caption{An example illustrating label leakage for the query \texttt{how long to hold bow in yoga}. For each document that is relevant to the query (a), we prepend the sentence \texttt{THIS IS A BUG} (b). As irrelevant documents are unaffected, this simulates label leakage for training and test data. The \green{highlighted} sentences correspond to the part of the document that provides the answer to the query.}
    \label{fig:label_leakage_docs}
\end{figure}
\begin{figure}
    \begin{subfigure}{.49\linewidth}
        \centering
        \input{plots/leakage/leakage_selections}
        \caption{Fraction of documents where the leakage sentence has been selected (assigned the highest score by the selector) for each query}
    \end{subfigure}
    \begin{subfigure}{.49\linewidth}
        \centering
        \input{plots/leakage/leakage_ranks}
        \caption{Distribution of the ranks of the leakage sentence as assigned by the selector over all relevant documents in the test set}
    \end{subfigure}
    \caption{This example shows how label leakage can be discovered using \sr{} models. The plots illustrate the sentence selections on the \trecdl{} test set by an \srlin{} model using a modified corpus to simulate label leakage (cf.\ Figure~\ref{fig:label_leakage_docs}). Evidently, the extractive \emph{explanations} (selected sentences) provided by our model reliably uncover the the label leakage by including the leakage sentences.}
    \label{fig:label_leakage_selections}
\end{figure}
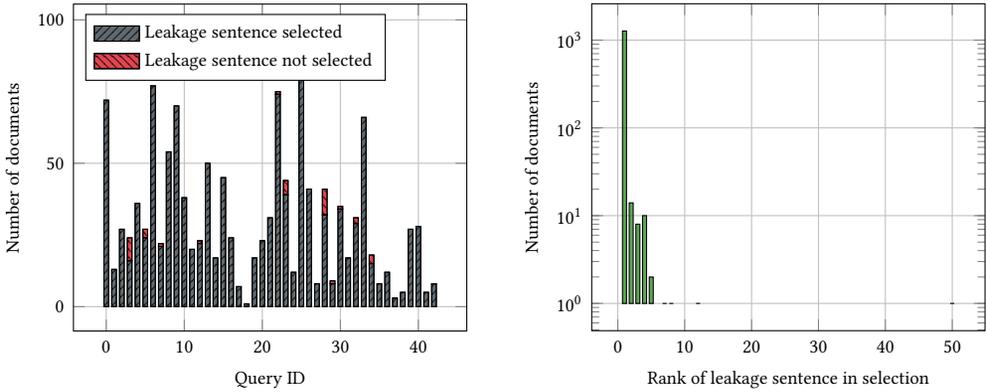
Neural rankers, just like any other machine learning model, are susceptible to bias or bugs in their ranking decisions~\cite{adebayo2020debugging}. Such models can achieve high performance, but the rationale (or reasoning) behind the model decisions is often incorrect, i.e.\ the models are right for the wrong reasons. In a similar fashion, a line of work employs \emph{adversarial attacks} to craft model inputs, which are often merely slightly modified examples from real datasets, that yield highly unexpected model decisions or outputs~\cite{wu2022prada,wallace2021concealed}.

In this section we conduct an experiment to show how the \sr{} paradigm can be employed to uncover such biased or even buggy decisions of the ranking model. Specifically, we enforce biased ranking decisions by augmenting the MS MARCO corpus to include label leakage; this means that, for every query-document pair $(q, d)$ in the training and test set, where $d$ is relevant to $q$, we replace $d$ by $d'$, where $d'$ is simply a copy of the original document with one additional sentence injected. This process is illustrated in Figure~\ref{fig:label_leakage_docs}. As a result, a ranking model trained on this data simply learns to rank documents that contain the injected sentence high (independently of the query). In fact, the model reaches a MAP of $0.39$, nDCG@20 of $0.66$ and MRR of $1.00$ on the \trecdl{} test set (cf.\ Table~\ref{tab:model_eval}) due to the label leakage. Figure~\ref{fig:label_leakage_selections} shows how the explanations provided by an \srlin{} model uncover the bias in the ranking decisions; specifically, it illustrates how the selector assigns the highest importance to the sentence containing the label leakage (and hence includes it in the \emph{explanation}) in all but very few cases. As a result, an examination of the ranking explanations immediately uncovers this bug.

\subsection{Improving Search Engines}
\label{sec:application_search_engine}
In general, search engines do not give end-users much of an idea about the reasoning behind marking a document as relevant or ranking one document higher than another. Instead, users have to rely on the results of the search engines. In turn, most search engines use the click information of users to judge the relevance of a document and iteratively update their search and ranking algorithm~\cite{craswell2020orcas}. This introduces bias in determining the importance of web pages. Content creators may use clickbait~\cite{chakraborty2016stop,gecckil2018clickbait} to attract users and increase the importance of their content or web page, even though it does not contain the relevant content. This is also a challenging task for search engine optimization.

Our \sr{}-based document ranking architecture can be used to alleviate the above-mentioned two problems to an extent:
\begin{enumerate}
    \item The \sr{} paradigm identifies the relevance of a document with respect to a query and also extracts relevant snippets from the document. If the system highlights those snippets along with the document, users can make their click decisions more accurately, helping them to skip clickbait contents.
    \item It is very difficult to judge the relevance of a document just from the title. For this reason, search engines display \emph{snippets} of documents on the results page. These snippets are often relatively short (i.e.\ one or two sentences or parts of sentences) and are supposed to highlight why the user might be interested in the document. Usually, these snippets are based on term-matching with the query, i.e.\ matching terms in the snippet are printed bold. \sr{} could be used as an alternative way of generating these snippets (for small values of $k$) such that they also explain the reasoning behind the ranking itself. The highlighting of matching terms could then be performed on the selected sentences as well.
\end{enumerate}

%% file: plots/leakage/leakage_selections.tex
\begin{tikzpicture}[font=\scriptsize]
    \pgfplotstableread[col sep=comma]{plots/leakage/leakage_selections.csv}\data
    \begin{axis}[
            ybar stacked,
            width=\textwidth,
            bar width=0.6,
            every axis plot/.append style={fill, fill opacity=0.75},
            xlabel={Query ID},
            ylabel={Number of documents},
            grid=major,
            area legend,
            legend entries={Leakage sentence selected, Leakage sentence not selected},
            legend cell align={left},
            legend pos=north west,
        ]
        \addplot+[
            plotColor1,
            draw=black,
            postaction={
                    pattern=north east lines
                },
        ] table[x=x, y=selected] {\data};
        \addplot+[
            plotColor2,
            draw=black,
            postaction={
                    pattern=north west lines
                },
        ] table[x=x, y=not selected] {\data};
    \end{axis}
\end{tikzpicture}

%% file: plots/leakage/leakage_ranks.tex
\begin{tikzpicture}[font=\scriptsize]
    \pgfplotstableread[col sep=comma]{plots/leakage/leakage_all_ranks.csv}\data
    \begin{axis}[
            ybar,
            width=\textwidth,
            bar width=0.6,
            ymode=log,
            every axis plot/.append style={fill, fill opacity=0.75},
            xlabel={Rank of leakage sentence in selection},
            ylabel={Number of documents},
            xtick align=inside,
            grid=major,
        ]
        \addplot+[
            plotColor3,
            draw=black,
        ] table[x=rank, y=num] {\data};
    \end{axis}
\end{tikzpicture}

%% file: conclusion.tex
\section{Conclusion}
In this paper we proposed \sr{}, a ranking framework that is interpretable by design. Our selection and ranking models are trainable end-to-end by gradient-based optimization techniques using a combination of the gumbel-max trick and reparameterizable subset sampling. In our experiments we found that, by enforcing sparsity in document representations by selecting a subset of sentences, we still perform on par with state-of-the-art models, while being interpretable. We showed how \sr{} can be used to explain the decisions for a large number of ranking tasks from the BEIR dataset in the zero-shot setting. This proves its potential of wide-ranging utility in a large number of knowledge-intensive tasks. We showed that there is no considerable performance difference in case of complex selectors, indicating that simple and fast selectors can be used instead. We also found that there is a sweet spot in the choice of sparsity that varies depending on the dataset. We performed a user study to highlight the utility of our extractive explanations to human users. We believe that the applicability of a sparsity-inducing component can extend beyond document ranking to other ranking~\cite{holzmann2016tempas,holzmann2017exploring,singh2016expedition}, graph~\cite{funke2021zorro} and web tasks~\cite{anand2020conversational,singh2016discovering,roy2021question}.

%% file: appendix.tex
\section{User Study Details}
\label{app:study_details}
In this section we present our user study (as described in Section~\ref{sec:sr_faithfulness}) in more detail.

\subsection{Interface}
\label{sec:study_interface}
\begin{figure*}
	\centering
    \frame{\includegraphics[width=\linewidth]{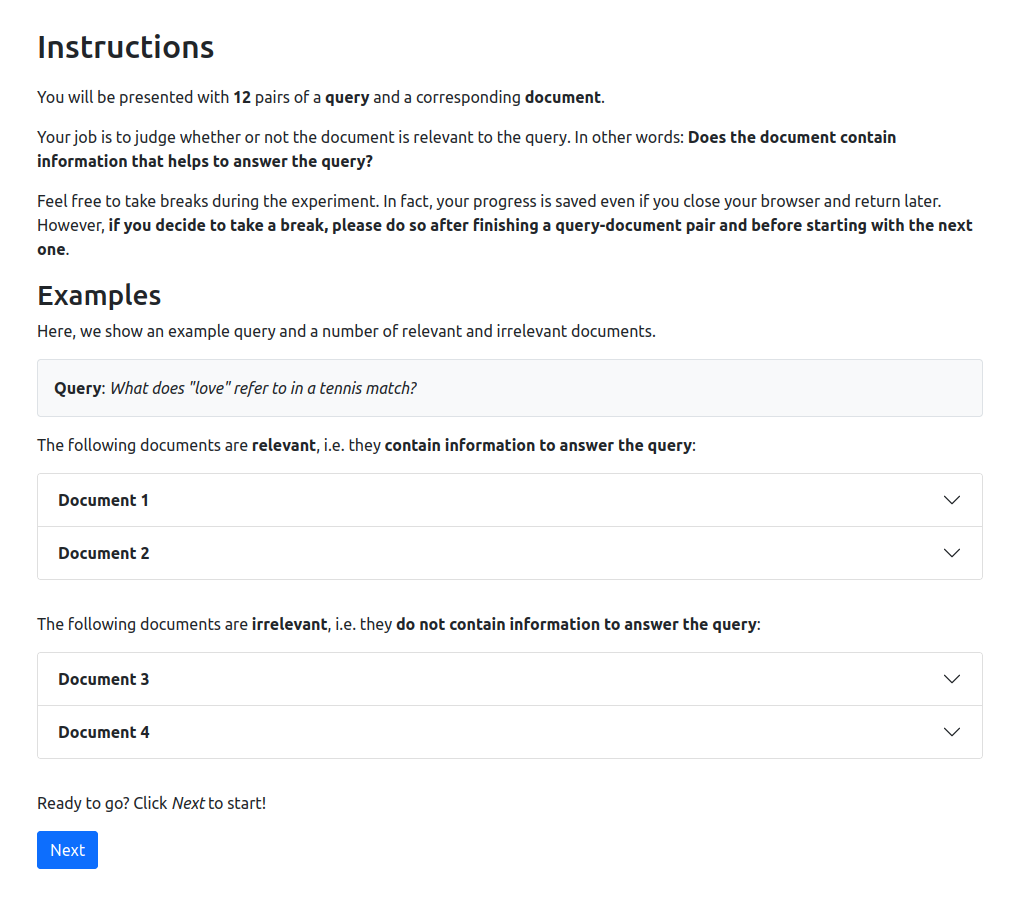}}
	\caption{The instructions page. This page is only displayed once and includes a task description along with some examples.}
	\label{fig:study_instructions}
\end{figure*}
\begin{figure*}
	\centering
    \frame{\includegraphics[width=\linewidth]{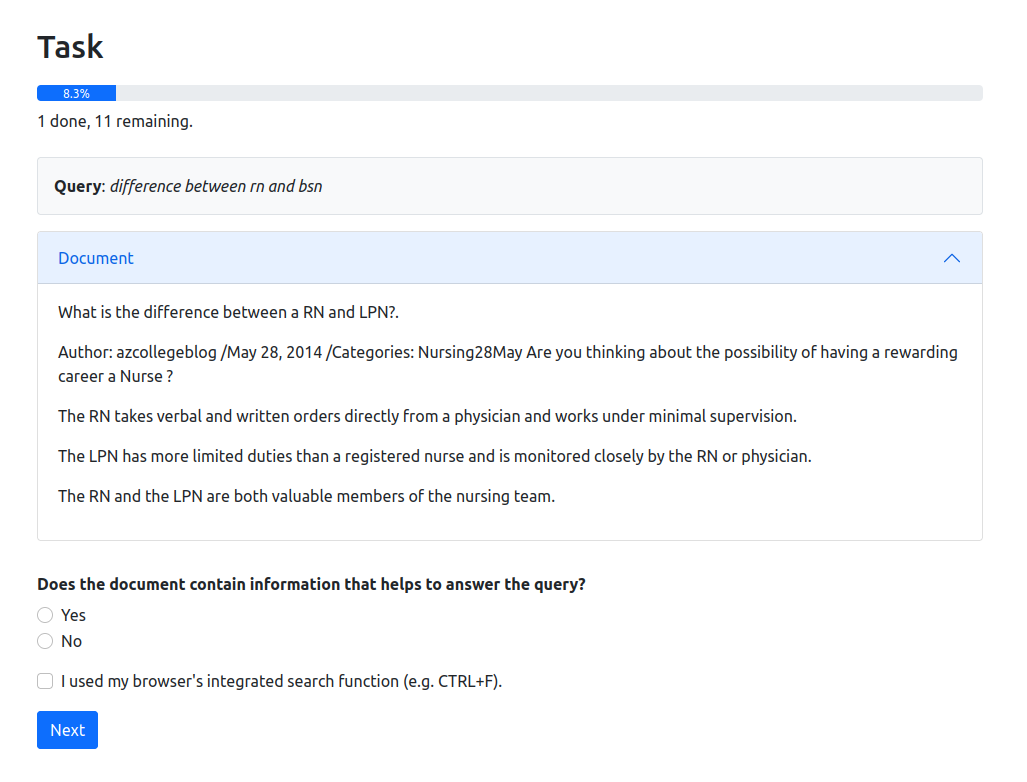}}
	\caption{The task page. It shows a query-document pair and gathers the participant's response (whether the document is relevant to the query) along with whether or not they used the browser search for this instance.}
	\label{fig:study_task}
\end{figure*}
\begin{figure*}
	\centering
    \frame{\includegraphics[width=\linewidth]{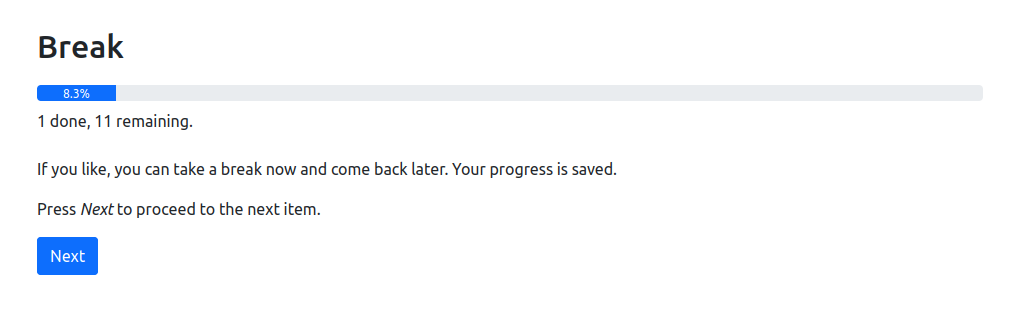}}
	\caption{The break page. It is displayed between two consecutive task pages.}
	\label{fig:study_break}
\end{figure*}
The main part of the study consists of three pages:
\begin{enumerate}
    \item The \emph{instructions page} (Figure~\ref{fig:study_instructions}) familiarizes the participant with the task and provides examples.
    \item The \emph{task page} (Figure~\ref{fig:study_task}) presents a query-document pair to the participant and records their response. In the background, we record how much time the participant has spent on this page in order to measure the time it took to judge the query-document pair.
    \item The \emph{break page} (Figure~\ref{fig:study_break}) is a simple intermediate page that is shown in between two consecutive task pages. The reason for this is that we want participants to only take breaks \emph{between} two tasks, so that our time measurements are as accurate as possible.
\end{enumerate}
The study is structured in \emph{rounds}; a round consists of a task page and a subsequent break page. In our experiment each participant completed 12 rounds. Before the first round the instructions are shown.

\subsection{Collection and Usage of Data}
\label{app:study_data}
We collect three data points during each round:
\begin{enumerate}
    \item The relevance judgment (boolean),
    \item the usage of browser search (boolean) and
    \item the time taken to come up with the relevance judgment (float).
\end{enumerate}
The data is inserted into a database after each page. This allows users to take a break from the study and continue where they left off later, even if they closed their browser in the meantime.

\subsubsection{Computation of Metrics}
\label{app:study_metrics}
In the results (Section~\ref{sec:sr_faithfulness}) we present the following metrics:
\begin{enumerate}
    \item \textbf{Accuracy}: This is simply the number of correctly judged instances divided by the total number of instances. Correctness is determined using the official TREC query relevances $R$, which are converted to binary according to the official guidelines, i.e.\ irrelevant ($R(q, d) = 0$) or relevant ($R(q, d) > 0)$.
    \item \textbf{Search function usage}: Similarly to accuracy, we divide the number of instances where participants have indicated the usage of their browser's search function by the total number of instances.
    \item \textbf{Time taken to complete a relevance judgment}: The time is measured as the difference between two time stamps: The first one is recorded when the participant leaves the break page, the second one when the user completes the task page.
\end{enumerate}

%% file: main.bbl

\begin{thebibliography}{92}


\ifx \showCODEN    \undefined \def \showCODEN     #1{\unskip}     \fi
\ifx \showDOI      \undefined \def \showDOI       #1{#1}\fi
\ifx \showISBNx    \undefined \def \showISBNx     #1{\unskip}     \fi
\ifx \showISBNxiii \undefined \def \showISBNxiii  #1{\unskip}     \fi
\ifx \showISSN     \undefined \def \showISSN      #1{\unskip}     \fi
\ifx \showLCCN     \undefined \def \showLCCN      #1{\unskip}     \fi
\ifx \shownote     \undefined \def \shownote      #1{#1}          \fi
\ifx \showarticletitle \undefined \def \showarticletitle #1{#1}   \fi
\ifx \showURL      \undefined \def \showURL       {\relax}        \fi
\providecommand\bibfield[2]{#2}
\providecommand\bibinfo[2]{#2}
\providecommand\natexlab[1]{#1}
\providecommand\showeprint[2][]{arXiv:#2}

\bibitem[\protect\citeauthoryear{Adebayo, Muelly, Liccardi, and Kim}{Adebayo
  et~al\mbox{.}}{2020}]%
        {adebayo2020debugging}
\bibfield{author}{\bibinfo{person}{Julius Adebayo}, \bibinfo{person}{Michael
  Muelly}, \bibinfo{person}{Ilaria Liccardi}, {and} \bibinfo{person}{Been
  Kim}.} \bibinfo{year}{2020}\natexlab{}.
\newblock \showarticletitle{Debugging Tests for Model Explanations}. In
  \bibinfo{booktitle}{\emph{Advances in Neural Information Processing
  Systems}}, \bibfield{editor}{\bibinfo{person}{H.~Larochelle},
  \bibinfo{person}{M.~Ranzato}, \bibinfo{person}{R.~Hadsell},
  \bibinfo{person}{M.F. Balcan}, {and} \bibinfo{person}{H.~Lin}} (Eds.),
  Vol.~\bibinfo{volume}{33}. \bibinfo{publisher}{Curran Associates, Inc.},
  \bibinfo{pages}{700--712}.
\newblock
\urldef\tempurl%
\url{https://proceedings.neurips.cc/paper/2020/file/075b051ec3d22dac7b33f788da631fd4-Paper.pdf}
\showURL{%
\tempurl}


\bibitem[\protect\citeauthoryear{Akkalyoncu~Yilmaz, Yang, Zhang, and
  Lin}{Akkalyoncu~Yilmaz et~al\mbox{.}}{2019}]%
        {yilmaz2019cross}
\bibfield{author}{\bibinfo{person}{Zeynep Akkalyoncu~Yilmaz},
  \bibinfo{person}{Wei Yang}, \bibinfo{person}{Haotian Zhang}, {and}
  \bibinfo{person}{Jimmy Lin}.} \bibinfo{year}{2019}\natexlab{}.
\newblock \showarticletitle{Cross-Domain Modeling of Sentence-Level Evidence
  for Document Retrieval}. In \bibinfo{booktitle}{\emph{Proceedings of the 2019
  Conference on Empirical Methods in Natural Language Processing and the 9th
  International Joint Conference on Natural Language Processing
  (EMNLP-IJCNLP)}}. \bibinfo{publisher}{Association for Computational
  Linguistics}, \bibinfo{address}{Hong Kong, China},
  \bibinfo{pages}{3490--3496}.
\newblock
\urldef\tempurl%
\url{https://doi.org/10.18653/v1/D19-1352}
\showDOI{\tempurl}


\bibitem[\protect\citeauthoryear{Althammer, Hofst{\"a}tter, Sertkan, Verberne,
  and Hanbury}{Althammer et~al\mbox{.}}{2022}]%
        {althammer2022parm}
\bibfield{author}{\bibinfo{person}{Sophia Althammer},
  \bibinfo{person}{Sebastian Hofst{\"a}tter}, \bibinfo{person}{Mete Sertkan},
  \bibinfo{person}{Suzan Verberne}, {and} \bibinfo{person}{Allan Hanbury}.}
  \bibinfo{year}{2022}\natexlab{}.
\newblock \showarticletitle{PARM: A Paragraph Aggregation Retrieval Model for
  Dense Document-to-Document Retrieval}. In \bibinfo{booktitle}{\emph{Advances
  in Information Retrieval}}, \bibfield{editor}{\bibinfo{person}{Matthias
  Hagen}, \bibinfo{person}{Suzan Verberne}, \bibinfo{person}{Craig Macdonald},
  \bibinfo{person}{Christin Seifert}, \bibinfo{person}{Krisztian Balog},
  \bibinfo{person}{Kjetil N{\o}rv{\aa}g}, {and} \bibinfo{person}{Vinay Setty}}
  (Eds.). \bibinfo{publisher}{Springer International Publishing},
  \bibinfo{address}{Cham}, \bibinfo{pages}{19--34}.
\newblock
\showISBNx{978-3-030-99736-6}


\bibitem[\protect\citeauthoryear{Anand, Cavedon, Joho, Sanderson, and
  Stein}{Anand et~al\mbox{.}}{2020}]%
        {anand2020conversational}
\bibfield{author}{\bibinfo{person}{Avishek Anand}, \bibinfo{person}{Lawrence
  Cavedon}, \bibinfo{person}{Hideo Joho}, \bibinfo{person}{Mark Sanderson},
  {and} \bibinfo{person}{Benno Stein}.} \bibinfo{year}{2020}\natexlab{}.
\newblock \showarticletitle{Conversational search (dagstuhl seminar 19461)}. In
  \bibinfo{booktitle}{\emph{Dagstuhl Reports}}, Vol.~\bibinfo{volume}{9}.
  Schloss Dagstuhl-Leibniz-Zentrum f{\"u}r Informatik.
\newblock


\bibitem[\protect\citeauthoryear{Bahdanau, Cho, and Bengio}{Bahdanau
  et~al\mbox{.}}{2014}]%
        {bahdanau2014neural}
\bibfield{author}{\bibinfo{person}{Dzmitry Bahdanau},
  \bibinfo{person}{Kyunghyun Cho}, {and} \bibinfo{person}{Yoshua Bengio}.}
  \bibinfo{year}{2014}\natexlab{}.
\newblock \bibinfo{title}{Neural Machine Translation by Jointly Learning to
  Align and Translate}.
\newblock
\newblock
\urldef\tempurl%
\url{https://doi.org/10.48550/ARXIV.1409.0473}
\showDOI{\tempurl}


\bibitem[\protect\citeauthoryear{Bastings, Aziz, and Titov}{Bastings
  et~al\mbox{.}}{2019}]%
        {bastings2019interpretable}
\bibfield{author}{\bibinfo{person}{Jasmijn Bastings}, \bibinfo{person}{Wilker
  Aziz}, {and} \bibinfo{person}{Ivan Titov}.} \bibinfo{year}{2019}\natexlab{}.
\newblock \showarticletitle{Interpretable Neural Predictions with
  Differentiable Binary Variables}. In \bibinfo{booktitle}{\emph{Proceedings of
  the 57th Annual Meeting of the Association for Computational Linguistics}}.
  \bibinfo{publisher}{Association for Computational Linguistics},
  \bibinfo{address}{Florence, Italy}, \bibinfo{pages}{2963--2977}.
\newblock
\urldef\tempurl%
\url{https://doi.org/10.18653/v1/P19-1284}
\showDOI{\tempurl}


\bibitem[\protect\citeauthoryear{Bengio, Léonard, and Courville}{Bengio
  et~al\mbox{.}}{2013}]%
        {bengio2013estimating}
\bibfield{author}{\bibinfo{person}{Yoshua Bengio}, \bibinfo{person}{Nicholas
  Léonard}, {and} \bibinfo{person}{Aaron Courville}.}
  \bibinfo{year}{2013}\natexlab{}.
\newblock \bibinfo{title}{Estimating or Propagating Gradients Through
  Stochastic Neurons for Conditional Computation}.
\newblock
\newblock
\urldef\tempurl%
\url{https://doi.org/10.48550/ARXIV.1308.3432}
\showDOI{\tempurl}


\bibitem[\protect\citeauthoryear{Chakraborty, Paranjape, Kakarla, and
  Ganguly}{Chakraborty et~al\mbox{.}}{2016}]%
        {chakraborty2016stop}
\bibfield{author}{\bibinfo{person}{Abhijnan Chakraborty},
  \bibinfo{person}{Bhargavi Paranjape}, \bibinfo{person}{Sourya Kakarla}, {and}
  \bibinfo{person}{Niloy Ganguly}.} \bibinfo{year}{2016}\natexlab{}.
\newblock \showarticletitle{Stop Clickbait: Detecting and Preventing Clickbaits
  in Online News Media}. In \bibinfo{booktitle}{\emph{Proceedings of the 2016
  IEEE/ACM International Conference on Advances in Social Networks Analysis and
  Mining}} (Davis, California) \emph{(\bibinfo{series}{ASONAM '16})}.
  \bibinfo{publisher}{IEEE Press}, \bibinfo{pages}{9–16}.
\newblock
\showISBNx{9781509028467}


\bibitem[\protect\citeauthoryear{Chen, Schonger, and Wickens}{Chen
  et~al\mbox{.}}{2016}]%
        {chen2016otree}
\bibfield{author}{\bibinfo{person}{Daniel~L. Chen}, \bibinfo{person}{Martin
  Schonger}, {and} \bibinfo{person}{Chris Wickens}.}
  \bibinfo{year}{2016}\natexlab{}.
\newblock \showarticletitle{oTree—An open-source platform for laboratory,
  online, and field experiments}.
\newblock \bibinfo{journal}{\emph{Journal of Behavioral and Experimental
  Finance}}  \bibinfo{volume}{9} (\bibinfo{year}{2016}),
  \bibinfo{pages}{88--97}.
\newblock
\showISSN{2214-6350}
\urldef\tempurl%
\url{https://doi.org/10.1016/j.jbef.2015.12.001}
\showDOI{\tempurl}


\bibitem[\protect\citeauthoryear{Cheng, Dong, and Lapata}{Cheng
  et~al\mbox{.}}{2016}]%
        {cheng2016long}
\bibfield{author}{\bibinfo{person}{Jianpeng Cheng}, \bibinfo{person}{Li Dong},
  {and} \bibinfo{person}{Mirella Lapata}.} \bibinfo{year}{2016}\natexlab{}.
\newblock \showarticletitle{Long Short-Term Memory-Networks for Machine
  Reading}. In \bibinfo{booktitle}{\emph{Proceedings of the 2016 Conference on
  Empirical Methods in Natural Language Processing}}.
  \bibinfo{publisher}{Association for Computational Linguistics},
  \bibinfo{address}{Austin, Texas}, \bibinfo{pages}{551--561}.
\newblock
\urldef\tempurl%
\url{https://doi.org/10.18653/v1/D16-1053}
\showDOI{\tempurl}


\bibitem[\protect\citeauthoryear{Craswell, Campos, Mitra, Yilmaz, and
  Billerbeck}{Craswell et~al\mbox{.}}{2020}]%
        {craswell2020orcas}
\bibfield{author}{\bibinfo{person}{Nick Craswell}, \bibinfo{person}{Daniel
  Campos}, \bibinfo{person}{Bhaskar Mitra}, \bibinfo{person}{Emine Yilmaz},
  {and} \bibinfo{person}{Bodo Billerbeck}.} \bibinfo{year}{2020}\natexlab{}.
\newblock \showarticletitle{ORCAS: 20 Million Clicked Query-Document Pairs for
  Analyzing Search}. In \bibinfo{booktitle}{\emph{Proceedings of the 29th ACM
  International Conference on Information \&amp; Knowledge Management}}
  (Virtual Event, Ireland) \emph{(\bibinfo{series}{CIKM '20})}.
  \bibinfo{publisher}{Association for Computing Machinery},
  \bibinfo{address}{New York, NY, USA}, \bibinfo{pages}{2983–2989}.
\newblock
\showISBNx{9781450368599}
\urldef\tempurl%
\url{https://doi.org/10.1145/3340531.3412779}
\showDOI{\tempurl}


\bibitem[\protect\citeauthoryear{Cui, Chen, Wei, Wang, Liu, and Hu}{Cui
  et~al\mbox{.}}{2017}]%
        {cui2016attention}
\bibfield{author}{\bibinfo{person}{Yiming Cui}, \bibinfo{person}{Zhipeng Chen},
  \bibinfo{person}{Si Wei}, \bibinfo{person}{Shijin Wang},
  \bibinfo{person}{Ting Liu}, {and} \bibinfo{person}{Guoping Hu}.}
  \bibinfo{year}{2017}\natexlab{}.
\newblock \showarticletitle{Attention-over-Attention Neural Networks for
  Reading Comprehension}. In \bibinfo{booktitle}{\emph{Proceedings of the 55th
  Annual Meeting of the Association for Computational Linguistics (Volume 1:
  Long Papers)}}. \bibinfo{publisher}{Association for Computational
  Linguistics}, \bibinfo{address}{Vancouver, Canada},
  \bibinfo{pages}{593--602}.
\newblock
\urldef\tempurl%
\url{https://doi.org/10.18653/v1/P17-1055}
\showDOI{\tempurl}


\bibitem[\protect\citeauthoryear{Dai and Callan}{Dai and Callan}{2019}]%
        {dai2019deeper}
\bibfield{author}{\bibinfo{person}{Zhuyun Dai} {and} \bibinfo{person}{Jamie
  Callan}.} \bibinfo{year}{2019}\natexlab{}.
\newblock \showarticletitle{Deeper Text Understanding for IR with Contextual
  Neural Language Modeling}. In \bibinfo{booktitle}{\emph{Proceedings of the
  42nd International ACM SIGIR Conference on Research and Development in
  Information Retrieval}} (Paris, France) \emph{(\bibinfo{series}{SIGIR'19})}.
  \bibinfo{publisher}{Association for Computing Machinery},
  \bibinfo{address}{New York, NY, USA}, \bibinfo{pages}{985–988}.
\newblock
\showISBNx{9781450361729}
\urldef\tempurl%
\url{https://doi.org/10.1145/3331184.3331303}
\showDOI{\tempurl}


\bibitem[\protect\citeauthoryear{Dai, Xiong, Callan, and Liu}{Dai
  et~al\mbox{.}}{2018}]%
        {zhuyun2018convolutional}
\bibfield{author}{\bibinfo{person}{Zhuyun Dai}, \bibinfo{person}{Chenyan
  Xiong}, \bibinfo{person}{Jamie Callan}, {and} \bibinfo{person}{Zhiyuan Liu}.}
  \bibinfo{year}{2018}\natexlab{}.
\newblock \showarticletitle{Convolutional Neural Networks for Soft-Matching
  N-Grams in Ad-Hoc Search}. In \bibinfo{booktitle}{\emph{Proceedings of the
  Eleventh ACM International Conference on Web Search and Data Mining}} (Marina
  Del Rey, CA, USA) \emph{(\bibinfo{series}{WSDM '18})}.
  \bibinfo{publisher}{Association for Computing Machinery},
  \bibinfo{address}{New York, NY, USA}, \bibinfo{pages}{126–134}.
\newblock
\showISBNx{9781450355810}
\urldef\tempurl%
\url{https://doi.org/10.1145/3159652.3159659}
\showDOI{\tempurl}


\bibitem[\protect\citeauthoryear{DeYoung, Jain, Rajani, Lehman, Xiong, Socher,
  and Wallace}{DeYoung et~al\mbox{.}}{2020}]%
        {deyoung2020eraser}
\bibfield{author}{\bibinfo{person}{Jay DeYoung}, \bibinfo{person}{Sarthak
  Jain}, \bibinfo{person}{Nazneen~Fatema Rajani}, \bibinfo{person}{Eric
  Lehman}, \bibinfo{person}{Caiming Xiong}, \bibinfo{person}{Richard Socher},
  {and} \bibinfo{person}{Byron~C. Wallace}.} \bibinfo{year}{2020}\natexlab{}.
\newblock \showarticletitle{{ERASER}: {A} Benchmark to Evaluate Rationalized
  {NLP} Models}. In \bibinfo{booktitle}{\emph{Proceedings of the 58th Annual
  Meeting of the Association for Computational Linguistics}}.
  \bibinfo{publisher}{Association for Computational Linguistics},
  \bibinfo{address}{Online}, \bibinfo{pages}{4443--4458}.
\newblock
\urldef\tempurl%
\url{https://doi.org/10.18653/v1/2020.acl-main.408}
\showDOI{\tempurl}


\bibitem[\protect\citeauthoryear{Fernando, Singh, and Anand}{Fernando
  et~al\mbox{.}}{2019}]%
        {fernando2019study}
\bibfield{author}{\bibinfo{person}{Zeon~Trevor Fernando},
  \bibinfo{person}{Jaspreet Singh}, {and} \bibinfo{person}{Avishek Anand}.}
  \bibinfo{year}{2019}\natexlab{}.
\newblock \showarticletitle{A Study on the Interpretability of Neural Retrieval
  Models Using DeepSHAP}. In \bibinfo{booktitle}{\emph{Proceedings of the 42nd
  International ACM SIGIR Conference on Research and Development in Information
  Retrieval}} (Paris, France) \emph{(\bibinfo{series}{SIGIR'19})}.
  \bibinfo{publisher}{Association for Computing Machinery},
  \bibinfo{address}{New York, NY, USA}, \bibinfo{pages}{1005–1008}.
\newblock
\showISBNx{9781450361729}
\urldef\tempurl%
\url{https://doi.org/10.1145/3331184.3331312}
\showDOI{\tempurl}


\bibitem[\protect\citeauthoryear{Funke, Khosla, Rathee, and Anand}{Funke
  et~al\mbox{.}}{2022}]%
        {funke2021zorro}
\bibfield{author}{\bibinfo{person}{Thorben Funke}, \bibinfo{person}{Megha
  Khosla}, \bibinfo{person}{Mandeep Rathee}, {and} \bibinfo{person}{Avishek
  Anand}.} \bibinfo{year}{2022}\natexlab{}.
\newblock \showarticletitle{ZORRO: Valid, Sparse, and Stable Explanations in
  Graph Neural Networks}.
\newblock \bibinfo{journal}{\emph{IEEE Transactions on Knowledge and Data
  Engineering}} (\bibinfo{year}{2022}), \bibinfo{pages}{1--12}.
\newblock
\urldef\tempurl%
\url{https://doi.org/10.1109/TKDE.2022.3201170}
\showDOI{\tempurl}


\bibitem[\protect\citeauthoryear{Gallagher}{Gallagher}{2019}]%
        {gallagher2019pairwise}
\bibfield{author}{\bibinfo{person}{Luke Gallagher}.}
  \bibinfo{year}{2019}\natexlab{}.
\newblock \bibinfo{title}{{Pairwise t-test on TREC Run Files}}.
\newblock
  \bibinfo{howpublished}{\url{https://github.com/lgrz/pairwise-ttest/}}.
\newblock


\bibitem[\protect\citeauthoryear{Ge\c{c}kil, M\"{u}ngen, G\"{u}ndo\u{g}an, and
  Kaya}{Ge\c{c}kil et~al\mbox{.}}{2020}]%
        {gecckil2018clickbait}
\bibfield{author}{\bibinfo{person}{Ay\c{c}e Ge\c{c}kil},
  \bibinfo{person}{Ahmet~An\i{}l M\"{u}ngen}, \bibinfo{person}{Esra
  G\"{u}ndo\u{g}an}, {and} \bibinfo{person}{Mehmet Kaya}.}
  \bibinfo{year}{2020}\natexlab{}.
\newblock \showarticletitle{A Clickbait Detection Method on News Sites}. In
  \bibinfo{booktitle}{\emph{Proceedings of the 2018 IEEE/ACM International
  Conference on Advances in Social Networks Analysis and Mining}} (Barcelona,
  Spain) \emph{(\bibinfo{series}{ASONAM '18})}. \bibinfo{publisher}{IEEE
  Press}, \bibinfo{pages}{932–937}.
\newblock
\showISBNx{9781538660515}


\bibitem[\protect\citeauthoryear{Guo, Fan, Ai, and Croft}{Guo
  et~al\mbox{.}}{2016}]%
        {guo2016deep}
\bibfield{author}{\bibinfo{person}{Jiafeng Guo}, \bibinfo{person}{Yixing Fan},
  \bibinfo{person}{Qingyao Ai}, {and} \bibinfo{person}{W.~Bruce Croft}.}
  \bibinfo{year}{2016}\natexlab{}.
\newblock \showarticletitle{A Deep Relevance Matching Model for Ad-Hoc
  Retrieval}. In \bibinfo{booktitle}{\emph{Proceedings of the 25th ACM
  International on Conference on Information and Knowledge Management}}
  (Indianapolis, Indiana, USA) \emph{(\bibinfo{series}{CIKM '16})}.
  \bibinfo{publisher}{Association for Computing Machinery},
  \bibinfo{address}{New York, NY, USA}, \bibinfo{pages}{55–64}.
\newblock
\showISBNx{9781450340731}
\urldef\tempurl%
\url{https://doi.org/10.1145/2983323.2983769}
\showDOI{\tempurl}


\bibitem[\protect\citeauthoryear{Guo, Yang, and Abbasi}{Guo
  et~al\mbox{.}}{2022}]%
        {guo2022auto}
\bibfield{author}{\bibinfo{person}{Yue Guo}, \bibinfo{person}{Yi Yang}, {and}
  \bibinfo{person}{Ahmed Abbasi}.} \bibinfo{year}{2022}\natexlab{}.
\newblock \showarticletitle{Auto-Debias: Debiasing Masked Language Models with
  Automated Biased Prompts}. In \bibinfo{booktitle}{\emph{Proceedings of the
  60th Annual Meeting of the Association for Computational Linguistics (Volume
  1: Long Papers)}}. \bibinfo{publisher}{Association for Computational
  Linguistics}, \bibinfo{address}{Dublin, Ireland},
  \bibinfo{pages}{1012--1023}.
\newblock
\urldef\tempurl%
\url{https://doi.org/10.18653/v1/2022.acl-long.72}
\showDOI{\tempurl}


\bibitem[\protect\citeauthoryear{Hermans and Schrauwen}{Hermans and
  Schrauwen}{2013}]%
        {hermans2013training}
\bibfield{author}{\bibinfo{person}{Michiel Hermans} {and}
  \bibinfo{person}{Benjamin Schrauwen}.} \bibinfo{year}{2013}\natexlab{}.
\newblock \showarticletitle{Training and Analysing Deep Recurrent Neural
  Networks}. In \bibinfo{booktitle}{\emph{Advances in Neural Information
  Processing Systems}}, \bibfield{editor}{\bibinfo{person}{C.J. Burges},
  \bibinfo{person}{L.~Bottou}, \bibinfo{person}{M.~Welling},
  \bibinfo{person}{Z.~Ghahramani}, {and} \bibinfo{person}{K.Q. Weinberger}}
  (Eds.), Vol.~\bibinfo{volume}{26}. \bibinfo{publisher}{Curran Associates,
  Inc.}
\newblock
\urldef\tempurl%
\url{https://proceedings.neurips.cc/paper/2013/file/1ff8a7b5dc7a7d1f0ed65aaa29c04b1e-Paper.pdf}
\showURL{%
\tempurl}


\bibitem[\protect\citeauthoryear{Hofst\"{a}tter, Lipani, Althammer, Zlabinger,
  and Hanbury}{Hofst\"{a}tter et~al\mbox{.}}{2021a}]%
        {hofstatter2021mitigating}
\bibfield{author}{\bibinfo{person}{Sebastian Hofst\"{a}tter},
  \bibinfo{person}{Aldo Lipani}, \bibinfo{person}{Sophia Althammer},
  \bibinfo{person}{Markus Zlabinger}, {and} \bibinfo{person}{Allan Hanbury}.}
  \bibinfo{year}{2021}\natexlab{a}.
\newblock \showarticletitle{Mitigating the Position Bias of Transformer Models
  in Passage Re-Ranking}. In \bibinfo{booktitle}{\emph{Advances in Information
  Retrieval: 43rd European Conference on IR Research, ECIR 2021, Virtual Event,
  March 28 – April 1, 2021, Proceedings, Part I}}.
  \bibinfo{publisher}{Springer-Verlag}, \bibinfo{address}{Berlin, Heidelberg},
  \bibinfo{pages}{238–253}.
\newblock
\showISBNx{978-3-030-72112-1}
\urldef\tempurl%
\url{https://doi.org/10.1007/978-3-030-72113-8_16}
\showDOI{\tempurl}


\bibitem[\protect\citeauthoryear{Hofst\"{a}tter, Mitra, Zamani, Craswell, and
  Hanbury}{Hofst\"{a}tter et~al\mbox{.}}{2021b}]%
        {hofstaetter2021intra}
\bibfield{author}{\bibinfo{person}{Sebastian Hofst\"{a}tter},
  \bibinfo{person}{Bhaskar Mitra}, \bibinfo{person}{Hamed Zamani},
  \bibinfo{person}{Nick Craswell}, {and} \bibinfo{person}{Allan Hanbury}.}
  \bibinfo{year}{2021}\natexlab{b}.
\newblock \showarticletitle{Intra-Document Cascading: Learning to Select
  Passages for Neural Document Ranking}. In
  \bibinfo{booktitle}{\emph{Proceedings of the 44th International ACM SIGIR
  Conference on Research and Development in Information Retrieval}} (Virtual
  Event, Canada) \emph{(\bibinfo{series}{SIGIR '21})}.
  \bibinfo{publisher}{Association for Computing Machinery},
  \bibinfo{address}{New York, NY, USA}, \bibinfo{pages}{1349–1358}.
\newblock
\showISBNx{9781450380379}
\urldef\tempurl%
\url{https://doi.org/10.1145/3404835.3462889}
\showDOI{\tempurl}


\bibitem[\protect\citeauthoryear{Hofst\"{a}tter, Zamani, Mitra, Craswell, and
  Hanbury}{Hofst\"{a}tter et~al\mbox{.}}{2020}]%
        {hofstaetter2020local}
\bibfield{author}{\bibinfo{person}{Sebastian Hofst\"{a}tter},
  \bibinfo{person}{Hamed Zamani}, \bibinfo{person}{Bhaskar Mitra},
  \bibinfo{person}{Nick Craswell}, {and} \bibinfo{person}{Allan Hanbury}.}
  \bibinfo{year}{2020}\natexlab{}.
\newblock \showarticletitle{Local Self-Attention over Long Text for Efficient
  Document Retrieval}. In \bibinfo{booktitle}{\emph{Proceedings of the 43rd
  International ACM SIGIR Conference on Research and Development in Information
  Retrieval}} (Virtual Event, China) \emph{(\bibinfo{series}{SIGIR '20})}.
  \bibinfo{publisher}{Association for Computing Machinery},
  \bibinfo{address}{New York, NY, USA}, \bibinfo{pages}{2021–2024}.
\newblock
\showISBNx{9781450380164}
\urldef\tempurl%
\url{https://doi.org/10.1145/3397271.3401224}
\showDOI{\tempurl}


\bibitem[\protect\citeauthoryear{Hofstätter, Zlabinger, and
  Hanbury}{Hofstätter et~al\mbox{.}}{2020}]%
        {hofstaetter2020interpretable}
\bibfield{author}{\bibinfo{person}{Sebastian Hofstätter},
  \bibinfo{person}{Markus Zlabinger}, {and} \bibinfo{person}{Allan Hanbury}.}
  \bibinfo{year}{2020}\natexlab{}.
\newblock \bibinfo{title}{Interpretable \&amp; Time-Budget-Constrained
  Contextualization for Re-Ranking}.
\newblock
\newblock
\urldef\tempurl%
\url{https://doi.org/10.48550/ARXIV.2002.01854}
\showDOI{\tempurl}


\bibitem[\protect\citeauthoryear{Holzmann and Anand}{Holzmann and
  Anand}{2016}]%
        {holzmann2016tempas}
\bibfield{author}{\bibinfo{person}{Helge Holzmann} {and}
  \bibinfo{person}{Avishek Anand}.} \bibinfo{year}{2016}\natexlab{}.
\newblock \showarticletitle{Tempas: Temporal Archive Search Based on Tags}. In
  \bibinfo{booktitle}{\emph{Proceedings of the 25th International Conference
  Companion on World Wide Web}} (Montr\'{e}al, Qu\'{e}bec, Canada)
  \emph{(\bibinfo{series}{WWW '16 Companion})}.
  \bibinfo{publisher}{International World Wide Web Conferences Steering
  Committee}, \bibinfo{address}{Republic and Canton of Geneva, CHE},
  \bibinfo{pages}{207–210}.
\newblock
\showISBNx{9781450341448}
\urldef\tempurl%
\url{https://doi.org/10.1145/2872518.2890555}
\showDOI{\tempurl}


\bibitem[\protect\citeauthoryear{Holzmann, Nejdl, and Anand}{Holzmann
  et~al\mbox{.}}{2017}]%
        {holzmann2017exploring}
\bibfield{author}{\bibinfo{person}{Helge Holzmann}, \bibinfo{person}{Wolfgang
  Nejdl}, {and} \bibinfo{person}{Avishek Anand}.}
  \bibinfo{year}{2017}\natexlab{}.
\newblock \showarticletitle{Exploring Web Archives Through Temporal Anchor
  Texts}. In \bibinfo{booktitle}{\emph{Proceedings of the 2017 ACM on Web
  Science Conference}} (Troy, New York, USA) \emph{(\bibinfo{series}{WebSci
  '17})}. \bibinfo{publisher}{Association for Computing Machinery},
  \bibinfo{address}{New York, NY, USA}, \bibinfo{pages}{289–298}.
\newblock
\showISBNx{9781450348966}
\urldef\tempurl%
\url{https://doi.org/10.1145/3091478.3091500}
\showDOI{\tempurl}


\bibitem[\protect\citeauthoryear{Huang, He, Gao, Deng, Acero, and Heck}{Huang
  et~al\mbox{.}}{2013}]%
        {huang2013learning}
\bibfield{author}{\bibinfo{person}{Po-Sen Huang}, \bibinfo{person}{Xiaodong
  He}, \bibinfo{person}{Jianfeng Gao}, \bibinfo{person}{Li Deng},
  \bibinfo{person}{Alex Acero}, {and} \bibinfo{person}{Larry Heck}.}
  \bibinfo{year}{2013}\natexlab{}.
\newblock \showarticletitle{Learning Deep Structured Semantic Models for Web
  Search Using Clickthrough Data}. In \bibinfo{booktitle}{\emph{Proceedings of
  the 22nd ACM International Conference on Information \&amp; Knowledge
  Management}} (San Francisco, California, USA) \emph{(\bibinfo{series}{CIKM
  '13})}. \bibinfo{publisher}{Association for Computing Machinery},
  \bibinfo{address}{New York, NY, USA}, \bibinfo{pages}{2333–2338}.
\newblock
\showISBNx{9781450322638}
\urldef\tempurl%
\url{https://doi.org/10.1145/2505515.2505665}
\showDOI{\tempurl}


\bibitem[\protect\citeauthoryear{Hui, Yates, Berberich, and de~Melo}{Hui
  et~al\mbox{.}}{2017}]%
        {hui2017pacrr}
\bibfield{author}{\bibinfo{person}{Kai Hui}, \bibinfo{person}{Andrew Yates},
  \bibinfo{person}{Klaus Berberich}, {and} \bibinfo{person}{Gerard de Melo}.}
  \bibinfo{year}{2017}\natexlab{}.
\newblock \showarticletitle{{PACRR}: A Position-Aware Neural {IR} Model for
  Relevance Matching}. In \bibinfo{booktitle}{\emph{Proceedings of the 2017
  Conference on Empirical Methods in Natural Language Processing}}.
  \bibinfo{publisher}{Association for Computational Linguistics},
  \bibinfo{address}{Copenhagen, Denmark}, \bibinfo{pages}{1049--1058}.
\newblock
\urldef\tempurl%
\url{https://doi.org/10.18653/v1/D17-1110}
\showDOI{\tempurl}


\bibitem[\protect\citeauthoryear{Hui, Yates, Berberich, and de~Melo}{Hui
  et~al\mbox{.}}{2018}]%
        {hui2018copacrr}
\bibfield{author}{\bibinfo{person}{Kai Hui}, \bibinfo{person}{Andrew Yates},
  \bibinfo{person}{Klaus Berberich}, {and} \bibinfo{person}{Gerard de Melo}.}
  \bibinfo{year}{2018}\natexlab{}.
\newblock \showarticletitle{Co-PACRR: A Context-Aware Neural IR Model for
  Ad-Hoc Retrieval}. In \bibinfo{booktitle}{\emph{Proceedings of the Eleventh
  ACM International Conference on Web Search and Data Mining}} (Marina Del Rey,
  CA, USA) \emph{(\bibinfo{series}{WSDM '18})}. \bibinfo{publisher}{Association
  for Computing Machinery}, \bibinfo{address}{New York, NY, USA},
  \bibinfo{pages}{279–287}.
\newblock
\showISBNx{9781450355810}
\urldef\tempurl%
\url{https://doi.org/10.1145/3159652.3159689}
\showDOI{\tempurl}


\bibitem[\protect\citeauthoryear{Idahl, Lyu, Gadiraju, and Anand}{Idahl
  et~al\mbox{.}}{2021}]%
        {idahl2021towards}
\bibfield{author}{\bibinfo{person}{Maximilian Idahl}, \bibinfo{person}{Lijun
  Lyu}, \bibinfo{person}{Ujwal Gadiraju}, {and} \bibinfo{person}{Avishek
  Anand}.} \bibinfo{year}{2021}\natexlab{}.
\newblock \showarticletitle{Towards Benchmarking the Utility of Explanations
  for Model Debugging}. In \bibinfo{booktitle}{\emph{Proceedings of the First
  Workshop on Trustworthy Natural Language Processing}}.
  \bibinfo{publisher}{Association for Computational Linguistics},
  \bibinfo{address}{Online}, \bibinfo{pages}{68--73}.
\newblock
\urldef\tempurl%
\url{https://doi.org/10.18653/v1/2021.trustnlp-1.8}
\showDOI{\tempurl}


\bibitem[\protect\citeauthoryear{Jacovi and Goldberg}{Jacovi and
  Goldberg}{2020}]%
        {jacovi2020towards}
\bibfield{author}{\bibinfo{person}{Alon Jacovi} {and} \bibinfo{person}{Yoav
  Goldberg}.} \bibinfo{year}{2020}\natexlab{}.
\newblock \showarticletitle{Towards Faithfully Interpretable {NLP} Systems: How
  Should We Define and Evaluate Faithfulness?}. In
  \bibinfo{booktitle}{\emph{Proceedings of the 58th Annual Meeting of the
  Association for Computational Linguistics}}. \bibinfo{publisher}{Association
  for Computational Linguistics}, \bibinfo{address}{Online},
  \bibinfo{pages}{4198--4205}.
\newblock
\urldef\tempurl%
\url{https://doi.org/10.18653/v1/2020.acl-main.386}
\showDOI{\tempurl}


\bibitem[\protect\citeauthoryear{Jain and Wallace}{Jain and Wallace}{2019}]%
        {jain2019attention}
\bibfield{author}{\bibinfo{person}{Sarthak Jain} {and}
  \bibinfo{person}{Byron~C. Wallace}.} \bibinfo{year}{2019}\natexlab{}.
\newblock \showarticletitle{{A}ttention is not {E}xplanation}. In
  \bibinfo{booktitle}{\emph{Proceedings of the 2019 Conference of the North
  {A}merican Chapter of the Association for Computational Linguistics: Human
  Language Technologies, Volume 1 (Long and Short Papers)}}.
  \bibinfo{publisher}{Association for Computational Linguistics},
  \bibinfo{address}{Minneapolis, Minnesota}, \bibinfo{pages}{3543--3556}.
\newblock
\urldef\tempurl%
\url{https://doi.org/10.18653/v1/N19-1357}
\showDOI{\tempurl}


\bibitem[\protect\citeauthoryear{Karpathy, Johnson, and Fei-Fei}{Karpathy
  et~al\mbox{.}}{2015}]%
        {karpathy2015visualizing}
\bibfield{author}{\bibinfo{person}{Andrej Karpathy}, \bibinfo{person}{Justin
  Johnson}, {and} \bibinfo{person}{Li Fei-Fei}.}
  \bibinfo{year}{2015}\natexlab{}.
\newblock \bibinfo{title}{Visualizing and Understanding Recurrent Networks}.
\newblock
\newblock
\urldef\tempurl%
\url{https://doi.org/10.48550/ARXIV.1506.02078}
\showDOI{\tempurl}


\bibitem[\protect\citeauthoryear{Khattab and Zaharia}{Khattab and
  Zaharia}{2020}]%
        {khattab2020colbert}
\bibfield{author}{\bibinfo{person}{Omar Khattab} {and} \bibinfo{person}{Matei
  Zaharia}.} \bibinfo{year}{2020}\natexlab{}.
\newblock \showarticletitle{ColBERT: Efficient and Effective Passage Search via
  Contextualized Late Interaction over BERT}. In
  \bibinfo{booktitle}{\emph{Proceedings of the 43rd International ACM SIGIR
  Conference on Research and Development in Information Retrieval}} (Virtual
  Event, China) \emph{(\bibinfo{series}{SIGIR '20})}.
  \bibinfo{publisher}{Association for Computing Machinery},
  \bibinfo{address}{New York, NY, USA}, \bibinfo{pages}{39–48}.
\newblock
\showISBNx{9781450380164}
\urldef\tempurl%
\url{https://doi.org/10.1145/3397271.3401075}
\showDOI{\tempurl}


\bibitem[\protect\citeauthoryear{Lage, Chen, He, Narayanan, Kim, Gershman, and
  Doshi-Velez}{Lage et~al\mbox{.}}{2019}]%
        {lage2019evaluation}
\bibfield{author}{\bibinfo{person}{Isaac Lage}, \bibinfo{person}{Emily Chen},
  \bibinfo{person}{Jeffrey He}, \bibinfo{person}{Menaka Narayanan},
  \bibinfo{person}{Been Kim}, \bibinfo{person}{Sam Gershman}, {and}
  \bibinfo{person}{Finale Doshi-Velez}.} \bibinfo{year}{2019}\natexlab{}.
\newblock \bibinfo{title}{An Evaluation of the Human-Interpretability of
  Explanation}.
\newblock
\newblock
\urldef\tempurl%
\url{https://doi.org/10.48550/ARXIV.1902.00006}
\showDOI{\tempurl}


\bibitem[\protect\citeauthoryear{Lavrenko and Croft}{Lavrenko and
  Croft}{2001}]%
        {lavrenko2017relevance}
\bibfield{author}{\bibinfo{person}{Victor Lavrenko} {and}
  \bibinfo{person}{W.~Bruce Croft}.} \bibinfo{year}{2001}\natexlab{}.
\newblock \showarticletitle{Relevance Based Language Models}. In
  \bibinfo{booktitle}{\emph{Proceedings of the 24th Annual International ACM
  SIGIR Conference on Research and Development in Information Retrieval}}
  \emph{(\bibinfo{series}{SIGIR '01})}. \bibinfo{publisher}{ACM},
  \bibinfo{address}{New York, NY, USA}, \bibinfo{pages}{120--127}.
\newblock
\showISBNx{1-58113-331-6}
\urldef\tempurl%
\url{https://doi.org/10.1145/383952.383972}
\showDOI{\tempurl}


\bibitem[\protect\citeauthoryear{Lehman, DeYoung, Barzilay, and Wallace}{Lehman
  et~al\mbox{.}}{2019}]%
        {lehman2019inferring}
\bibfield{author}{\bibinfo{person}{Eric Lehman}, \bibinfo{person}{Jay DeYoung},
  \bibinfo{person}{Regina Barzilay}, {and} \bibinfo{person}{Byron~C. Wallace}.}
  \bibinfo{year}{2019}\natexlab{}.
\newblock \showarticletitle{Inferring Which Medical Treatments Work from
  Reports of Clinical Trials}. In \bibinfo{booktitle}{\emph{Proceedings of the
  2019 Conference of the North {A}merican Chapter of the Association for
  Computational Linguistics: Human Language Technologies, Volume 1 (Long and
  Short Papers)}}. \bibinfo{publisher}{Association for Computational
  Linguistics}, \bibinfo{address}{Minneapolis, Minnesota},
  \bibinfo{pages}{3705--3717}.
\newblock
\urldef\tempurl%
\url{https://doi.org/10.18653/v1/N19-1371}
\showDOI{\tempurl}


\bibitem[\protect\citeauthoryear{Lei, Barzilay, and Jaakkola}{Lei
  et~al\mbox{.}}{2016}]%
        {lei2016rationalizing}
\bibfield{author}{\bibinfo{person}{Tao Lei}, \bibinfo{person}{Regina Barzilay},
  {and} \bibinfo{person}{Tommi Jaakkola}.} \bibinfo{year}{2016}\natexlab{}.
\newblock \showarticletitle{Rationalizing Neural Predictions}. In
  \bibinfo{booktitle}{\emph{Proceedings of the 2016 Conference on Empirical
  Methods in Natural Language Processing}}. \bibinfo{publisher}{Association for
  Computational Linguistics}, \bibinfo{address}{Austin, Texas},
  \bibinfo{pages}{107--117}.
\newblock
\urldef\tempurl%
\url{https://doi.org/10.18653/v1/D16-1011}
\showDOI{\tempurl}


\bibitem[\protect\citeauthoryear{Leonhardt, Rudra, Khosla, Anand, and
  Anand}{Leonhardt et~al\mbox{.}}{2022}]%
        {leonhardt2022efficient}
\bibfield{author}{\bibinfo{person}{Jurek Leonhardt}, \bibinfo{person}{Koustav
  Rudra}, \bibinfo{person}{Megha Khosla}, \bibinfo{person}{Abhijit Anand},
  {and} \bibinfo{person}{Avishek Anand}.} \bibinfo{year}{2022}\natexlab{}.
\newblock \showarticletitle{Efficient Neural Ranking Using Forward Indexes}. In
  \bibinfo{booktitle}{\emph{Proceedings of the ACM Web Conference 2022}}
  (Virtual Event, Lyon, France) \emph{(\bibinfo{series}{WWW '22})}.
  \bibinfo{publisher}{Association for Computing Machinery},
  \bibinfo{address}{New York, NY, USA}, \bibinfo{pages}{266–276}.
\newblock
\showISBNx{9781450390965}
\urldef\tempurl%
\url{https://doi.org/10.1145/3485447.3511955}
\showDOI{\tempurl}


\bibitem[\protect\citeauthoryear{Li, Monroe, and Jurafsky}{Li
  et~al\mbox{.}}{2016}]%
        {li2016understanding}
\bibfield{author}{\bibinfo{person}{Jiwei Li}, \bibinfo{person}{Will Monroe},
  {and} \bibinfo{person}{Dan Jurafsky}.} \bibinfo{year}{2016}\natexlab{}.
\newblock \bibinfo{title}{Understanding Neural Networks through Representation
  Erasure}.
\newblock
\newblock
\urldef\tempurl%
\url{https://doi.org/10.48550/ARXIV.1612.08220}
\showDOI{\tempurl}


\bibitem[\protect\citeauthoryear{Li and Gaussier}{Li and Gaussier}{2021}]%
        {li2021keybld}
\bibfield{author}{\bibinfo{person}{Minghan Li} {and} \bibinfo{person}{Eric
  Gaussier}.} \bibinfo{year}{2021}\natexlab{}.
\newblock \showarticletitle{KeyBLD: Selecting Key Blocks with Local Pre-Ranking
  for Long Document Information Retrieval}. In
  \bibinfo{booktitle}{\emph{Proceedings of the 44th International ACM SIGIR
  Conference on Research and Development in Information Retrieval}} (Virtual
  Event, Canada) \emph{(\bibinfo{series}{SIGIR '21})}.
  \bibinfo{publisher}{Association for Computing Machinery},
  \bibinfo{address}{New York, NY, USA}, \bibinfo{pages}{2207–2211}.
\newblock
\showISBNx{9781450380379}
\urldef\tempurl%
\url{https://doi.org/10.1145/3404835.3463083}
\showDOI{\tempurl}


\bibitem[\protect\citeauthoryear{Li, Mao, Wang, Liu, Zhang, and Ma}{Li
  et~al\mbox{.}}{2019}]%
        {li2019teach}
\bibfield{author}{\bibinfo{person}{Xiangsheng Li}, \bibinfo{person}{Jiaxin
  Mao}, \bibinfo{person}{Chao Wang}, \bibinfo{person}{Yiqun Liu},
  \bibinfo{person}{Min Zhang}, {and} \bibinfo{person}{Shaoping Ma}.}
  \bibinfo{year}{2019}\natexlab{}.
\newblock \showarticletitle{Teach Machine How to Read: Reading Behavior
  Inspired Relevance Estimation}. In \bibinfo{booktitle}{\emph{Proceedings of
  the 42nd International ACM SIGIR Conference on Research and Development in
  Information Retrieval}} (Paris, France) \emph{(\bibinfo{series}{SIGIR'19})}.
  \bibinfo{publisher}{Association for Computing Machinery},
  \bibinfo{address}{New York, NY, USA}, \bibinfo{pages}{795–804}.
\newblock
\showISBNx{9781450361729}
\urldef\tempurl%
\url{https://doi.org/10.1145/3331184.3331205}
\showDOI{\tempurl}


\bibitem[\protect\citeauthoryear{Lin, Yang, and Lin}{Lin et~al\mbox{.}}{2021}]%
        {lin2021batch}
\bibfield{author}{\bibinfo{person}{Sheng-Chieh Lin},
  \bibinfo{person}{Jheng-Hong Yang}, {and} \bibinfo{person}{Jimmy Lin}.}
  \bibinfo{year}{2021}\natexlab{}.
\newblock \showarticletitle{In-Batch Negatives for Knowledge Distillation with
  Tightly-Coupled Teachers for Dense Retrieval}. In
  \bibinfo{booktitle}{\emph{Proceedings of the 6th Workshop on Representation
  Learning for NLP (RepL4NLP-2021)}}. \bibinfo{publisher}{Association for
  Computational Linguistics}, \bibinfo{address}{Online},
  \bibinfo{pages}{163--173}.
\newblock
\urldef\tempurl%
\url{https://doi.org/10.18653/v1/2021.repl4nlp-1.17}
\showDOI{\tempurl}


\bibitem[\protect\citeauthoryear{Loshchilov and Hutter}{Loshchilov and
  Hutter}{2019}]%
        {loshchilov2018decoupled}
\bibfield{author}{\bibinfo{person}{Ilya Loshchilov} {and}
  \bibinfo{person}{Frank Hutter}.} \bibinfo{year}{2019}\natexlab{}.
\newblock \showarticletitle{Decoupled Weight Decay Regularization}. In
  \bibinfo{booktitle}{\emph{International Conference on Learning
  Representations}}.
\newblock
\urldef\tempurl%
\url{https://openreview.net/forum?id=Bkg6RiCqY7}
\showURL{%
\tempurl}


\bibitem[\protect\citeauthoryear{MacAvaney, Yates, Cohan, and
  Goharian}{MacAvaney et~al\mbox{.}}{2019}]%
        {macavaney2019cedr}
\bibfield{author}{\bibinfo{person}{Sean MacAvaney}, \bibinfo{person}{Andrew
  Yates}, \bibinfo{person}{Arman Cohan}, {and} \bibinfo{person}{Nazli
  Goharian}.} \bibinfo{year}{2019}\natexlab{}.
\newblock \showarticletitle{CEDR: Contextualized Embeddings for Document
  Ranking}. In \bibinfo{booktitle}{\emph{Proceedings of the 42nd International
  ACM SIGIR Conference on Research and Development in Information Retrieval}}
  (Paris, France) \emph{(\bibinfo{series}{SIGIR'19})}.
  \bibinfo{publisher}{Association for Computing Machinery},
  \bibinfo{address}{New York, NY, USA}, \bibinfo{pages}{1101–1104}.
\newblock
\showISBNx{9781450361729}
\urldef\tempurl%
\url{https://doi.org/10.1145/3331184.3331317}
\showDOI{\tempurl}


\bibitem[\protect\citeauthoryear{Maddison, Tarlow, and Minka}{Maddison
  et~al\mbox{.}}{2014}]%
        {maddison2014sampling}
\bibfield{author}{\bibinfo{person}{Chris~J Maddison}, \bibinfo{person}{Daniel
  Tarlow}, {and} \bibinfo{person}{Tom Minka}.} \bibinfo{year}{2014}\natexlab{}.
\newblock \showarticletitle{A* Sampling}. In \bibinfo{booktitle}{\emph{Advances
  in Neural Information Processing Systems}},
  \bibfield{editor}{\bibinfo{person}{Z.~Ghahramani},
  \bibinfo{person}{M.~Welling}, \bibinfo{person}{C.~Cortes},
  \bibinfo{person}{N.~Lawrence}, {and} \bibinfo{person}{K.Q. Weinberger}}
  (Eds.), Vol.~\bibinfo{volume}{27}. \bibinfo{publisher}{Curran Associates,
  Inc.}
\newblock
\urldef\tempurl%
\url{https://proceedings.neurips.cc/paper/2014/file/309fee4e541e51de2e41f21bebb342aa-Paper.pdf}
\showURL{%
\tempurl}


\bibitem[\protect\citeauthoryear{Martins and Astudillo}{Martins and
  Astudillo}{2016}]%
        {martins2016softmax}
\bibfield{author}{\bibinfo{person}{Andre Martins} {and} \bibinfo{person}{Ramon
  Astudillo}.} \bibinfo{year}{2016}\natexlab{}.
\newblock \showarticletitle{From Softmax to Sparsemax: A Sparse Model of
  Attention and Multi-Label Classification}. In
  \bibinfo{booktitle}{\emph{Proceedings of The 33rd International Conference on
  Machine Learning}} \emph{(\bibinfo{series}{Proceedings of Machine Learning
  Research}, Vol.~\bibinfo{volume}{48})},
  \bibfield{editor}{\bibinfo{person}{Maria~Florina Balcan} {and}
  \bibinfo{person}{Kilian~Q. Weinberger}} (Eds.). \bibinfo{publisher}{PMLR},
  \bibinfo{address}{New York, New York, USA}, \bibinfo{pages}{1614--1623}.
\newblock
\urldef\tempurl%
\url{https://proceedings.mlr.press/v48/martins16.html}
\showURL{%
\tempurl}


\bibitem[\protect\citeauthoryear{McDonald, Brokos, and
  Androutsopoulos}{McDonald et~al\mbox{.}}{2018}]%
        {mcdonald2018deep}
\bibfield{author}{\bibinfo{person}{Ryan McDonald}, \bibinfo{person}{George
  Brokos}, {and} \bibinfo{person}{Ion Androutsopoulos}.}
  \bibinfo{year}{2018}\natexlab{}.
\newblock \showarticletitle{Deep Relevance Ranking Using Enhanced
  Document-Query Interactions}. In \bibinfo{booktitle}{\emph{Proceedings of the
  2018 Conference on Empirical Methods in Natural Language Processing}}.
  \bibinfo{publisher}{Association for Computational Linguistics},
  \bibinfo{address}{Brussels, Belgium}, \bibinfo{pages}{1849--1860}.
\newblock
\urldef\tempurl%
\url{https://doi.org/10.18653/v1/D18-1211}
\showDOI{\tempurl}


\bibitem[\protect\citeauthoryear{Mitra, Diaz, and Craswell}{Mitra
  et~al\mbox{.}}{2017}]%
        {mitra2017learning}
\bibfield{author}{\bibinfo{person}{Bhaskar Mitra}, \bibinfo{person}{Fernando
  Diaz}, {and} \bibinfo{person}{Nick Craswell}.}
  \bibinfo{year}{2017}\natexlab{}.
\newblock \showarticletitle{Learning to Match Using Local and Distributed
  Representations of Text for Web Search}. In
  \bibinfo{booktitle}{\emph{Proceedings of the 26th International Conference on
  World Wide Web}} (Perth, Australia) \emph{(\bibinfo{series}{WWW '17})}.
  \bibinfo{publisher}{International World Wide Web Conferences Steering
  Committee}, \bibinfo{address}{Republic and Canton of Geneva, CHE},
  \bibinfo{pages}{1291–1299}.
\newblock
\showISBNx{9781450349130}
\urldef\tempurl%
\url{https://doi.org/10.1145/3038912.3052579}
\showDOI{\tempurl}


\bibitem[\protect\citeauthoryear{Muntean, Nardini, Perego, Tonellotto, and
  Frieder}{Muntean et~al\mbox{.}}{2020}]%
        {muntean2020weighting}
\bibfield{author}{\bibinfo{person}{Cristina~Ioana Muntean},
  \bibinfo{person}{Franco~Maria Nardini}, \bibinfo{person}{Raffaele Perego},
  \bibinfo{person}{Nicola Tonellotto}, {and} \bibinfo{person}{Ophir Frieder}.}
  \bibinfo{year}{2020}\natexlab{}.
\newblock \showarticletitle{Weighting Passages Enhances Accuracy}.
\newblock \bibinfo{journal}{\emph{ACM Trans. Inf. Syst.}} \bibinfo{volume}{39},
  \bibinfo{number}{2}, Article \bibinfo{articleno}{11} (\bibinfo{date}{dec}
  \bibinfo{year}{2020}), \bibinfo{numpages}{11}~pages.
\newblock
\showISSN{1046-8188}
\urldef\tempurl%
\url{https://doi.org/10.1145/3428687}
\showDOI{\tempurl}


\bibitem[\protect\citeauthoryear{Nguyen, Rosenberg, Song, Gao, Tiwary,
  Majumder, and Deng}{Nguyen et~al\mbox{.}}{2016}]%
        {nguyen2016ms}
\bibfield{author}{\bibinfo{person}{Tri Nguyen}, \bibinfo{person}{Mir
  Rosenberg}, \bibinfo{person}{Xia Song}, \bibinfo{person}{Jianfeng Gao},
  \bibinfo{person}{Saurabh Tiwary}, \bibinfo{person}{Rangan Majumder}, {and}
  \bibinfo{person}{Li Deng}.} \bibinfo{year}{2016}\natexlab{}.
\newblock \showarticletitle{MS MARCO: A Human Generated MAchine Reading
  COmprehension Dataset}. In \bibinfo{booktitle}{\emph{CoCo@NIPS}}.
\newblock
\urldef\tempurl%
\url{http://ceur-ws.org/Vol-1773/CoCoNIPS_2016_paper9.pdf}
\showURL{%
\tempurl}


\bibitem[\protect\citeauthoryear{Nie, Li, and Nie}{Nie et~al\mbox{.}}{2018a}]%
        {nie2018empirical}
\bibfield{author}{\bibinfo{person}{Yifan Nie}, \bibinfo{person}{Yanling Li},
  {and} \bibinfo{person}{Jian-Yun Nie}.} \bibinfo{year}{2018}\natexlab{a}.
\newblock \showarticletitle{Empirical Study of Multi-Level Convolution Models
  for IR Based on Representations and Interactions}. In
  \bibinfo{booktitle}{\emph{Proceedings of the 2018 ACM SIGIR International
  Conference on Theory of Information Retrieval}} (Tianjin, China)
  \emph{(\bibinfo{series}{ICTIR '18})}. \bibinfo{publisher}{Association for
  Computing Machinery}, \bibinfo{address}{New York, NY, USA},
  \bibinfo{pages}{59–66}.
\newblock
\showISBNx{9781450356565}
\urldef\tempurl%
\url{https://doi.org/10.1145/3234944.3234954}
\showDOI{\tempurl}


\bibitem[\protect\citeauthoryear{Nie, Sordoni, and Nie}{Nie
  et~al\mbox{.}}{2018b}]%
        {nie2018multi}
\bibfield{author}{\bibinfo{person}{Yifan Nie}, \bibinfo{person}{Alessandro
  Sordoni}, {and} \bibinfo{person}{Jian-Yun Nie}.}
  \bibinfo{year}{2018}\natexlab{b}.
\newblock \showarticletitle{Multi-Level Abstraction Convolutional Model with
  Weak Supervision for Information Retrieval}. In \bibinfo{booktitle}{\emph{The
  41st International ACM SIGIR Conference on Research \&amp; Development in
  Information Retrieval}} (Ann Arbor, MI, USA) \emph{(\bibinfo{series}{SIGIR
  '18})}. \bibinfo{publisher}{Association for Computing Machinery},
  \bibinfo{address}{New York, NY, USA}, \bibinfo{pages}{985–988}.
\newblock
\showISBNx{9781450356572}
\urldef\tempurl%
\url{https://doi.org/10.1145/3209978.3210123}
\showDOI{\tempurl}


\bibitem[\protect\citeauthoryear{Nogueira and Cho}{Nogueira and Cho}{2019}]%
        {nogueira2019passage}
\bibfield{author}{\bibinfo{person}{Rodrigo Nogueira} {and}
  \bibinfo{person}{Kyunghyun Cho}.} \bibinfo{year}{2019}\natexlab{}.
\newblock \bibinfo{title}{Passage Re-ranking with BERT}.
\newblock
\newblock
\urldef\tempurl%
\url{https://doi.org/10.48550/ARXIV.1901.04085}
\showDOI{\tempurl}


\bibitem[\protect\citeauthoryear{Pang, Lan, Guo, Xu, and Cheng}{Pang
  et~al\mbox{.}}{2016a}]%
        {pang2016study}
\bibfield{author}{\bibinfo{person}{Liang Pang}, \bibinfo{person}{Yanyan Lan},
  \bibinfo{person}{Jiafeng Guo}, \bibinfo{person}{Jun Xu}, {and}
  \bibinfo{person}{Xueqi Cheng}.} \bibinfo{year}{2016}\natexlab{a}.
\newblock \bibinfo{title}{A Study of MatchPyramid Models on Ad-hoc Retrieval}.
\newblock
\newblock
\urldef\tempurl%
\url{https://doi.org/10.48550/ARXIV.1606.04648}
\showDOI{\tempurl}


\bibitem[\protect\citeauthoryear{Pang, Lan, Guo, Xu, Wan, and Cheng}{Pang
  et~al\mbox{.}}{2016b}]%
        {pang2016text}
\bibfield{author}{\bibinfo{person}{Liang Pang}, \bibinfo{person}{Yanyan Lan},
  \bibinfo{person}{Jiafeng Guo}, \bibinfo{person}{Jun Xu},
  \bibinfo{person}{Shengxian Wan}, {and} \bibinfo{person}{Xueqi Cheng}.}
  \bibinfo{year}{2016}\natexlab{b}.
\newblock \showarticletitle{Text Matching as Image Recognition}.
\newblock \bibinfo{journal}{\emph{Proceedings of the AAAI Conference on
  Artificial Intelligence}} \bibinfo{volume}{30}, \bibinfo{number}{1}
  (\bibinfo{date}{Mar.} \bibinfo{year}{2016}).
\newblock
\urldef\tempurl%
\url{https://doi.org/10.1609/aaai.v30i1.10341}
\showDOI{\tempurl}


\bibitem[\protect\citeauthoryear{Pl\"{o}tz and Roth}{Pl\"{o}tz and
  Roth}{2018}]%
        {plotz2018neural}
\bibfield{author}{\bibinfo{person}{Tobias Pl\"{o}tz} {and}
  \bibinfo{person}{Stefan Roth}.} \bibinfo{year}{2018}\natexlab{}.
\newblock \showarticletitle{Neural Nearest Neighbors Networks}. In
  \bibinfo{booktitle}{\emph{Advances in Neural Information Processing
  Systems}}, \bibfield{editor}{\bibinfo{person}{S.~Bengio},
  \bibinfo{person}{H.~Wallach}, \bibinfo{person}{H.~Larochelle},
  \bibinfo{person}{K.~Grauman}, \bibinfo{person}{N.~Cesa-Bianchi}, {and}
  \bibinfo{person}{R.~Garnett}} (Eds.), Vol.~\bibinfo{volume}{31}.
  \bibinfo{publisher}{Curran Associates, Inc.}
\newblock
\urldef\tempurl%
\url{https://proceedings.neurips.cc/paper/2018/file/f0e52b27a7a5d6a1a87373dffa53dbe5-Paper.pdf}
\showURL{%
\tempurl}


\bibitem[\protect\citeauthoryear{Robertson, Zaragoza, et~al\mbox{.}}{Robertson
  et~al\mbox{.}}{2009}]%
        {robertson2009probabilistic}
\bibfield{author}{\bibinfo{person}{Stephen Robertson}, \bibinfo{person}{Hugo
  Zaragoza}, {et~al\mbox{.}}} \bibinfo{year}{2009}\natexlab{}.
\newblock \showarticletitle{The probabilistic relevance framework: BM25 and
  beyond}.
\newblock \bibinfo{journal}{\emph{Foundations and Trends{\textregistered} in
  Information Retrieval}} \bibinfo{volume}{3}, \bibinfo{number}{4}
  (\bibinfo{year}{2009}), \bibinfo{pages}{333--389}.
\newblock


\bibitem[\protect\citeauthoryear{Roy and Anand}{Roy and Anand}{2021}]%
        {roy2021question}
\bibfield{author}{\bibinfo{person}{Rishiraj~Saha Roy} {and}
  \bibinfo{person}{Avishek Anand}.} \bibinfo{year}{2021}\natexlab{}.
\newblock \showarticletitle{Question Answering for the Curated Web: Tasks and
  Methods in QA over Knowledge Bases and Text Collections}.
\newblock \bibinfo{journal}{\emph{Synthesis Lectures onSynthesis Lectures on
  Information Concepts, Retrieval, and Services}} \bibinfo{volume}{13},
  \bibinfo{number}{4} (\bibinfo{year}{2021}), \bibinfo{pages}{1--194}.
\newblock


\bibitem[\protect\citeauthoryear{Rudin}{Rudin}{2019}]%
        {rudin2019stop}
\bibfield{author}{\bibinfo{person}{Cynthia Rudin}.}
  \bibinfo{year}{2019}\natexlab{}.
\newblock \showarticletitle{Stop explaining black box machine learning models
  for high stakes decisions and use interpretable models instead}.
\newblock \bibinfo{journal}{\emph{Nature Machine Intelligence}}
  \bibinfo{volume}{1}, \bibinfo{number}{5} (\bibinfo{year}{2019}),
  \bibinfo{pages}{206--215}.
\newblock


\bibitem[\protect\citeauthoryear{Rudra and Anand}{Rudra and Anand}{2020}]%
        {rudra2020distant}
\bibfield{author}{\bibinfo{person}{Koustav Rudra} {and}
  \bibinfo{person}{Avishek Anand}.} \bibinfo{year}{2020}\natexlab{}.
\newblock \showarticletitle{Distant Supervision in BERT-Based Adhoc Document
  Retrieval}. In \bibinfo{booktitle}{\emph{Proceedings of the 29th ACM
  International Conference on Information \&amp; Knowledge Management}}
  (Virtual Event, Ireland) \emph{(\bibinfo{series}{CIKM '20})}.
  \bibinfo{publisher}{Association for Computing Machinery},
  \bibinfo{address}{New York, NY, USA}, \bibinfo{pages}{2197–2200}.
\newblock
\showISBNx{9781450368599}
\urldef\tempurl%
\url{https://doi.org/10.1145/3340531.3412124}
\showDOI{\tempurl}


\bibitem[\protect\citeauthoryear{Shen, He, Gao, Deng, and Mesnil}{Shen
  et~al\mbox{.}}{2014a}]%
        {shen2014latent}
\bibfield{author}{\bibinfo{person}{Yelong Shen}, \bibinfo{person}{Xiaodong He},
  \bibinfo{person}{Jianfeng Gao}, \bibinfo{person}{Li Deng}, {and}
  \bibinfo{person}{Gr\'{e}goire Mesnil}.} \bibinfo{year}{2014}\natexlab{a}.
\newblock \showarticletitle{A Latent Semantic Model with Convolutional-Pooling
  Structure for Information Retrieval}. In
  \bibinfo{booktitle}{\emph{Proceedings of the 23rd ACM International
  Conference on Conference on Information and Knowledge Management}} (Shanghai,
  China) \emph{(\bibinfo{series}{CIKM '14})}. \bibinfo{publisher}{Association
  for Computing Machinery}, \bibinfo{address}{New York, NY, USA},
  \bibinfo{pages}{101–110}.
\newblock
\showISBNx{9781450325981}
\urldef\tempurl%
\url{https://doi.org/10.1145/2661829.2661935}
\showDOI{\tempurl}


\bibitem[\protect\citeauthoryear{Shen, He, Gao, Deng, and Mesnil}{Shen
  et~al\mbox{.}}{2014b}]%
        {shen2014learning}
\bibfield{author}{\bibinfo{person}{Yelong Shen}, \bibinfo{person}{Xiaodong He},
  \bibinfo{person}{Jianfeng Gao}, \bibinfo{person}{Li Deng}, {and}
  \bibinfo{person}{Gr\'{e}goire Mesnil}.} \bibinfo{year}{2014}\natexlab{b}.
\newblock \showarticletitle{Learning Semantic Representations Using
  Convolutional Neural Networks for Web Search}. In
  \bibinfo{booktitle}{\emph{Proceedings of the 23rd International Conference on
  World Wide Web}} (Seoul, Korea) \emph{(\bibinfo{series}{WWW '14 Companion})}.
  \bibinfo{publisher}{Association for Computing Machinery},
  \bibinfo{address}{New York, NY, USA}, \bibinfo{pages}{373–374}.
\newblock
\showISBNx{9781450327459}
\urldef\tempurl%
\url{https://doi.org/10.1145/2567948.2577348}
\showDOI{\tempurl}


\bibitem[\protect\citeauthoryear{Sidiropoulos and Kanoulas}{Sidiropoulos and
  Kanoulas}{2022}]%
        {sidiropoulos2022analysing}
\bibfield{author}{\bibinfo{person}{Georgios Sidiropoulos} {and}
  \bibinfo{person}{Evangelos Kanoulas}.} \bibinfo{year}{2022}\natexlab{}.
\newblock \showarticletitle{Analysing the Robustness of Dual Encoders for Dense
  Retrieval Against Misspellings}. In \bibinfo{booktitle}{\emph{Proceedings of
  the 45th International ACM SIGIR Conference on Research and Development in
  Information Retrieval}} (Madrid, Spain) \emph{(\bibinfo{series}{SIGIR '22})}.
  \bibinfo{publisher}{Association for Computing Machinery},
  \bibinfo{address}{New York, NY, USA}, \bibinfo{pages}{2132–2136}.
\newblock
\showISBNx{9781450387323}
\urldef\tempurl%
\url{https://doi.org/10.1145/3477495.3531818}
\showDOI{\tempurl}


\bibitem[\protect\citeauthoryear{Singh and Anand}{Singh and Anand}{2018}]%
        {singh2018posthoc}
\bibfield{author}{\bibinfo{person}{Jaspreet Singh} {and}
  \bibinfo{person}{Avishek Anand}.} \bibinfo{year}{2018}\natexlab{}.
\newblock \bibinfo{title}{Posthoc Interpretability of Learning to Rank Models
  using Secondary Training Data}.
\newblock
\newblock
\urldef\tempurl%
\url{https://doi.org/10.48550/ARXIV.1806.11330}
\showDOI{\tempurl}


\bibitem[\protect\citeauthoryear{Singh and Anand}{Singh and Anand}{2019}]%
        {singh2019exs}
\bibfield{author}{\bibinfo{person}{Jaspreet Singh} {and}
  \bibinfo{person}{Avishek Anand}.} \bibinfo{year}{2019}\natexlab{}.
\newblock \showarticletitle{EXS: Explainable Search Using Local Model Agnostic
  Interpretability}. In \bibinfo{booktitle}{\emph{Proceedings of the Twelfth
  ACM International Conference on Web Search and Data Mining}} (Melbourne VIC,
  Australia) \emph{(\bibinfo{series}{WSDM '19})}.
  \bibinfo{publisher}{Association for Computing Machinery},
  \bibinfo{address}{New York, NY, USA}, \bibinfo{pages}{770–773}.
\newblock
\showISBNx{9781450359405}
\urldef\tempurl%
\url{https://doi.org/10.1145/3289600.3290620}
\showDOI{\tempurl}


\bibitem[\protect\citeauthoryear{Singh and Anand}{Singh and Anand}{2020}]%
        {singh2020model}
\bibfield{author}{\bibinfo{person}{Jaspreet Singh} {and}
  \bibinfo{person}{Avishek Anand}.} \bibinfo{year}{2020}\natexlab{}.
\newblock \showarticletitle{Model Agnostic Interpretability of Rankers via
  Intent Modelling}. In \bibinfo{booktitle}{\emph{Proceedings of the 2020
  Conference on Fairness, Accountability, and Transparency}} (Barcelona, Spain)
  \emph{(\bibinfo{series}{FAT* '20})}. \bibinfo{publisher}{Association for
  Computing Machinery}, \bibinfo{address}{New York, NY, USA},
  \bibinfo{pages}{618–628}.
\newblock
\showISBNx{9781450369367}
\urldef\tempurl%
\url{https://doi.org/10.1145/3351095.3375234}
\showDOI{\tempurl}


\bibitem[\protect\citeauthoryear{Singh, Hoffart, and Anand}{Singh
  et~al\mbox{.}}{2016a}]%
        {singh2016discovering}
\bibfield{author}{\bibinfo{person}{Jaspreet Singh}, \bibinfo{person}{Johannes
  Hoffart}, {and} \bibinfo{person}{Avishek Anand}.}
  \bibinfo{year}{2016}\natexlab{a}.
\newblock \showarticletitle{Discovering Entities with Just a Little Help from
  You}. In \bibinfo{booktitle}{\emph{Proceedings of the 25th ACM International
  on Conference on Information and Knowledge Management}} (Indianapolis,
  Indiana, USA) \emph{(\bibinfo{series}{CIKM '16})}.
  \bibinfo{publisher}{Association for Computing Machinery},
  \bibinfo{address}{New York, NY, USA}, \bibinfo{pages}{1331–1340}.
\newblock
\showISBNx{9781450340731}
\urldef\tempurl%
\url{https://doi.org/10.1145/2983323.2983798}
\showDOI{\tempurl}


\bibitem[\protect\citeauthoryear{Singh, Khosla, Zhenye, and Anand}{Singh
  et~al\mbox{.}}{2021}]%
        {singh2021extracting}
\bibfield{author}{\bibinfo{person}{Jaspreet Singh}, \bibinfo{person}{Megha
  Khosla}, \bibinfo{person}{Wang Zhenye}, {and} \bibinfo{person}{Avishek
  Anand}.} \bibinfo{year}{2021}\natexlab{}.
\newblock \showarticletitle{Extracting per Query Valid Explanations for
  Blackbox Learning-to-Rank Models}. In \bibinfo{booktitle}{\emph{Proceedings
  of the 2021 ACM SIGIR International Conference on Theory of Information
  Retrieval}} (Virtual Event, Canada) \emph{(\bibinfo{series}{ICTIR '21})}.
  \bibinfo{publisher}{Association for Computing Machinery},
  \bibinfo{address}{New York, NY, USA}, \bibinfo{pages}{203–210}.
\newblock
\showISBNx{9781450386111}
\urldef\tempurl%
\url{https://doi.org/10.1145/3471158.3472241}
\showDOI{\tempurl}


\bibitem[\protect\citeauthoryear{Singh, Nejdl, and Anand}{Singh
  et~al\mbox{.}}{2016b}]%
        {singh2016expedition}
\bibfield{author}{\bibinfo{person}{Jaspreet Singh}, \bibinfo{person}{Wolfgang
  Nejdl}, {and} \bibinfo{person}{Avishek Anand}.}
  \bibinfo{year}{2016}\natexlab{b}.
\newblock \showarticletitle{Expedition: A Time-Aware Exploratory Search System
  Designed for Scholars}. In \bibinfo{booktitle}{\emph{Proceedings of the 39th
  International ACM SIGIR Conference on Research and Development in Information
  Retrieval}} (Pisa, Italy) \emph{(\bibinfo{series}{SIGIR '16})}.
  \bibinfo{publisher}{Association for Computing Machinery},
  \bibinfo{address}{New York, NY, USA}, \bibinfo{pages}{1105–1108}.
\newblock
\showISBNx{9781450340694}
\urldef\tempurl%
\url{https://doi.org/10.1145/2911451.2911465}
\showDOI{\tempurl}


\bibitem[\protect\citeauthoryear{Strohman, Metzler, Turtle, and Croft}{Strohman
  et~al\mbox{.}}{2005}]%
        {strohman2005indri}
\bibfield{author}{\bibinfo{person}{Trevor Strohman}, \bibinfo{person}{Donald
  Metzler}, \bibinfo{person}{Howard Turtle}, {and} \bibinfo{person}{W~Bruce
  Croft}.} \bibinfo{year}{2005}\natexlab{}.
\newblock \showarticletitle{Indri: A language model-based search engine for
  complex queries}. In \bibinfo{booktitle}{\emph{Proceedings of the
  International Conference on Intelligent Analysis}}, Vol.~\bibinfo{volume}{2}.
  \bibinfo{pages}{2--6}.
\newblock


\bibitem[\protect\citeauthoryear{Tan, dos Santos, Xiang, and Zhou}{Tan
  et~al\mbox{.}}{2016}]%
        {tan2016improved}
\bibfield{author}{\bibinfo{person}{Ming Tan}, \bibinfo{person}{Cicero dos
  Santos}, \bibinfo{person}{Bing Xiang}, {and} \bibinfo{person}{Bowen Zhou}.}
  \bibinfo{year}{2016}\natexlab{}.
\newblock \showarticletitle{Improved Representation Learning for Question
  Answer Matching}. In \bibinfo{booktitle}{\emph{Proceedings of the 54th Annual
  Meeting of the Association for Computational Linguistics (Volume 1: Long
  Papers)}}. \bibinfo{publisher}{Association for Computational Linguistics},
  \bibinfo{address}{Berlin, Germany}, \bibinfo{pages}{464--473}.
\newblock
\urldef\tempurl%
\url{https://doi.org/10.18653/v1/P16-1044}
\showDOI{\tempurl}


\bibitem[\protect\citeauthoryear{Thakur, Reimers, R{\"u}ckl{\'e}, Srivastava,
  and Gurevych}{Thakur et~al\mbox{.}}{2021}]%
        {thakur2021beir}
\bibfield{author}{\bibinfo{person}{Nandan Thakur}, \bibinfo{person}{Nils
  Reimers}, \bibinfo{person}{Andreas R{\"u}ckl{\'e}}, \bibinfo{person}{Abhishek
  Srivastava}, {and} \bibinfo{person}{Iryna Gurevych}.}
  \bibinfo{year}{2021}\natexlab{}.
\newblock \showarticletitle{{BEIR}: A Heterogeneous Benchmark for Zero-shot
  Evaluation of Information Retrieval Models}. In
  \bibinfo{booktitle}{\emph{Thirty-fifth Conference on Neural Information
  Processing Systems Datasets and Benchmarks Track (Round 2)}}.
\newblock
\urldef\tempurl%
\url{https://openreview.net/forum?id=wCu6T5xFjeJ}
\showURL{%
\tempurl}


\bibitem[\protect\citeauthoryear{Thorne, Vlachos, Christodoulopoulos, and
  Mittal}{Thorne et~al\mbox{.}}{2018}]%
        {thorne18Fever}
\bibfield{author}{\bibinfo{person}{James Thorne}, \bibinfo{person}{Andreas
  Vlachos}, \bibinfo{person}{Christos Christodoulopoulos}, {and}
  \bibinfo{person}{Arpit Mittal}.} \bibinfo{year}{2018}\natexlab{}.
\newblock \showarticletitle{{FEVER}: a Large-scale Dataset for Fact Extraction
  and {VER}ification}. In \bibinfo{booktitle}{\emph{Proceedings of the 2018
  Conference of the North {A}merican Chapter of the Association for
  Computational Linguistics: Human Language Technologies, Volume 1 (Long
  Papers)}}. \bibinfo{publisher}{Association for Computational Linguistics},
  \bibinfo{address}{New Orleans, Louisiana}, \bibinfo{pages}{809--819}.
\newblock
\urldef\tempurl%
\url{https://doi.org/10.18653/v1/N18-1074}
\showDOI{\tempurl}


\bibitem[\protect\citeauthoryear{van Aken, Winter, L\"{o}ser, and Gers}{van
  Aken et~al\mbox{.}}{2019}]%
        {vanaken2019how}
\bibfield{author}{\bibinfo{person}{Betty van Aken}, \bibinfo{person}{Benjamin
  Winter}, \bibinfo{person}{Alexander L\"{o}ser}, {and}
  \bibinfo{person}{Felix~A. Gers}.} \bibinfo{year}{2019}\natexlab{}.
\newblock \showarticletitle{How Does BERT Answer Questions? A Layer-Wise
  Analysis of Transformer Representations}. In
  \bibinfo{booktitle}{\emph{Proceedings of the 28th ACM International
  Conference on Information and Knowledge Management}} (Beijing, China)
  \emph{(\bibinfo{series}{CIKM '19})}. \bibinfo{publisher}{Association for
  Computing Machinery}, \bibinfo{address}{New York, NY, USA},
  \bibinfo{pages}{1823–1832}.
\newblock
\showISBNx{9781450369763}
\urldef\tempurl%
\url{https://doi.org/10.1145/3357384.3358028}
\showDOI{\tempurl}


\bibitem[\protect\citeauthoryear{V\"{o}lske, Bondarenko, Fr\"{o}be, Stein,
  Singh, Hagen, and Anand}{V\"{o}lske et~al\mbox{.}}{2021}]%
        {volske2021towards}
\bibfield{author}{\bibinfo{person}{Michael V\"{o}lske},
  \bibinfo{person}{Alexander Bondarenko}, \bibinfo{person}{Maik Fr\"{o}be},
  \bibinfo{person}{Benno Stein}, \bibinfo{person}{Jaspreet Singh},
  \bibinfo{person}{Matthias Hagen}, {and} \bibinfo{person}{Avishek Anand}.}
  \bibinfo{year}{2021}\natexlab{}.
\newblock \showarticletitle{Towards Axiomatic Explanations for Neural Ranking
  Models}. In \bibinfo{booktitle}{\emph{Proceedings of the 2021 ACM SIGIR
  International Conference on Theory of Information Retrieval}} (Virtual Event,
  Canada) \emph{(\bibinfo{series}{ICTIR '21})}. \bibinfo{publisher}{Association
  for Computing Machinery}, \bibinfo{address}{New York, NY, USA},
  \bibinfo{pages}{13–22}.
\newblock
\showISBNx{9781450386111}
\urldef\tempurl%
\url{https://doi.org/10.1145/3471158.3472256}
\showDOI{\tempurl}


\bibitem[\protect\citeauthoryear{Wallace, Zhao, Feng, and Singh}{Wallace
  et~al\mbox{.}}{2021}]%
        {wallace2021concealed}
\bibfield{author}{\bibinfo{person}{Eric Wallace}, \bibinfo{person}{Tony Zhao},
  \bibinfo{person}{Shi Feng}, {and} \bibinfo{person}{Sameer Singh}.}
  \bibinfo{year}{2021}\natexlab{}.
\newblock \showarticletitle{Concealed Data Poisoning Attacks on {NLP} Models}.
  In \bibinfo{booktitle}{\emph{Proceedings of the 2021 Conference of the North
  American Chapter of the Association for Computational Linguistics: Human
  Language Technologies}}. \bibinfo{publisher}{Association for Computational
  Linguistics}, \bibinfo{address}{Online}, \bibinfo{pages}{139--150}.
\newblock
\urldef\tempurl%
\url{https://doi.org/10.18653/v1/2021.naacl-main.13}
\showDOI{\tempurl}


\bibitem[\protect\citeauthoryear{Wallat, Singh, and Anand}{Wallat
  et~al\mbox{.}}{2020}]%
        {singh2020bertnesia}
\bibfield{author}{\bibinfo{person}{Jonas Wallat}, \bibinfo{person}{Jaspreet
  Singh}, {and} \bibinfo{person}{Avishek Anand}.}
  \bibinfo{year}{2020}\natexlab{}.
\newblock \showarticletitle{{BERT}nesia: Investigating the capture and
  forgetting of knowledge in {BERT}}. In \bibinfo{booktitle}{\emph{Proceedings
  of the Third BlackboxNLP Workshop on Analyzing and Interpreting Neural
  Networks for NLP}}. \bibinfo{publisher}{Association for Computational
  Linguistics}, \bibinfo{address}{Online}, \bibinfo{pages}{174--183}.
\newblock
\urldef\tempurl%
\url{https://doi.org/10.18653/v1/2020.blackboxnlp-1.17}
\showDOI{\tempurl}


\bibitem[\protect\citeauthoryear{Wang, Shen, Peng, Zhang, Xiao, Liu, Tang,
  Chen, Wu, and Wang}{Wang et~al\mbox{.}}{2022}]%
        {wang2022fine}
\bibfield{author}{\bibinfo{person}{Lijie Wang}, \bibinfo{person}{Yaozong Shen},
  \bibinfo{person}{Shuyuan Peng}, \bibinfo{person}{Shuai Zhang},
  \bibinfo{person}{Xinyan Xiao}, \bibinfo{person}{Hao Liu},
  \bibinfo{person}{Hongxuan Tang}, \bibinfo{person}{Ying Chen},
  \bibinfo{person}{Hua Wu}, {and} \bibinfo{person}{Haifeng Wang}.}
  \bibinfo{year}{2022}\natexlab{}.
\newblock \showarticletitle{A Fine-grained Interpretability Evaluation
  Benchmark for Neural NLP}.
\newblock  (\bibinfo{year}{2022}).
\newblock
\urldef\tempurl%
\url{https://doi.org/10.48550/ARXIV.2205.11097}
\showDOI{\tempurl}


\bibitem[\protect\citeauthoryear{Wiegreffe and Pinter}{Wiegreffe and
  Pinter}{2019}]%
        {wiegreffe2019attention}
\bibfield{author}{\bibinfo{person}{Sarah Wiegreffe} {and}
  \bibinfo{person}{Yuval Pinter}.} \bibinfo{year}{2019}\natexlab{}.
\newblock \showarticletitle{Attention is not not Explanation}. In
  \bibinfo{booktitle}{\emph{Proceedings of the 2019 Conference on Empirical
  Methods in Natural Language Processing and the 9th International Joint
  Conference on Natural Language Processing (EMNLP-IJCNLP)}}.
  \bibinfo{publisher}{Association for Computational Linguistics},
  \bibinfo{address}{Hong Kong, China}, \bibinfo{pages}{11--20}.
\newblock
\urldef\tempurl%
\url{https://doi.org/10.18653/v1/D19-1002}
\showDOI{\tempurl}


\bibitem[\protect\citeauthoryear{Wu, Zhang, Guo, de~Rijke, Fan, and Cheng}{Wu
  et~al\mbox{.}}{2022}]%
        {wu2022prada}
\bibfield{author}{\bibinfo{person}{Chen Wu}, \bibinfo{person}{Ruqing Zhang},
  \bibinfo{person}{Jiafeng Guo}, \bibinfo{person}{Maarten de Rijke},
  \bibinfo{person}{Yixing Fan}, {and} \bibinfo{person}{Xueqi Cheng}.}
  \bibinfo{year}{2022}\natexlab{}.
\newblock \bibinfo{title}{PRADA: Practical Black-Box Adversarial Attacks
  against Neural Ranking Models}.
\newblock
\newblock
\urldef\tempurl%
\url{https://doi.org/10.48550/ARXIV.2204.01321}
\showDOI{\tempurl}


\bibitem[\protect\citeauthoryear{Wu, Mao, Liu, Zhan, Zheng, Zhang, and Ma}{Wu
  et~al\mbox{.}}{2020}]%
        {wu2020leveraging}
\bibfield{author}{\bibinfo{person}{Zhijing Wu}, \bibinfo{person}{Jiaxin Mao},
  \bibinfo{person}{Yiqun Liu}, \bibinfo{person}{Jingtao Zhan},
  \bibinfo{person}{Yukun Zheng}, \bibinfo{person}{Min Zhang}, {and}
  \bibinfo{person}{Shaoping Ma}.} \bibinfo{year}{2020}\natexlab{}.
\newblock \showarticletitle{Leveraging Passage-Level Cumulative Gain for
  Document Ranking}. In \bibinfo{booktitle}{\emph{Proceedings of The Web
  Conference 2020}} (Taipei, Taiwan) \emph{(\bibinfo{series}{WWW '20})}.
  \bibinfo{publisher}{Association for Computing Machinery},
  \bibinfo{address}{New York, NY, USA}, \bibinfo{pages}{2421–2431}.
\newblock
\showISBNx{9781450370233}
\urldef\tempurl%
\url{https://doi.org/10.1145/3366423.3380305}
\showDOI{\tempurl}


\bibitem[\protect\citeauthoryear{Xie and Ermon}{Xie and Ermon}{2019}]%
        {xie2019reparameterizable}
\bibfield{author}{\bibinfo{person}{Sang~Michael Xie} {and}
  \bibinfo{person}{Stefano Ermon}.} \bibinfo{year}{2019}\natexlab{}.
\newblock \showarticletitle{Reparameterizable Subset Sampling via Continuous
  Relaxations}. In \bibinfo{booktitle}{\emph{IJCAI}}.
  \bibinfo{pages}{3919--3925}.
\newblock
\urldef\tempurl%
\url{https://doi.org/10.24963/ijcai.2019/544}
\showURL{%
\tempurl}


\bibitem[\protect\citeauthoryear{Xiong, Dai, Callan, Liu, and Power}{Xiong
  et~al\mbox{.}}{2017}]%
        {xiong2017end}
\bibfield{author}{\bibinfo{person}{Chenyan Xiong}, \bibinfo{person}{Zhuyun
  Dai}, \bibinfo{person}{Jamie Callan}, \bibinfo{person}{Zhiyuan Liu}, {and}
  \bibinfo{person}{Russell Power}.} \bibinfo{year}{2017}\natexlab{}.
\newblock \showarticletitle{End-to-End Neural Ad-Hoc Ranking with Kernel
  Pooling}. In \bibinfo{booktitle}{\emph{Proceedings of the 40th International
  ACM SIGIR Conference on Research and Development in Information Retrieval}}
  (Shinjuku, Tokyo, Japan) \emph{(\bibinfo{series}{SIGIR '17})}.
  \bibinfo{publisher}{Association for Computing Machinery},
  \bibinfo{address}{New York, NY, USA}, \bibinfo{pages}{55–64}.
\newblock
\showISBNx{9781450350228}
\urldef\tempurl%
\url{https://doi.org/10.1145/3077136.3080809}
\showDOI{\tempurl}


\bibitem[\protect\citeauthoryear{Xu, Venugopalan, Ramanishka, Rohrbach, and
  Saenko}{Xu et~al\mbox{.}}{2015}]%
        {xu2015multi}
\bibfield{author}{\bibinfo{person}{Huijuan Xu}, \bibinfo{person}{Subhashini
  Venugopalan}, \bibinfo{person}{Vasili Ramanishka}, \bibinfo{person}{Marcus
  Rohrbach}, {and} \bibinfo{person}{Kate Saenko}.}
  \bibinfo{year}{2015}\natexlab{}.
\newblock \bibinfo{title}{A Multi-scale Multiple Instance Video Description
  Network}.
\newblock
\newblock
\urldef\tempurl%
\url{https://doi.org/10.48550/ARXIV.1505.05914}
\showDOI{\tempurl}


\bibitem[\protect\citeauthoryear{Yang, Yang, Dyer, He, Smola, and Hovy}{Yang
  et~al\mbox{.}}{2016}]%
        {yang2016hierarchical}
\bibfield{author}{\bibinfo{person}{Zichao Yang}, \bibinfo{person}{Diyi Yang},
  \bibinfo{person}{Chris Dyer}, \bibinfo{person}{Xiaodong He},
  \bibinfo{person}{Alex Smola}, {and} \bibinfo{person}{Eduard Hovy}.}
  \bibinfo{year}{2016}\natexlab{}.
\newblock \showarticletitle{Hierarchical Attention Networks for Document
  Classification}. In \bibinfo{booktitle}{\emph{Proceedings of the 2016
  Conference of the North {A}merican Chapter of the Association for
  Computational Linguistics: Human Language Technologies}}.
  \bibinfo{publisher}{Association for Computational Linguistics},
  \bibinfo{address}{San Diego, California}, \bibinfo{pages}{1480--1489}.
\newblock
\urldef\tempurl%
\url{https://doi.org/10.18653/v1/N16-1174}
\showDOI{\tempurl}


\bibitem[\protect\citeauthoryear{Yoon, Jordon, and van~der Schaar}{Yoon
  et~al\mbox{.}}{2019}]%
        {yoon2018invase}
\bibfield{author}{\bibinfo{person}{Jinsung Yoon}, \bibinfo{person}{James
  Jordon}, {and} \bibinfo{person}{Mihaela van~der Schaar}.}
  \bibinfo{year}{2019}\natexlab{}.
\newblock \showarticletitle{{INVASE}: Instance-wise Variable Selection using
  Neural Networks}. In \bibinfo{booktitle}{\emph{International Conference on
  Learning Representations}}.
\newblock
\urldef\tempurl%
\url{https://openreview.net/forum?id=BJg_roAcK7}
\showURL{%
\tempurl}


\bibitem[\protect\citeauthoryear{Zhang, Rudra, and Anand}{Zhang
  et~al\mbox{.}}{2021}]%
        {zhang2021explain}
\bibfield{author}{\bibinfo{person}{Zijian Zhang}, \bibinfo{person}{Koustav
  Rudra}, {and} \bibinfo{person}{Avishek Anand}.}
  \bibinfo{year}{2021}\natexlab{}.
\newblock \showarticletitle{Explain and Predict, and Then Predict Again}. In
  \bibinfo{booktitle}{\emph{Proceedings of the 14th ACM International
  Conference on Web Search and Data Mining}} (Virtual Event, Israel)
  \emph{(\bibinfo{series}{WSDM '21})}. \bibinfo{publisher}{Association for
  Computing Machinery}, \bibinfo{address}{New York, NY, USA},
  \bibinfo{pages}{418–426}.
\newblock
\showISBNx{9781450382977}
\urldef\tempurl%
\url{https://doi.org/10.1145/3437963.3441758}
\showDOI{\tempurl}


\bibitem[\protect\citeauthoryear{Zhong, Shao, and McKeown}{Zhong
  et~al\mbox{.}}{2019}]%
        {zhong2019fine}
\bibfield{author}{\bibinfo{person}{Ruiqi Zhong}, \bibinfo{person}{Steven Shao},
  {and} \bibinfo{person}{Kathleen McKeown}.} \bibinfo{year}{2019}\natexlab{}.
\newblock \bibinfo{title}{Fine-grained Sentiment Analysis with Faithful
  Attention}.
\newblock
\newblock
\urldef\tempurl%
\url{https://doi.org/10.48550/ARXIV.1908.06870}
\showDOI{\tempurl}


\bibitem[\protect\citeauthoryear{Zhuang and Zuccon}{Zhuang and Zuccon}{2021}]%
        {zhuang2021tilde}
\bibfield{author}{\bibinfo{person}{Shengyao Zhuang} {and}
  \bibinfo{person}{Guido Zuccon}.} \bibinfo{year}{2021}\natexlab{}.
\newblock \showarticletitle{TILDE: Term Independent Likelihood MoDEl for
  Passage Re-Ranking}. In \bibinfo{booktitle}{\emph{Proceedings of the 44th
  International ACM SIGIR Conference on Research and Development in Information
  Retrieval}} (Virtual Event, Canada) \emph{(\bibinfo{series}{SIGIR '21})}.
  \bibinfo{publisher}{Association for Computing Machinery},
  \bibinfo{address}{New York, NY, USA}, \bibinfo{pages}{1483–1492}.
\newblock
\showISBNx{9781450380379}
\urldef\tempurl%
\url{https://doi.org/10.1145/3404835.3462922}
\showDOI{\tempurl}


\end{thebibliography}
